\documentstyle[11pt,epsfig,twoside]{article}
\epsfverbosetrue
\newcommand{\beq}{\begin{equation}}
\newcommand{\eeq}{\end{equation}}
\newcommand{\beqn}{\begin{eqnarray}}
\newcommand{\eeqn}{\end{eqnarray}}
\newcommand{\beqns}{\begin{eqnarray*}}
\newcommand{\eeqns}{\end{eqnarray*}}
\newcommand{\vs}{\\[0.3cm]\noindent}

\newcommand{\hssm}{\hspace{-0.025cm}}
\newcommand{\intl}{\int\limits}

\newcommand{\mc}{\multicolumn}

\newcommand{\hmm}{\hspace{-0.2cm}}
\newcommand{\hmmm}{\hspace{-0.3cm}}
\newcommand{\ft}{\footnotesize}
\def\rs{\raisebox{1.3ex}[-1.5ex]}

\newcommand{\GF}{G_{\rm F}}

\def\MSbar{\overline{\rm MS}}
\def\ea{{\em et al.}}
\def\CoL{Collaboration}
\def\tho{{\rm theo}}
\def\exp{{\rm exp}}
\def\mod{{\rm mod}}
\def\opt{{\rm opt}}
\def\QCD{{\rm QCD}}
\def\min{{\rm min}}
\def\max{{\rm max}}
\def\Min{{\rm Min}}
\def\Max{{\rm Max}}
\def\syst{{\rm syst}}

\def\rhoeta{(\bar\rho,\,\bar\eta)}
\def\arhoeta{\{\bar\rho,\,\bar\eta\}}
\def\rhobar{\bar\rho}
\def\etabar{\bar\eta}

\def\sta{{\rm sin}2\alpha}
\def\stb{{\rm sin}2\beta}

\def\stbwa{{\rm sin}2\beta_{\rm WA}}
\def\ttg{{\rm tan}\gamma}
\def\stastb{(\sta,\,\stb)}
\def\astastb{\{\sta,\,\stb\}}
\def\stagam{(\sta,\,\gamma)}
\def\stbgam{(\stb,\,\gamma)}

\def\xo{x_0}
\def\xobar{\bar\xo}
\def\xexp{x_{\exp}}
\def\xthe{x_{\tho}}
\def\ymod{y_{\mod}}
\def\ymodopt{y_{\mod}^{\opt}}
\def\ythe{y_{\tho}}
\def\yQCD{y_{\QCD}}
\def\yNP{y_{\rm NP}}

\def\sigexp{\sigma_{\exp}}
\def\sigxo{\sigma_o}
\def\Lik{{\cal L}}
\def\Hatsyst{{\cal L}_{\rm syst}}
\def\expHatsyst{_{\rm exp}{\cal L}_{\rm syst}}
\def\F{{\cal F}}
\def\Ndof{N_{\rm dof}}
\def\VCKM{{V}}

\def\Likexp{{\cal L}_{\exp}}
\def\Likthe{{\cal L}_{\tho}}
\def\J{{J}}
\def\a{a}
\def\Mu{\mu}
\def\ChiMinGlob{\chi^2_{\min ;\ymod}}
\def\Chi2MinaMu{\chi^2_{\min ;\a,\Mu}}
\def\Chi2MinMu{\chi^2_{\min ;\Mu}(a)}
\def\Prob{{\cal P}}
\def\ProbCERN{{\rm Prob}}

\def\Nthe{N_{\tho}}
\def\Nmod{N_{\mod}}
\def\NQCD{N_{\QCD}}
\def\Ncst{N_{\rm cst}}
\def\Neff{N_{\exp}^{\rm eff}}
\def\Na{N_{\a}}
\def\Nmu{N_{\Mu}}
\def\Nexp{N_{\exp}}

\def\NNP{N_{\rm NP}}
\def\Vud{|V_{ud}|}
\def\Vus{|V_{us}|}
\def\Vub{|V_{ub}|}
\def\Vcb{|V_{cb}|}
\def\Vcd{|V_{cd}|}
\def\Vubcb{\left|\frac{V_{ub}}{V_{cb}}\right|}
\def\VuboverVcb{|V_{ub}/V_{cb}|}
\def\Vubthe{|V_{ub}^\tho|}
\def\Vcbthe{|V_{cb}^\tho|}
\def\dmd{\Delta m_d}
\def\dms{\Delta m_s}
\def\DmBd{\dmd}
\def\DmBs{\dms}

\def\fbd{f_{B_d}}
\def\fbs{f_{B_s}}
\def\fbdbd{f_{B_d}\sqrt{B_d}}
\def\fbsbs{f_{B_s}\sqrt{B_s}}
\def\epsk{\epsilon_K}
\def\epe{\epsilon^\prime/\epsilon_K}
\def\kdo{{K^0-\bar{K^0}}}
\def\bdo{{B_d^0-\bar{B_d^0}}}
\def\bso{{B_s^0-\bar{B_s^0}}}
\def\CL{{\rm CL}}
\def\CLs{{\rm CLs}}
\def\CLcont{{\rm CL}_{\rm cont}}
\def\CLcut{{\rm CL}_{\rm cut}}

\def\corcof{{c}}
\def\babar{$\mbox{\sl B\hspace{-0.4em} {\small\sl A}\hspace{-0.37em} \sl B\hspace{-0.4em} {\small\sl A\hspace{-0.02em}R}}$}
\def\babarem{$\mbox{\sl B\hspace{-0.4em} {\small\sl A}\hspace{-0.37em} \sl B\hspace{-0.4em} {\small\sl A\hspace{-0.02em}R}}$}
\def\tinyfig{7.5cm}
\def\smallfig{11.2cm}
\def\mediumfig{11.8cm}
\def\largefig{13.3cm}

\def\ckmfit{{\em R}fit}
\def\ckmfoot{{\em ER}fit}
\def\CkmFitter{{\em CkmFitter}}
\def\MINUIT{{\em MINUIT}}

\def\ie{{\em i.e.}} 
\def\cf{{\em c.f.}} 
\def\eg{{\em e.g.}} 
\def\via{{\em via} }

\def\babar{{B\footnotesize\hssm A\normalsize\hssm B\footnotesize\hssm AR}}
\def\babarEm{{\em B\footnotesize\em\hssm A\normalsize\em\hssm B\footnotesize\em\hssm AR}}

\begin{document}

\begin{titlepage}
\setcounter{page}{1}

\begin{flushright} 
LAL 01-06 \\
hep-ph/0104062 \\
% \babar\ Analysis Document \#153\\
April 2001
\end{flushright} 

\begin{center}
\vspace{1.5cm}
{\large\bf
A NEW APPROACH TO A GLOBAL FIT 
%\\[0.1cm] 
OF THE CKM MATRIX  \\[1.0CM]
%{\footnotesize 
%	RESULTS 
%	INCLUDING AN INTERPRETATION
%	OF THE NEW \boldmath$\stb$ MEASUREMENTS}
%}\\[0.8cm] 
}
\vspace{0.5cm}
\begin{large}
A.~H\"ocker$^{\rm a}$, H.~Lacker$^{\rm a}$, 
S.~Laplace$^{\rm a}$ and
F.~Le Diberder$^{\rm b}$ \\
\end{large}
\vspace{1.cm}
{\small \em $^{\rm a}$Laboratoire de l'Acc\'el\'erateur Lin\'eaire,\\
IN2P3-CNRS et Universit\'e de Paris-Sud, BP 34, F-91898 Orsay Cedex, France}\\
\vspace{0.2cm}
{\small \em $^{\rm b}$Laboratoire de Physique Nucl\'eaire et 
		des Hautes Energies, \\
	4, place Jussieu, Tour 33 r.d.c., 
	F-75005 Paris, Cedex 05, France}
\vspace{4.5cm}

{\small{\bf Abstract}}
\end{center}
{%\small
\vspace{-0.2cm}
We report on a new approach to a global CKM matrix analysis 
taking into account most recent experimental and theoretical 
results. The statistical framework (\ckmfit) developed in this 
paper advocates frequentist statistics. Other approaches, 
such as Bayesian statistics or the 95\%~CL scan method are
also discussed. We emphasize the distinction of
a model testing and a model dependent, metrological phase 
in which the various parameters of the theory are estimated.
Measurements and theoretical parameters entering the global
fit are thoroughly discussed, in particular with respect
to their theoretical uncertainties. 
Graphical results for confidence levels are drawn 
in various one and two-dimensional 
parameter spaces. Numerical results are provided for
all relevant CKM parameterizations, the CKM elements
and theoretical input parameters. Predictions for
branching ratios of rare $K$ and $B$ meson decays
are obtained. A simple, predictive SUSY extension of 
the Standard Model is discussed.
}

\vspace{3.5cm}
\vfill
\centerline{\small\em (Accepted for publication in The European Physical Journal C)}
\vspace{1cm}
\thispagestyle{empty}

\end{titlepage}

\newpage\thispagestyle{empty}{\tiny\beqns\eeqns}\newpage
\setcounter{page}{1}

{
%\scriptsize
\normalsize
\tableofcontents
\hfill
}

%
% --------------------- Introduction ---------------------------------
%
\section{Introduction}

Within the Standard Model (SM), CP violation is generated by
a single non-vanishing phase in the unitary 
Cabibbo-Kobayashi-Maskawa (CKM) quark mixing matrix 
$\VCKM$~\cite{cabibbo,kmmatrix}.
A useful parameterization~\cite{wolfenstein,buras} follows from the 
observation that the elements of $\VCKM$ exhibit a hierarchy
in terms of the parameter $\lambda=|V_{us}|$. Other
parameters of this representation are $A$, $\rho$ and $\eta$, 
where CP violation necessarily requires $\eta\ne0$.
%\footnote
%{
%  The Wolfenstein CKM matrix approximation used in this 
%  analysis is valid to the order $O(|\lambda|^6)\simeq0.01\%$.
%}.
%The allowed region in $\rho-\eta$ space can be elegantly 
%displayed using the {\em Unitarity Triangle} (UT) described
%by the (rescaled) unitarity relation between the first and 
%the third column
%\beq
%\label{eq_utriangle}
%   \frac{V_{ud}V_{ub}^*}{V_{cd}V_{cb}^*} 
%	+ 1 
%	+ \frac{V_{td}V_{tb}^*}{V_{cd}V_{cb}^*} = 0~.
%\eeq
%
The parameters $\lambda$ and $A$ are 
obtained from measurements of semileptonic decay rates of 
$K$ mesons and $B$ meson decays involving $b\rightarrow c$ 
transitions, respectively. Constraints on $\rho$ and $\eta$ 
are obtained from measurements of semileptonic 
$B$ decays yielding $|V_{ub}|$ and the ratio 
$|V_{ub}/V_{cb}|$. Standard Model predictions of $B^0_d$ and 
$B^0_s$ oscillations, and of indirect CP violation in the 
neutral kaon sector, depend on CKM parameters; therefore
measurements of these observables provide constraints in the 
$\rho-\eta$ plane, albeit being limited by theoretical 
uncertainties coming mainly from long distance QCD effects.
Finally, in the era of the B-factories, it will be possible, for 
the first time, to assess the CP-violating angles $\alpha$, 
$\beta$ and $\gamma$ of the  {\em Unitarity Triangle} (UT)
expressing the unitarity relation between the first and 
the third column of $\VCKM$.
\vs
The first goal of a global CKM fit is to {\it probe the 
validity of the SM}, that is to quantify the agreement between 
the SM and the experimental information.
Furthermore, one intends to perform a detailed {\it Metrology}, 
that is to find allowed ranges for CKM matrix elements
and related quantities, assuming the SM to be correct. 
Finally, within an extended theoretical framework, one may 
search for specific signals of {\em new physics}, by estimating 
the additional theoretical parameters.
\vs
Analyzing data in a well defined theoretical framework 
ceases to be a straightforward task when one 
moves away from Gaussian statistics. This is the case
for the theoretically limited precision on the SM
predictions of the neutral $K$ and $B$ mixing 
observables and, to a lesser extent, for the 
semileptonic decay rates
of $B$ decays to charmed and charmless final states.
The statistical approach (\ckmfit) developed in this analysis
allows a non-Bayesian treatment of the, {\em a priori}
unknown (\ie, not statistically distributed), theoretical
parameters and theoretical systematics
of measurements.  The ensemble of the
statistical analysis is realized in the program package 
\CkmFitter\footnote
{
  \CkmFitter\ is a framework package that hosts several statistical 
  approaches to a global CKM fit, such as \ckmfit, Bayesian 
  techniques and the 95\% CL Scan
  method. It is available as public share ware. Please, contact
  the authors for more information.
}. 
A detailed description of the methods it uses, with emphasis 
on the new method denoted \ckmfit\ which is proposed here, and 
the presentation of state-of-the-art results are the subject of 
this paper\footnote
{
   Visit the \CkmFitter\ web page to find plots,
   reference links, detailed descriptions and more:\\
   http://www.slac.stanford.edu/$\sim$laplace/ckmfitter.html
}.
\vs
The paper is organized as follows: after recalling the most
common CKM parameterizations, we comprehensively discuss the
statistical framework of the analysis, starting with the
introduction of the relevant likelihoods in 
Section~\ref{TheLikelihoodFunction}, followed by a 
definition of the three analysis steps: metrology, model
testing and probing for new physics. We then recall the principles 
of the 95\%~CL Scan scheme~\cite{schune} and of the 
Bayesian approach~\cite{Achille1,Achille2}
in Section~\ref{AlternativeStatisticalTreatments}
(see also Refs.~\cite{otherCkm} and references therein 
for a tentative collection of publications on the 
CKM matrix and related topics). We work out their limitations 
and motivate further going ideas, while never leaving
non-Bayesian grounds. This is followed by a discussion
of the treatment of experimental and theoretical 
systematics in Section~\ref{sec_likelihoodsAndSysErrors}.
In Section~\ref{sec_fitIngredients} we present a 
compendium of the input measurements, their predictions
in the framework of the SM, and discuss the theoretical 
parameters and their uncertainties used in the analysis. 
We display our fit results as confidence 
levels in various parameter spaces in 
Section~\ref{sec_constrainedFits} and produce  
tables of constraints on all relevant CKM parameters,
constraining measurements and theoretical inputs,
and predictions of rare $K$ and $B$ decays.
Within our statistical approach, we perform a test of
the goodness of the theory and discuss the effect of
a simple, predictive Minimal Supersymmetric extension
of the SM. Deepening statistical discussions on some
crucial issues of the analysis are given in the appendix.

%\vfill
%\pagebreak
%
% ---------------- The CKM Matrix --------------------------
%
\section{The CKM Matrix}

Invariance under local gauge transformation prevents the bare 
masses of the leptons and quarks to appear in the 
$SU(3)\times SU(2)\times U(1)$ Lagrange density of the SM.
Instead, the spontaneous breakdown of electroweak symmetry 
dynamically generates masses for the fermions due to the Yukawa 
coupling of the fermion fields to the Higgs 
doublet. Since the latter has a non-vanishing vacuum 
expectation value, the Yukawa couplings $g$ give rise to 
the $3\times3$ mass matrices
\beq
\label{eq_mumd}
   M_i=\frac{v g_i}{\sqrt{2}}~,
\eeq
with $i=u(d)$ for up(down)-type quarks and $i=e$ for 
the massive leptons. The transformation of the $M_i$ 
from the basis of the flavor eigenstates to the basis 
of the mass eigenstates is realized by unitary 
rotation matrices $U_i$, where
\beq
	U_{u(d,e)} M_{u(d,e)} U_{u(d,e)}^\dag 
	= {\rm diag}\left(m_{u(d,e)},
			  m_{c(s,\mu)},
			  m_{t(b,\tau)}
		     \right)~.
\eeq
For the Lagrange density in the basis of the mass-eigenstates
the neutral-current part remains unchanged (\ie, there are 
no flavor changing neutral currents present at tree level), 
whereas the charged current part of the quark sector is 
modified by the product of the up-type and down-type quark 
mass matrices,
\beq
	\VCKM = U_u U_d^\dag~,
\eeq
which is the CKM mixing matrix.
By convention, $\VCKM$ operates on the $-1/3$ charged 
down-type quark mass eigenstates
\beq
\label{eq_ckm1}
\VCKM = \left(
	\begin{array}{ccc}
	V_{ud} & V_{us} & V_{ub} \\
	V_{cd} & V_{cs} & V_{cb} \\
	V_{td} & V_{ts} & V_{tb} \\
	\end{array}
	\right)
\eeq
and, being the product of unitary matrices, $\VCKM$ itself 
is unitary:
\beq
\label{eq_unitarity}
	\VCKM \VCKM^\dag={\rm Id}~.
\eeq
There exists a hierarchy between the elements of $\VCKM$ both for 
their value (the diagonal elements dominate) and their errors
(since they dominate, they are better known). Unitarity and 
the phase arbitrariness of fields reduce the initially nine
complex parameters of $\VCKM$ to three real numbers 
and one phase, where the latter accounts for CP violation.
It is therefore interesting to over-constrain $\VCKM$
since deviations from unitarity would reveal the existence of new 
generation(s) or new couplings.
\vs
The charged current couplings among left-handed quark fields 
are proportional to the elements of $\VCKM$. For right-handed
quarks there exist no $W$ boson interaction in the SM and
the $Z$, photon and gluon couplings are flavor diagonal.
For left-handed leptons the analysis proceeds similar
to the quarks.
% with the notable difference that, since
%the neutrinos are (almost) massless, one can choose to 
%make the same unitary transformation on the left-handed
%charged leptons and neutrinos so that the analog of $\VCKM$
%in the lepton sector becomes the unit matrix.
\vs
There are many ways of parameterizing the CKM matrix in terms of four 
parameters. It is the purpose of this section to summarize the
most popular representations.

\subsection{The Standard Parameterization}

The Standard Parameterization of $\VCKM$ is 
taken to be the one proposed by Chau and Keung~\cite{chau}, 
and advocated by the PDG~\cite{pdg2000}.
It is obtained by the product
of three complex rotation matrices, where the rotations are 
characterized by the Euler angles $\theta_{12},~\theta_{13}$ 
and $\theta_{23}$, which are the mixing angles
between the generations, and an overall phase $\delta$:
\beq
\label{eq_ckmPdg}
\VCKM = \left(
	\begin{array}{ccc}
	c_{12}c_{13}	
		&    s_{12}c_{13}   
			&   s_{13}e^{-i\delta}  \\
	-s_{12}c_{23}-c_{12}s_{23}s_{13}e^{i\delta} 
		&  c_{12}c_{23}-s_{12}s_{23}s_{13}e^{i\delta} 
			& s_{23}c_{13} \\
	s_{12}s_{23}-c_{12}c_{23}s_{13}e^{i\delta}  
		&  -c_{12}s_{23}-s_{12}c_{23}s_{13}e^{i\delta} 
			& c_{23}c_{13} 
	\end{array}
	\right)
\eeq
where $c_{ij}={\rm cos}\theta_{ij}$, 
$s_{ij}={\rm sin}\theta_{ij}$ for $i<j=1,2,3$. 
This parameterization has the considerable advantage of being 
exact in the sense that it strictly satisfies the unitarity 
relation~(\ref{eq_unitarity}). 

\subsection{The Wolfenstein Parameterization}

Following the observation of a hierarchy between the different
matrix elements, Wolfenstein~\cite{wolfenstein} proposed a simple 
expansion of the CKM matrix in terms of the four parameters $\lambda$, 
$A$, $\rho$ and $\eta$ ($\lambda=|V_{us}|\sim 0.22$ being the expansion 
parameter), which is widely used in contemporary literature. Using
the convention of Ref.~\cite{buras} one has
{\small
\beq
\label{eq_ckmWolf}
\VCKM \simeq \left(
	\begin{array}{ccc}
	1-\frac{1}{2}\lambda^2-\frac{1}{8}\lambda^4  
		& \lambda 
			& A\lambda^3(\rho-i\eta) \\
	-\lambda\left[ 1+A^2\lambda^4
			\left(\rho+i\eta -\frac{1}{2}\right)
                 \right] 
		& 1-\frac{1}{2}\lambda^2-\frac{1}{8}\lambda^4(1+4A^2) 
			& A\lambda^2 \\
	A\lambda^3 \left[ 1 - \left(\rho+i\eta\right)
			      \left(1-\frac{1}{2}\lambda^2\right)
                   \right]
		& -A\lambda^2 \left[ 1+\lambda^2
         		\left(\rho+i\eta -\frac{1}{2}\right)
                               \right]
			& 1-\frac{1}{2}A^2\lambda^4
	\end{array}
	\right)
\eeq
}
It is obtained from Eq.~(\ref{eq_ckmPdg}) \via\ the definitions
\beqn
	s_{12} 		   &=& \lambda~,\nonumber \\ 
	s_{23} 		   &=& A\lambda^2~, \\
	s_{13}e^{-i\delta} &=& A\lambda^3(\rho -i\eta)~,\nonumber
\eeqn
and is valid to the order $O(|\lambda|^6)\simeq0.01\%$.

\subsection{Phase Invariance}

It was shown by Jarlskog~\cite{jarlskog} that the
determinant of the commutator of the up-type and 
down-type unitary mass matrices~(\ref{eq_mumd}) reads
\beq
\label{eq_commu}
	{\rm det}[M_u,M_d] = -2 i F_u F_d J~,
\eeq
with $F_u$, $F_d$, being
\beq
	F_{u(d)} = (m_{t(b)}-m_{c(s)})(m_{t(b)}-m_{u(d)})
  		       (m_{c(s)}-m_{u(d)})/m_{t(b)}^3~.
\eeq
The phase-convention independent measure of CP violation, 
$J$, is given by
\beq
\label{eq_jarlskog}
{\rm Im}\left[V_{ij}V_{kl}V_{il}^*V_{kj}^*\right]
	= J \sum_{m,n=1}^3 \epsilon_{ikm}\epsilon_{jln}~,
\eeq
with the CKM matrix elements $V_{ij}$ and $\epsilon_{ikm}$ 
being the total antisymmetric tensor. One 
representation of Eq.~(\ref{eq_jarlskog}) reads, for 
instance, $J={\rm Im}[V_{ud}V_{cs}V_{us}^*V_{cd}^*]$. 
A non-vanishing CKM phase and hence CP violation 
necessarily requires $J\ne0$. 
The Jarlskog parameter expressed in the Standard 
Parameterization~(\ref{eq_ckmPdg}) reads
\beq
J = c_{12}c_{23}c_{13}^2s_{12}s_{23}s_{13}{\rm sin}\delta~,
\eeq
and, using the Wolfenstein approximation~(\ref{eq_ckmWolf}), 
valid to the order $O(|\lambda|^{10})$, one finds
\beq
	J = A^2 \lambda^6 \eta~\sim10^{-5}~.
\eeq
The empirical value of $J$ is small compared to its 
maximum of $1/(6\sqrt{3})\simeq 0.1$ showing
that CP violation is suppressed as a consequence of the strong 
hierarchy exhibited by the CKM matrix elements.
It is the remarkable outcome of Eq.~(\ref{eq_commu}) that 
CP violation requires not only $J$ to be non-zero, but 
also the existence of a non-degenerated mass hierarchy. Equal 
masses between at least two generations of up-type or down-type 
quarks would necessarily remove the CKM phase.
\vs
Phase convention invariance of the $\VCKM$-transformed quark
wave functions is a requirement for physically meaningful 
quantities. Such invariants are the moduli $|V_{ij}|^2$ and
the quadri-products $V_{ij}V_{kl}V_{il}^*V_{kj}^*$ (\cf, 
the Jarlskog invariant $J$~(\ref{eq_jarlskog})). Non-trivial
higher invariants can be reformulated as functions of moduli 
and quadri-products~(see, \eg, Ref.~\cite{CPV-TheBook}).
Indeed, Eq.~(\ref{eq_jarlskog}) expresses the fact that, owing to 
the orthogonality of any pair of different rows or columns
of $\VCKM$, the imaginary parts of all quadri-products are 
equal up to their sign. We will use phase-invariant representations
and formulae throughout this paper~\footnote
{
	We are indebted to K.~Schubert for drawing our attention
	to this point.
}.

\subsection{The Unitarity Triangle}

The allowed region in $\rho-\eta$ space can be elegantly 
displayed using the unitarity triangle (UT) described
by the {\em rescaled} unitarity relation between the first 
and the third column of the CKM matrix
\beq
\label{eq_utriangle}
   \frac{V_{ud}V_{ub}^*}{V_{cd}V_{cb}^*} 
	+ \frac{V_{cd}V_{cb}^*}{V_{cd}V_{cb}^*} 
	+ \frac{V_{td}V_{tb}^*}{V_{cd}V_{cb}^*} = 0~.
\eeq
Note that twice the area of the {\em non-rescaled}
\begin{figure}[t]
  \epsfxsize\tinyfig
  \centerline{\epsffile{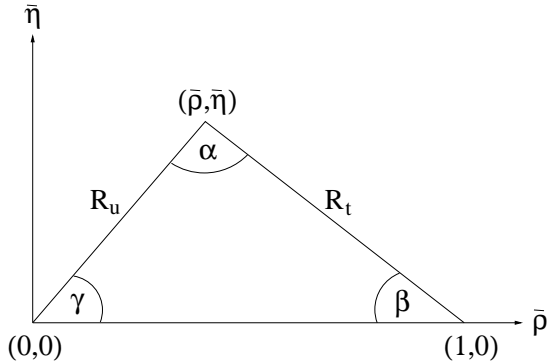}}
  \vspace{-0.0cm}
  \caption[.]{\label{fig_utriangle}\em
	The rescaled Unitarity Triangle in the Wolfenstein 
	parameterization.}
\end{figure}
UT corresponds to the Jarlskog parameter $J$. This identity
provides a geometrical interpretation of the phase convention
invariance of $J$: a rotation of the CKM matrix rotates the 
UT accordingly while leaving its area, and thus $J$, invariant.
It is the remarkable property of the UT~(\ref{eq_utriangle})
that its three sides are governed by the same power of $\lambda$
and $A$
\beq
	\frac{A\lambda^3}{A\lambda^3}
	+ 1
	+ \frac{A\lambda^3}{A\lambda^3} \simeq 0~,
\eeq
which predicts large CP asymmetries in the $B$ sector~\cite{sandabook}. 
As a comparison, the corresponding UT for the kaon sector is 
almost flat
\beq
   0 = \frac{V_{ud}V_{us}^*}{V_{cd}V_{cs}^*} 
	+ \frac{V_{cd}V_{cs}^*}{V_{cd}V_{cs}^*} 
	+ \frac{V_{td}V_{ts}^*}{V_{cd}V_{cs}^*} 
	\simeq
	\frac{\lambda}{\lambda}
	+ 1
	+ \frac{A^2\lambda^5}{\lambda}~,
\eeq
exhibiting small CP asymmetries.
\begin{figure}[t]
  \epsfxsize\mediumfig
  \centerline{\epsffile{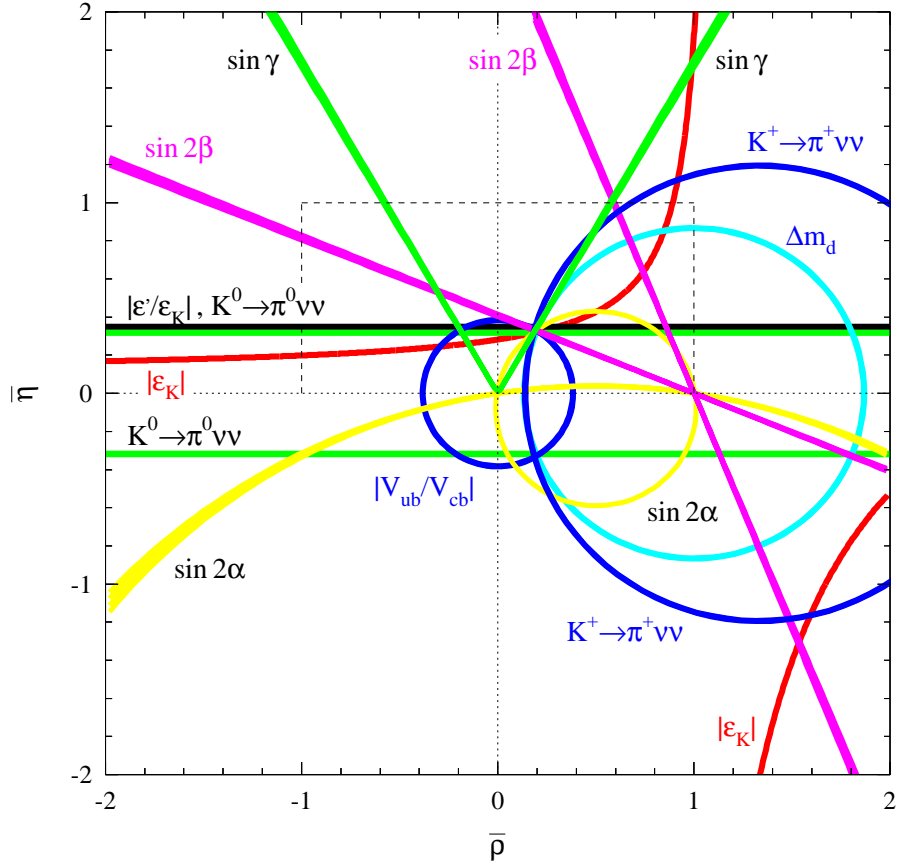}}
  \vspace{-0.0cm}
  \caption[.]{\label{fig_rhoetaall}\em
	Constraints in the $\rhoeta$ plane for the most relevant
	observables. The theoretical parameters used correspond to some
	``standard'' set chosen to reproduce compatibility. The
	dashed lines indicate the rectangle on which we concentrate
	in the following for the global fit.}
\end{figure}
The UT~(\ref{eq_utriangle}) is sketched in Fig.~\ref{fig_utriangle}
in the complex $\rhoeta$ plane ($\rhobar=\rho(1-\lambda^2/2)$, 
$\etabar=\eta(1-\lambda^2/2)$) of the Wolfenstein 
parameterization\footnote
{
   The length of the vector to the triangle apex is 
   given by
   $
%   \beq
	\left|1+\frac{V_{ud}V_{ub}^*}{V_{cd}V_{cb}^*}\right|
	= \sqrt{\rho^2+\eta^2}
	  \left(1-\frac{\lambda^2}{2}\right)
	  + O(|\lambda|^4)~,
%	 \nonumber
%   \eeq
   $
   so that the replacements $\rho\rightarrow\rho(1-\lambda^2/2)$ 
   and $\eta\rightarrow\eta(1-\lambda^2/2)$, where
   $V_{ud} = 1-\lambda^2/2-O(\lambda^4)$, improves the 
   precision of the apex coordinate in the Wolfenstein
   approximation~\cite{buras}.
}.
The sides $R_u$ and $R_t$ (the third side being normalized to unity) 
are given by
\beqn
R_u &=& 
	\left|\frac{V_{ud}V_{ub}^*}{V_{cd}V_{cb}^*} \right|
		\;=\; \sqrt{\rhobar^2+\etabar^2}~, \\
\label{eq_rt}
R_t &=& 
	\left|\frac{V_{td}V_{tb}^*}{V_{cd}V_{cb}^*}\right| 
		\;=\; \sqrt{(1-\rhobar)^2+\etabar^2}~. 
\eeqn
and the three angles, $\alpha,~\beta,~\gamma$, read
\beq
\alpha = {\rm arg}\left[ - \frac{V_{td}V_{tb}^*}{V_{ud}V_{ub}^*} \right] 
~,\hspace{0.5cm}
\beta  = {\rm arg}\left[ - \frac{V_{cd}V_{cb}^*}{V_{td}V_{tb}^*} \right] 
~,\hspace{0.5cm}
\gamma = {\rm arg}\left[ - \frac{V_{ud}V_{ub}^*}{V_{cd}V_{cb}^*} \right]~,
\eeq
where $\gamma\equiv\delta$ in the Standard Parameterization.
The angles and sides of the UT obey the trigonometric relation,
${\rm sin}\alpha:{\rm sin}\beta:{\rm sin}\gamma=1:R_u:R_t$.

The relations between the angles and the $\rhobar$, $\etabar$ 
coordinates are given by
%\footnote
%{
% It is instructive to display the SM constraints
% in the correlation planes of the measured UT angles.
% The inverse of the relations~(\ref{eq_sta})-(\ref{eq_gam}),
% valid in the relevant parts of the quadrants $-1\le\sta\le1$, 
% $0\le\stb\le1$ and $0\le\tg\le\infty$, are given by
% \beqn
% \label{eq_inv_sasb}
% \rhobar_{\pm}(\sa,\,\sb)
% 	&=& \frac{1}{2} \pm\left(\frac{1}{4}
% 		+ \etabar(\sa,\,\sb)\,
%	\left(\frac{1}{\sa} - \etabar(\sa,\,\sb)\right) 
%		- \frac{1}{\sa}\sqrt{(1 - \sasq)\,\etabar^2(\sa,\,\sb)}
%	\right)^{\!\frac{1}{2}}	\\
% \etabar(\sa,\,\sb) 
%	&=& \frac{1}{2\sa} 
%	+ \frac{1}{2}
%	\left(\sb - \frac{1}{\sa}
%	\left(\sqrt{1-\sbsq} + \sqrt{(1-\sasq)(2-\sbsq-2\sqrt{1-\sbsq})}
%	\right)
%	\right) 				\nonumber 	\\
% \label{eq_inv_satg}
% \rhobar(\sa,\,\tg)
% 	&=& \frac{ \sa + \tg\,(1 - \sqrt{1 - \sasq}) }
%		 { \sa \,(1 + \tgsq) }      	\\
% \etabar(\sa,\,\tg)
%	&=& \rhobar(\sa,\,\tg) \,\tg 	\nonumber	\\
% \label{eq_inv_sbtg}
% \rhobar(\sb,\,\tg)
%	&=& \frac{ \sb + \tg\,(1 - \sqrt{1 - \sbsq}) }
%		 { \sb \,(1 + \tgsq) + 2\tg}      	\\
% \etabar(\sb,\,\tg)
%	&=& \rhobar(\sb,\,\tg)\, \tg 	\nonumber
% \eeqn
% with $\sa\equiv\sta$, $\sb\equiv\stb$ and $\tg\equiv\ttg$. The sign
% in Eq.~(\ref{eq_inv_sasb}) is positive, if $\sa<0$ and 
% $\sb>{\rm cos}\alpha$, while it is negative in all other cases.
%}
\beqn
\label{eq_sta}
\sta 	&=& 
	\frac{2\etabar(\etabar^2-\rhobar(1-\rhobar))}
	     {(\etabar^2+(1-\rhobar)^2)(\etabar^2+\rhobar^2)}~, \\
\label{eq_stb}
\stb 	&=& 
	\frac{2\etabar(1-\rhobar)}{\etabar^2+(1-\rhobar)^2}~,\\
\label{eq_gam}
\ttg 			&=& 
	\frac{\etabar}{\rhobar}~.
\eeqn

A graphical compilation of the most relevant present and future
constraints sensitive to the CP violating phase $\delta$ is
displayed in Fig.~\ref{fig_rhoetaall}. 
We simplify the representation by assuming 
a measurement of $\sta$ whereas, in principle, the UT angle 
$\alpha$ can be directly determined from $B\rightarrow3\pi$ 
decays. For the third UT angle $\gamma$, we assume a 
measurement of ${\rm sin}\gamma$, even though charmless $B$ decays
may allow a non-ambiguous determination of $\gamma$.
A more detailed elaboration of future measurements is given
in Ref.~\cite{heikobcp4}. Some ``standard'' set 
of theoretical parameters is used for this exercise
in order to reproduce compatibility between the constraints. The 
present experimental values for the observables and their 
dependence on the CKM matrix elements in the framework of the 
SM are discussed in Section~\ref{sec_fitIngredients}.
\vs
Over-constraining 
the unitary CKM matrix aims at validating 
or not the SM with three generations. The interpretation of these
constraints requires a robust statistical framework which protects
against misleading conclusions. The next section describes to some
detail the statistical approach applied for the analysis reported
in this work.

%
% --------------------- The Statistical Framework -------------------
%
\section{The Statistical Framework}
\label{sec_statFrame}

We are considering an analysis involving a set of $\Nexp$ 
measurements collectively denoted by $\xexp=\{\xexp(1), \dots, $ $ \xexp(\Nexp)\}$, 
described by a set of corresponding theoretical expressions
$\xthe=\{\xthe(1),$ $\dots,\xthe(\Nexp)\}$. 
The theoretical expressions $\xthe$ are functions of a set 
of $\Nmod$ parameters $\ymod=\{\ymod(1),\dots,\ymod(\Nmod)\}$.
Their precise definition is irrelevant for the present discussion
(\cf, Section~\ref{sec_fitIngredients} for details)
beside the fact that:
\begin{itemize}

\item{} a subset of $\Nthe$ parameters within the $\ymod$ 
        set are fundamental, and free, parameters 
        of the theory (\ie, the four CKM unknowns in the 
        SM, the top quark mass, etc.) and are denoted 
        $\ythe$, where $\ythe=\{\ythe(1),\dots,\ythe(\Nthe)\}$.

\item{} the remaining $\NQCD=\Nmod-\Nthe$ parameters,
        which appear due to our present inability to compute precisely 
        strong interaction quantities (\eg, $\fbd$, $B_d$, etc.), are 
        denoted $\yQCD$, where $\yQCD=\{\yQCD(1),\dots,\yQCD(\NQCD)\}$.

\end{itemize}

\noindent There are three quite different goals the analysis aims at:
\begin{enumerate}
\item   Within the SM,
        to achieve the best estimate of the $\ythe$ parameters:
        that is to say to perform a careful metrology of the theoretical 
        parameters, for later use.

\item   Within the SM,
        to set a confidence level (\CL) which quantifies the agreement 
        between data and the theory, as a whole.

\item   Within an extended theoretical framework, \eg, Supersymmetry,
        to search for specific signs of new physics by pinning down 
        additional fundamental, and free, parameters of the theory.
\end{enumerate}

These three goals imply three statistical treatments all of which
rely on a likelihood function meant to gauge the agreement 
between data and theory. 

\subsection{The Likelihood Function}
\label{TheLikelihoodFunction}

We adopt a $\chi^2$-like notation and denote
\beq
\label{eq_chi2Function}
\chi^2(\ymod)\equiv-2\ln(\Lik(\ymod))~,
\eeq
where $\Lik$, the likelihood function
(it is defined below) 
receives contributions of two types
\beq
\label{eq_likFunction}
\Lik(\ymod)=\Likexp(\xexp-\xthe(\ymod))\ \Likthe(\yQCD)~.
\eeq
The first term, the experimental likelihood $\Likexp$, 
measures the agreement between $\xexp$ and $\xthe$,
while the second term, the theoretical likelihood $\Likthe$, 
expresses our present knowledge on the $\yQCD$ parameters.
\vs
It has to be recognized from the outset that the notation 
$\chi^2$ of Eq.~(\ref{eq_chi2Function})
is a commodity which can be misleading. 
In general, 
denoting "$\ProbCERN$" the well known routine from the CERN library,
one {\em cannot} infer a \CL\ from the above $\chi^2$ value
using
\begin{eqnarray}
\label{eq_probcern}
\CL      &=&     \ProbCERN(\chi^2(\ymod),\Ndof)~, \\
        &=&     \frac{1}{\sqrt{2^{\Ndof}}\Gamma({\Ndof}/2)}
                \intl_{\chi^2(\ymod)}^\infty 
                \hmmm e^{-t/2}t^{\Ndof/2-1}\, dt~.
\end{eqnarray}
This is because neither $\Likexp$ nor $\Likthe$
(they are further discussed in the sections below)
are built from 
purely Gaussian measurements:
\begin{itemize}

\item{} In most cases $\Likexp$ should handle experimental 
        systematics, and, in some instance,
        it has to account for inconsistent measurements.

\item{} In practice, $\Likthe$ relies on hard-to-quantify educated guessworks,
        akin to the ones used to define experimental systematics,
        but in most cases even less well-defined.

\end{itemize}
The first limitation is not specific to the present analysis
and is not the main source of concern, here.
The second limitation is the most challenging:
its impact on the analysis is particularly strong
with the data presently available.
The statistical treatment advocated below,
denoted \ckmfit, 
is designed to cope with both of the above limitations.
Notwithstanding its attractive features,
the \ckmfit\ scheme does not offer a treatment of the problem at hand
free from any assumption: 
an ill-defined problem cannot be dealt with rigorously.
However, 
the \ckmfit\ scheme extracts the most
out of simple and clear-cut {\it a priori} assumptions.
%which,
%admitedly,
%are Bayesian in nature.
\vs
The alternative statistical treatments discussed in 
Section~\ref{AlternativeStatisticalTreatments}
differ from the \ckmfit\ scheme
by the procedure used to define the \CL\ from the above $\chi^2$,
or by the content and interpretation of $\Likthe$.

\subsubsection{The Experimental Likelihood}
\label{sec_TheExperimentalLikelihood}

The experimental component of the likelihood 
is given by the product
\beq
\label{eq_likExp}
\Likexp(\xexp-\xthe(\ymod))=\prod_{i=1}^{\Nexp}\Likexp(i)~,
\eeq
where the individual likelihood components account for 
independent measurements\footnote
{
        Features marked by $^{(*)}$ in the following 
        item list are
        not issued in the analysis presented in this work,
        but may become important in future CKM profiles.
}.

\subparagraph{The Likelihood Components:} 

Ideally, the individual likelihood components $\Likexp(i)$ 
would be pure Gaussians
\beq
\label{eq_thegaussian}
\Likexp(i)= 
        {1\over\sqrt{2\pi}\sigexp(i)}\,
        {\rm exp}\left[-{1\over 2}
        \left({\xexp(i)-\xthe(i)\over\sigexp(i)}\right)^2\right]~,
\eeq
each with a standard deviation given by the experimental 
statistical uncertainty $\sigexp(i)$ of the $i^{\rm th}$ 
measurement. However, in practice, one has to deal with 
additional experimental and theoretical systematic uncertainties. 

\subparagraph{Experimental Systematics:}
\label{sec_ExperimentalSystematics}

An experimental systematics is assumed to take the form of a possible biasing offset,
the measurement could be corrected for, were it be known.
%when it can be expressed as a dependence on not perfectly 
%known parameters, 
Their precise treatment is discussed in 
Section~\ref{sec_likelihoodsAndSysErrors}.
There, a natural extension of the usual method 
of adding linearly or in quadrature statistical and systematic 
uncertainties is proposed. 

\subparagraph{Theoretical Systematics:}
\label{sec_TheoreticalSystematics}

Theoretical systematics, 
when they imply small effects,
are treated as the experimental ones.
However,
because most theoretical systematics imply large effects
and affect in a non-linear way the $\xthe$ prediction, 
most of them are dealt with through the theoretical likelihood 
component $\Likthe$
(\cf, Section~\ref{sec_theoreticalLikelihood}). 

\subparagraph{Model Dependent Measurements$^{(*)}$:}

When theoretical systematics cannot be expressed as a dependence 
on not perfectly known parameters, but are expressed as a set of 
measurements based on alternative models, labelled by the index 
$m$, (\eg, the exclusive ${\VuboverVcb}$ measurement exhibits 
such a model dependence) $\Likexp(i)$ is defined by
\beq
        \Likexp(i)=\Likexp(i,m)~,
\eeq
and $m$ is treated as an additional $\yQCD$ parameter,
taking only discrete values.
%\beq
%\Likexp(i)=\sum_m\w_m\Likexp(i,m)~,
%\eeq
%where the sum of weights is normalized to unity: 
%$\sum\w_m=1$.
%In the absence of theoretical prejudices, 
%the weights are taken to be the same.

\subparagraph{Identical Observables and Consistency:}

When several measurements refer to the same observable
(\eg, various measurements of $\DmBd$) they have to be 
consistent, 
independently of the theoretical framework used for the analysis. 
\vs
Similarly, when several measurements refer to different observables
which are linked to the same $\ythe$ parameter, \eg., 
$\Vud$ and $\Vus$, or determinations of $\Vub$ stemming from 
different observables, or measurements of $\stb$ obtained from 
similar $B$ decays, one may {\it decide} to overrule possible 
disagreement by imposing the measurements to be consistent.
By doing so, one is deliberately blinding oneself from possible 
new physics effects which may have revealed themselves otherwise.
Clearly, such overruling should be applied with great caution,
and it should be well advertized whenever it occurs.
\vs
The method to deal with this {\it imposed} consistency is 
to account for the measurements at once, 
by merging them into a single component, 
usually obtained from their weighted mean. 
A more refined treatment is needed when this set of measurements 
is clearly inconsistent. 
A general method to handle a set of measurements, 
whether they are consistent or not, 
is proposed in Ref.~\cite{combagost,combiner}. 
Yet, for ``not too large'' inconsistencies, 
the proposed method yields similar results 
as the $\chi^2$ rescaling approach adopted by the PDG~\cite{pdg2000}.
%As a consequence,  
To clarify the presentation of the \ckmfit\ scheme, we use the 
latter: when N inconsistent measurements appear in this analysis, 
the error obtained for their weighted mean
is rescaled by the factor $\sqrt{\chi^2_{\rm wm}/(N-1)}$,
where $\chi^2_{\rm wm}$ is the weighted mean $\chi^2$.

\subparagraph{Related Observables and Consistency$^{(*)}$:}
\label{sec_relatedObservables}

In some instances, several observables, 
although not identical, 
are functionally related in a way independent of the theoretical 
framework used for the analysis.
The number of such instances is denoted $\Ncst$,
and the effective number of measurements,
the one to be used to compute degrees of freedom, 
is defined by 
\beq
\label{EqNeff}
	\Neff=\Nexp-\Ncst~.
\eeq
An example is provided by the set of measurements 
yielding separately $\Vub$, $\Vcb$ and ${\VuboverVcb}$. 
The ratio of the first two should be compatible with the 
third measurement, whether or not the SM is valid.
Since the measurements are not referring to a unique observable
%(then, the functional relation is a mere identity) 
they cannot be merged simply into a single component, 
as above.
One should normalize their contribution 
to ensure that they do not contribute to the overall $\chi^2$ value, 
if they are in the best possible mutual agreement, 
independently of the theoretical framework used for the analysis. 
This normalization is in fact what is done in the case 
of identical observables.
It is irrelevant for the metrological phase of the analysis,
and for the third phase, 
where one searches for specific sign of new physics: 
then, 
any constant can be added to the $\chi^2$ without affecting the result. 
However,
it is relevant for the second phase, 
where one probes the SM: 
a statistical fluctuation in the set $\Vub$, 
$\Vcb$ and ${\VuboverVcb}$ 
which makes them violate their functional relation should 
not trigger a claim for new physics. 
In this example, 
the normalization constant is obtained as the maximal value 
of the function of the two variables $\Vubthe$, $\Vcbthe$
\beqn
\Likexp(\Vubthe,\Vcbthe)&=&
        ~\phantom{\times}\ \ \Likexp\left(\Vub-\Vubthe\right)        \nonumber\\
        &&~\times\,\Likexp\left(\Vcb-\Vcbthe\right)\nonumber\\
        &&~\times\,\Likexp\left(\Vubcb-\frac{\Vubthe}{\Vcbthe}\right)~.
\eeqn
Here as well,
care should be taken not to normalize that way 
the contributions of observables the functional connection of which is model dependent. 
For instance, 
the measurements leading to ${\rm sin}\gamma$ and 
$\pi-\beta-\alpha$ are sensitive 
to new physics because their measured values may violate 
the SM functional relation 
${\rm sin}\gamma={\rm sin}(\pi-\beta-\alpha)$
%it is precisely the goal of the second phase of the 
%analysis to detect such functional violations
\footnote
{
        It is worth pointing out that an apparent functional 
        violation is present (since long ago)
        in the available data 
        (\cf. Section~\ref{TheCKMMatrixElements}): $\Vud^2+\Vus^2+\Vub^2<1$.
}.

\subparagraph{Normalization:}
\label{sec_Normalization}

More generally,
the normalization of each individual likelihood component is 
chosen such that its maximal value is equal to one.
This is not important for the analysis, but it is 
convenient: it ensures that a measurement does not 
contribute numerically to the overall $\chi^2$ value
if it is in the best possible agreement with theory, 
and that the (so-called) $\chi^2$ takes only positive values.
In the pure Gaussian case, it implies simply to drop the 
normalization constant of Eq.~(\ref{eq_thegaussian}):
one is thus recovering the standard $\chi^2$ definition.

\subsubsection{The Theoretical Likelihood}
\label{sec_theoreticalLikelihood}

The theoretical component of the likelihood is given by 
the product
\beq
\label{eq_likThe}
        \Likthe(\yQCD)=\prod_{i=1}^{\NQCD} \Likthe(i)~,
\eeq
where the individual likelihood components $\Likthe(i)$ 
account for the partial knowledge available on the $\yQCD$ 
parameters (\eg, $f_{B_d}$) including more or less accurately 
known correlations between them (\eg, $f_{B_d}/f_{B_s}$).
Ideally, one should incorporate in $\Likexp$ measurements 
from which constraints on $\yQCD$ parameters can be 
derived.
%~\cite{Bigi}. 
By doing so, one could remove altogether the theoretical component 
of the likelihood. However, this is not what is done, 
because usually there is no such measurement: 
the {\it a priori} knowledge on the $\yQCD$ stems rather from 
educated guesswork\footnote
{
   The same remark applies to experimental systematics, but,
   since these are usually not the dominant part of the 
   experimental uncertainties, the problem is less acute.
}.
As a result, the $\Likthe(i)$ components are incorporated by 
hand in Eq.~(\ref{eq_likThe})
and they can hardly be considered as issued from 
probability distribution functions (PDF). 
In effect,
their mere presence in the discussion is a clear sign that the problem
at hand is ill-defined.
It demonstrates that a (here critical) piece of information is coming 
neither from experimental,
nor from statistically limited computations,
but from somewhere else:
from the mind of physicists.
At present,
these components play a leading role in the analysis
and it is mandatory to handle them with the greatest caution.

\subparagraph{The Default Scheme - Range Fit (\ckmfit):}
\label{sec_CkmFitScheme}

In the scheme we propose,
the theoretical likelihoods $\Likthe(i)$ do not contribute to the 
$\chi^2$ of the fit while the corresponding $\yQCD$ parameters take 
values within ranges, thereafter termed ``allowed ranges'' and 
denoted $[\yQCD]$. The numerical derivation of these ranges is 
discussed in Sections~\ref{sec_likelihoodsAndSysErrors} and 
\ref{sec_fitIngredients}. Most of them are identified to the 
ranges $[\yQCD-\sigma_\syst\ ,\yQCD+\sigma_\syst]$,
where $\sigma_\syst$ is the theoretical systematics evaluated 
for $\yQCD$. Hence $\yQCD$ values are treated on an equal footing, 
irrespective of how close they are from the edges of the allowed range.
Instances where even only one of the $\yQCD$ trespasses 
its range are not considered\footnote{
In the case of model dependence (\cf, 
Section~\ref{sec_TheExperimentalLikelihood})
the allowed values for the discrete parameter $m$ labelling the models
correspond to the set of models deemed acceptable.
\ckmfit\ is allowed to select at will any one within this set,
in the same way that it is allowed to select a $\yQCD$ parameter 
at will within 
its allowed range.
In practice,
when $\yQCD$ parameters cannot be handled beforehand 
as explained in Section~\ref{sec_likelihoodsAndSysErrors},
the actuation of the allowed range in \CkmFitter\  is obtained using 
the {\tt Set Limit} option of the \MINUIT\  package.
It is equivalent to set
the component $\Likthe(i)$ to unity 
when the corresponding $\yQCD(i)$ parameter is within $[\yQCD(i)]$,
and to zero otherwise.
}. 
\vs
This is the unique, 
simple, 
and clear-cut assumption made in the \ckmfit \
scheme:
$\yQCD$ parameters are bound to remain within {\it predefined} allowed ranges.
The \ckmfit \ scheme departs from a perfect frequentist analysis only because 
the allowed ranges $[\yQCD]$ do not extend to the whole physical space 
where the parameters could {\it a priori} take their values\footnote{
Not all $\yQCD$ parameters need to be 
        given an {\it a priori} allowed range:
        \eg, values taken by final state strong interaction 
        phases (FSI) appearing in $B$ decays
        are not necessarily theoretically constrained.}.
This should {\it not} be understood as implying that 
a uniform PDF is ascribed to each $\yQCD$ parameter.
This important remark is further discussed in 
Section \ref{sec_TheBayesianTreatment} and in 
Appendix~\ref{TheBayesianMethodcaughtunder-conservative}.
\vs
This unique and minimal assumption, 
is nevertheless a strong assumption:
all the results obtained should be understood as valid
only if all the assumed allowed ranges 
contain the true values of their $\yQCD$ parameters.
But there is no guarantee that this is the case,
and this restriction should be kept in mind.
On the other hand, also the contrary is true: if the ranges
are chosen too big, one may miss a discovery.

%\subparagraph{Scan Scheme:}
%
%The above \ckmfit\ scheme is conceptually close to the 
%{\em 95\%~CL Scan Method}~\cite{schune}.
%However,
%significant differences exist between the two treatments
%as it is briefly discussed in Section~\ref{sec_scan}
%and in Appendix~\ref{DrawbacksoftheSCANmethod}.
%
%\subparagraph{Extended Scheme - Extended Range Fit (\ckmfoot):}
%
%An extended treatment of \ckmfit\ , 
%where trespassing the allowed ranges is not strictly prohibited,
%is presented in Section~\ref{TheExtendedConservativeMethod}.
%
%
%\subparagraph{Bayesian Scheme:}
%
%The Bayesian method implies the definition of a probability 
%density function (PDF)
%for each $\yQCD$.
%It is not the method advocated here. 
%For the sake of completeness, 
%it is discussed in Section~\ref{sec_TheBayesianTreatment}
%and in Appendix~\ref{TheBayesianMethodcaughtunder-conservative}.

\subsection{Metrology}
\label{sec_metrology}

For metrology, 
one is not interested in the quality of the agreement between 
data and the theory as a whole. Rather,
taking for granted that the theory as a whole is correct, 
one is only interested in the 
%relative 
quality of the 
agreement between data and various realizations of the theory, 
specified by distinct sets of $\ymod$ values.
More precisely,
as discussed in Section~\ref{sec_RelevantandIrrelevantParameters},
the realizations of the theory one considers are
under-specified by various subsets of so-called 
relevant parameters values.
In the following we denote
\begin{equation}
        \ChiMinGlob~,
\end{equation}
the absolute minimal value of the $\chi^2$ function of 
Eq.~(\ref{eq_chi2Function}), obtained when letting all $\Nmod$ 
parameters free to vary. 
\vs
In principle,
this absolute minimal value does not correspond to a unique $\ymod$ location.
This is because measurements 
(resp. theoretical predictions)
entering in the analysis  are all affected
by more or less important experimental (resp. theoretical) systematics.
These systematics being handled by means of allowed ranges,
%within which all values are treated on the same footing,
there is always a multi-dimensional degeneracy for any value of
$\chi^2$.
\vs
In practice, with the presently available observables,
theoretical systematics play a prominent role.
If one does not incorporate significant $\stb$ measurements in the analysis,
the domain where $\chi^2=\ChiMinGlob$ is noticeably wide.
For convenience,
in the following we refer to this domain as $\ymodopt$,
as if a unique point in the $\ymod$ space were leading to $\ChiMinGlob$.
The projections of the $\ymodopt$ domain onto one dimensional
spaces result in finite intervals within which the data analysis cannot 
make distinction (and similarly for two-dimensional spaces).
When $\stb$ is incorporated in the analysis,
one adds a measurement with negligible systematics
which lifts partially the degeneracy and
makes some projections of $\ymodopt$ become point-like.
\vs
This degeneracy should be treated carefully when one is exploring
a sub-space $\a$ of $\ymod$:
points widely apart in $\a$ can lead to the same $\chi^2$, 
provided the other parameter values are changed accordingly.
However,
except for numerical accidents,
identical $\chi^2(\ymod)=\ChiMinGlob$ values imply identical 
$\Likexp$ components,
and hence identical predictions: $\xthe(\ymod)$ values are constant
within $\ymodopt$.
%Stated differently.
%the degenerate optimal realizations of SM are experimentally undistinguishable.
\vs
Ideally, for metrological purposes, 
one should attempt to estimate as best as possible 
the complete $\ymod$ set. In that case,
one should use the offset-corrected $\chi^2$
\beq
        \Delta\chi^2(\ymod)=\chi^2(\ymod)-\ChiMinGlob~,
\eeq
where $\chi^2(\ymod)$ is the $\chi^2$ for a given set 
of model parameters $\ymod$. 
The minimal value of $\Delta\chi^2(\ymod)$ is zero, 
by construction.
This ensures that,
to be consistent with the assumption that the SM is correct, 
CLs equal to unity are obtained when exploring the $\ymod$ space
(namely, once $\ymod$ enters the $\ymodopt$ domain).
%The quantity of interest would be $\Delta\chi^2(\ymod)$.
In a Gaussian situation, one would then directly obtain the 
\CL\ for a particular set of $\ymod$ values as
\beq
\label{eq_gaussianCase}
        \Prob(\ymod)
        =\ProbCERN(\Delta\chi^2(\ymod),\Ndof)~,
\eeq
with $\Ndof=\Min(\Neff,\Nmod)$,
where $\Neff$ is defined in Eq.~(\ref{EqNeff}).

\subsubsection{Relevant and Less Relevant Parameters}
\label{sec_RelevantandIrrelevantParameters}

However, 
one is not necessarily interested in all the $\ymod$ values, 
but only in a subset of them.
This can be for two distinct reasons:
\begin{itemize}

\item{} The other parameters being deemed less relevant.
        For instance, in the SM,
        CP violation can be summarized by the value taken by the 
        Jarlskog parameter $\J$, or by the value taken by $\eta$ 
        (in the Wolfenstein parameterization):
%\footnote
%{
%       The CP measure $\J$ is independent of the parameterization
%       and constitutes thus a physical quantity. 
%       This is not the case for $\eta$ or $\delta$.
%}:
        the other CKM parameters and the $\yQCD$ parameters may 
        thus conceivably be considered of lower interest. 
        More generally, one is rarely considering a CKM fit as a 
        means to pin down anything else than CKM parameters,
        least of all $\yQCD$ parameters\footnote{Although it can be argued that,
        while theoretical uncertainties dominate,
        pinning down $\yQCD$ parameters might turn out to be the main (and not
        so interesting) achievement of a CKM analysis...}. 
        
\item{} Parameters that cannot be significantly constrained
        by the input data of the CKM fit.
        This is the case for most of the non-CKM parameters:
        $\yQCD$ parameters, but also the top quark mass, etc.

\end{itemize}

In practice,
the $\ymod$ parameters usually retained as relevant for 
the discussion are $\rhobar$ and $\etabar$.
The other parameters $\lambda$, $A$, the top quark mass (etc.)
and all the $\yQCD$ are considered as subsidiary parameters, 
merely to be taken into account in the analysis,
but irrelevant for the discussion. In that case,
the aim of the metrological stage of the analysis is to 
set \CLs\ in the $\rhoeta$ plane.
\vs
We denote by $\a$ the subset of $\Na$ parameters under discussion 
(\eg, $\a=\arhoeta$ and $\Mu$ 
the $\Nmu$ remaining $\ymod$ parameters\footnote{
It is worth to stress that this splitting is arbitrary and 
that it can be changed at will:
for instance one may decide to focus only on $\a=\{\J\}$,
or to consider $\a=\astastb$, etc.}
. 
\vs
\centerline{\em The goal is to set {\rm \CLs} in 
                the $\a$ space, irrespective of the $\Mu$ values.
           }	  
\vs
The smaller the region in the $\a$ space where the 
\CL\ is sizeable (above $\CLcut=0.05$, say)
the stronger the constraint is.
The ultimate (and unattainable)
goal being to make this allowed region shrink to a point:
it would then correspond to the 'true' $\a$. 
This means that one seeks to exclude the 
largest possible region of the $\a$ space.
To do so, for a fixed value of $\a$, 
one has to find the $\Mu$ values which maximize 
the agreement between data and theory, 
and set the \CL\ on $\a$ at the value corresponding to 
this optimized $\Mu$
\beq
\label{TheIncredibleStatement}
{\rm CL}(a)={\rm Max}_\mu\{ {\rm CL}(a,\mu)\}~.
\eeq
Proceeding that way,
one uses the most conservative estimate for a given $\a$ point:
this point will be engulfed in the excluded region only if 
$\CL(\a,\Mu)<\CLcut$, $\forall\Mu$. 
%In fact,
%without any additional knowledge on $\Mu$,
%no other choice is left.
Stated differently, the CLs one is interested in are upper bounds
of confidence levels.
In effect, as discussed in section~\ref{sec_Illustrations},
this is the standard procedure one uses to obtain
\CLs~for a sub-set of fitted parameters.

\subsubsection{Metrology of Relevant Parameters}
\label{sec_MetrologyofRelevantParameters}

According to the above discussion,
we denote 
\begin{equation}
        \chi^2_{\min ;\,\Mu}(a)~,
\end{equation}
the minimal value of the $\chi^2$ function of Eq.~(\ref{eq_chi2Function}),
for a fixed value of $\a$,
when letting all $\Mu$ parameters free to vary.
For metrological purposes, one uses the offset-corrected $\chi^2$
\beq
\label{eq_metrologicalOffset}
\Delta\chi^2(a)
        =\chi^2_{\min ;\,\Mu}(a)-\ChiMinGlob~,
\eeq
the minimal value of which is zero, by construction:
it is reached when $\a$ enters the $\ymodopt$ domain.
%
%Hence, the CL~(\ref{eq_metrologicalOffset}) has to be
%interpreted as {\em upper bound for the best
%theoretical model at given space point} $\a$.
\vs 
Since only the minimal value of the 
$\chi^2$ with respect to $\Mu$ enters the \ckmfit\ analysis,
when $\Mu$ contains a $\yQCD(j)$ parameter which
appears only in one measurement $i$,
it is advisable to absorb its effect by computing beforehand
        \begin{equation}
        \label{digest}
                \Likexp(i)_{\max :\yQCD(j)}=
                \Max_{\yQCD(j)}\{\Likexp(i)\Likthe(\yQCD)\}~,
        \end{equation}
to clarify the analysis\footnote
{
                This should be done for instance for FSI phases,
                most notably for the determination of the angles
                $\alpha$ and $\gamma$.
                But this cannot be done for the product $\fbdbd$, 
                because it appears in $\DmBd$ but also indirectly in $\DmBs$,
                since $\fbdbd$ and $\fbsbs$ are theoretically linked.
%                However, this is not a requested step since the \CkmFitter\ package 
%                will do it, but repeatedly, and thus in an inefficient way.
}
(see Section~\ref{sec_likelihoodsAndSysErrors}).

\subparagraph{Gaussian Case:}
\label{sec_GaussianCase}

In a purely Gaussian situation
%(no systematics and a nice parabolic shape for the $\chi^2$)
one would directly obtain the \CL\ for $\a$ as 
\beq
\label{eq_conLev}
        \Prob(\a)=\ProbCERN(\Delta\chi^2(\a),\Ndof)~,
\eeq
where $\Ndof=\Min(\Neff-\Nmu,\Na)$.
Equivalently,
one may derive the same \CL\ from the covariance matrix
obtained from the fit leading to the absolute minimum,
%since the validity of Eq.~(\ref{eq_conLev}) implies that 
if in the $\a$-region under consideration,
the $\chi^2$ is parabolic.

\subparagraph{Non-Gaussian Case:}
\label{sec_NonGaussianCase}

In a non-Gaussian situation, 
one has to consider $\Delta\chi^2(a)$ as a test statistics,
and one has to rely on a Monte Carlo simulation to obtain 
its expected distribution in order to compute $\Prob(a)$. 
As further discussed in Section~\ref{sec_probingTheSM},
this does not imply taking a Bayesian approach and to make 
use of PDFs for the unknown theoretical parameters $\Mu$.
\vs 
For the sake of simplicity, 
we use Eq.~(\ref{eq_conLev}) in the present work.
This implies that the experimental component 
		$\Likexp(\xexp-\xthe(\ymod))$ is free 
		from non Gaussian contributions and inconsistent measurements.
		However, the $\Delta\chi^2(\a)$ function
		itself does not have to be parabolic.
		What matters is that the $\Likexp$ components 
		are derived from Gaussian measurements
		(\cf, Section~\ref{sec_Illustrations} for an example),
		being understood that no $\Likthe$ components are present.

\subsubsection{Illustrations}
\label{sec_Illustrations}

To illustrate the above definitions we consider two specific examples 
in this section\footnote
{
	A more involved example is discussed in 
	Appendix~\ref{sec_RatiosOfBranchingFractions}
}.

\subparagraph{Standard Situation:}

Consider an analysis which depends on only two quantities:
the first is a fundamental parameter $\a$, 
and the second is a QCD parameter $\yQCD$.
We assume here that the situation is a standard one,
where it turns out that both quantities are simultaneously measurable:
the full $\chi^2(\a,\yQCD)$ function takes the form
\begin{eqnarray}
\chi^2(\a,\yQCD)&=&\left({\a-\a^0\over\sigma[\a]}\right)^2
                +\left({\yQCD-\yQCD^0\over\sigma[\yQCD]}\right)^2
		\\ \nonumber
		& &-2{\corcof}
		\left({\a-\a^0\over\sigma[\a]}\right)
		\left({\yQCD-\yQCD^0\over\sigma[\yQCD]}\right)
		+\chi^2_{\min;\,\a,\yQCD}
		\, ,
\end{eqnarray}
where $\corcof$ is a correlation coefficient.
Applying the \ckmfit\ scheme, 
the $95\%~\CL$ interval for $\a$ is obtained as follows.
One first computes the offset-corrected $\chi^2$
\begin{eqnarray}
\Delta\chi^2(\a)
&=&\chi^2_{\min;\,\yQCD}(\a)-\chi^2_{\min;\,\a,\yQCD}~, \\ 
&=&\left({\a-\a^0\over\sigma[\a]}\right)^2(1-\corcof^2)~.
\end{eqnarray}
The limits $\a_\pm$ of the $95\%~\CL$ interval are such that
\begin{equation}
\ProbCERN(\Delta\chi^2(\a_\pm),1)=0.05 
\rightarrow \Delta\chi^2(\a_\pm)=3.84~,
\end{equation}
and hence
\begin{equation}
\label{Correctapm}
a_\pm=\a^0\pm 1.96\ {\sigma[\a]\over \sqrt{1-\corcof^2}}~,
\end{equation}
which is just the standard answer for the $95\%~\CL$ interval of $\a$,
if one disregards information on $\yQCD$.

\subparagraph{Measurement of \boldmath$\stb$:}

If one uses $\a=\arhoeta$,
a measurement of $\stb$ alone yields a double infinite degeneracy 
corresponding to the solutions of Eq.~(\ref{eq_stb}),
namely
\begin{equation}
\label{stbtwolines}
\etabar=(1-\rhobar){1\pm\sqrt{1-(\stb_\exp)^2}\over\stb_\exp}~.
\end{equation}
Along the two above straight lines in the $\rhoeta$ plane,
$\chi^2(a)=\ChiMinGlob=0$.
There is no $\Mu$ parameters here,
and hence
\begin{equation}
\label{stbchi2}
\Delta\chi^2(\a)=
\chi^2_{\min ;\,\Mu}(\a)-\ChiMinGlob
=\left({\stb_\exp - \stb_\tho \over \sigma[\stb]}\right)^2~.
\end{equation}
Using Eq.~(\ref{eq_conLev}) one gets the \CL\ in the $\rhoeta$ plane
\begin{equation}
\label{probstb}
\Prob(\rhobar,\etabar)
	=\ProbCERN(\Delta\chi^2(\a),1)~.
\end{equation}
While the double infinite degeneracy of $\Delta\chi^2(\a)=0$ 
clearly precludes a parabolic behavior for this function,
Eq.~(\ref{probstb}) remains exact due to the right hand 
side of Eq.~(\ref{stbchi2}).
%\vs
%As alluded to in Section~\ref{sec_Remarks},
%the $\stb$ measurement is actually not straightforward.
%The CP-violating observable is not directly $\stb$:
%it depends on $\DmBd$ in an intricate way.
%For the fits discussed in Section~\ref{sec_constrainedFits},
%the measurements of $\stb$ are used ignoring this complication:
%they are treated as a stand-alone $\Likexp$ component,
%where a fixed value of $\DmBd$ is implicitly used.
%To be correct one should incorporate properly the correlation
%between $\stb$ and $\DmBd$ by merging both in a common $\Likexp$ 
%component, or by making explicit the $\stb$ dependence on $\DmBd$.
%This care is not needed with the present level of precision 
%on $\stb$, but it will be relevant for the forthcoming 
%high-statistics measurements.
%
%
% --------------------- Probing the SM -----------------------------
%
\subsection{Probing the SM}
\label{sec_probingTheSM}

By construction, the metrological phase is unable to detect a 
failure of the SM to describe the data.
This is because Eq.~(\ref{eq_metrologicalOffset}) wipes out
the information contained in $\ChiMinGlob$.
%irrespectively of the size of this minimal value.
This value is a measure (a test statistics) 
of the best possible agreement between data and theory. 
Ideally, 
in a pure Gaussian case, 
this quantity could be turned into a \CL\ referring to the 
SM as a whole in a straightforward way
\beq
\label{eq_naiveGaussian}
\Prob({\rm SM})\le
	\ProbCERN(\ChiMinGlob,\Ndof)~,
\eeq
with $\Ndof=\Neff-\Nmod$, were it be a positive number. 
%Here,
%since the $\yQCD$ parameters are let free to vary at will,
%one must use: $\Ndof=\Neff-\Nthe$. Furthermore,
The whole Standard Model being at stake
one should not rely on Eq.~(\ref{eq_naiveGaussian}),
but use a Monte Carlo simulation to obtain the 
expected distribution of $\ChiMinGlob$.
The Monte Carlo simulation is built as follows\footnote
{
	For the sake of generality, 
	the theoretical likelihood is not assumed to be necessarily
	the trivial \ckmfit\ one (\cf, 
	Section~\ref{sec_theoreticalLikelihood}).
}:

\begin{itemize}

\item   One selects a set of $\ymod$ values 
	within $\ymodopt$ and assumes it to be 
	the true one\footnote{
        As discussed above, the various optimal realizations
	yield identical theoretical predictions,
	the choice made for a particular $\ymod$ within $\ymodopt$ is
	thus irrelevant. 
 	It was explicitly checked that the outcome of the analysis does 
	not depend on this choice.}.
%	At any rate, by construction, they are 
%	leading to the smallest $\chi^2$.
	
\item   Then one generates all $\xexp(i)$, following the 
	distribution of individual experimental likelihood 
	component $\Likexp(i)$, having reset their central
	values to the values $\xexp(i)=\xthe(i)$ computed with 	
	the above $\ymod$ set. In case of significant experimental
	systematics, this implies the use of appropriate PDFs
	as discussed in Section~\ref{sec_likelihoodsAndSysErrors}.
	
\item   In contrast to the above, one does not modify the 
	$\Likthe$ component of the likelihood: their central 
	values are kept to their original settings. This is 
	because these central values are not random numbers,
	but parameters contributing to the definition of $\Lik$.

\item   Then one computes the minimum of the $\chi^2$ by 
	letting all $\ymod$ free to vary, as is done in the 
	actual data analysis.

\item   From this sample of Monte Carlo simulations, one 
	builds $\F(\chi^2)$, the distribution of 
	$\ChiMinGlob$, normalized to unity. 

\end{itemize}

The \CL\ referring to the SM as a whole is then
\beq
\label{eq_monteCarlo}
\Prob({\rm SM})\;\le\hmm\hmm\intl_{\chi^2\ge\ChiMinGlob}
	\hmmm\hmmm\F(\chi^2) \ d\chi^2~,
\eeq
which is the upper bound to the CL one may set on the SM.

\subsection{Probing New Physics}
\label{sec_probingNewPhysics}

If the above analysis establishes that the SM cannot accommodate 
the data, the next step is to probe for the new physics revealed 
by the observed discrepancy. The goal is akin to metrology: 
it is to measure new physical parameters $\yNP$ 
(whose values, for example, are zero if the SM holds) 
complementing the set of $\ythe$ parameters of the SM.
The treatment is identical to the one of Section~\ref{sec_metrology}, 
using $\a=\{\yNP\}$. 
The outcome of the analysis is for example a $95\%~\CL$ domain 
of allowed values for $\yNP$ defined, in a first approximation,
from Eq.~(\ref{eq_conLev})
\beq
\label{eq_newPhysicsCL}
\Prob(\yNP)=\ProbCERN(\Delta\chi^2(\yNP),\NNP)\ge 0.05~.
\eeq
%assuming $\Neff-\Nthe-\NNP>0$.
Even if the SM cannot be said to be in significant 
disagreement with data, 
it remains worthwhile to perform this metrology of new physics, 
for two reasons:

\begin{itemize}

\item   It might be able to faster detect first signs 
	of discrepancy between data and the SM,
	if the theoretical extension used in the analysis 
	turns out to be the right one. The two approaches 
	are complementary, the first (\cf, 
	Section~\ref{sec_probingTheSM}) leads to a general 
	statement about the validity of the SM,
	independently of any assumption for the new physics,
	the second is specific to a particular extension of the 
	SM. In that sense it is less satisfactory.
	Being complementary, the two approaches can disagree:
	the first may conclude that the SM is in 
	acceptable agreement with data, while the second may 
	exclude the SM value $\yNP=0$, and, conversely, 
	the first may invalidate the SM, while the second 
	may lead to a fairly good value of $\Prob(\yNP=0)$,
	if the extension of the SM under consideration is not 
	on the right track.

\item   The most sensitive observables, and the precision 
	to be aimed at for their determination cannot be derived 
	by any other means than by this type of analysis. When 
	considering new experiments, it is therefore particularly 
	valuable to have a sensitive model of new physics, to 
	prioritize the efforts and set the precisions to be achieved.

\end{itemize}

%
% ------------------- Alternative Statistical Treatments ------------------
%
\section{Alternative Statistical Treatments}
\label{AlternativeStatisticalTreatments}

Several alternative statistical treatments are available.
Three of them are briefly discussed below: 
however not all variations are considered. 
The relative merits and limitations of the three treatments 
will not be discussed extensively here,
except to point out features of the \ckmfit\ scheme.
%together with the main drawbacks of each method.
%as perceived by the authors of this note (no polemic is intended).

\subsection{Reminder: The \ckmfit\ Scheme}
\label{sec_reminderCkmFitSchem}

Let us briefly re-sketch the main steps of
an analysis in the \ckmfit\ scheme: for a given point $\a$
in the parameter space
(\eg, $\a=\arhoeta$) 
\ckmfit\ proceeds to:

\begin{itemize}
\item{} Find the overall minimal $\ChiMinGlob$ with 
	respect to all theoretical parameters. 

\item{}	Perform a discrete, although fine scan of the $\a$ space, 
	and minimize $\chi^2_{\min ;\,\Mu}(a)$ with respect to
	the remaining parameters $\mu$, for each point:
	the $\yQCD$ parameters being allowed to vary 
        freely within their $[\yQCD]$ ranges. 

\item{}	Calculate the offset-corrected CL ($\Prob(\a)$ of
	Eq.~(\ref{eq_conLev})), for each point. 
        It is the upper bound of the confidence levels
	one may set on $\a$,
	which corresponds to 
        the best possible set of theoretical parameters $\Mu$.
		
\item{} Compute the CL of the overall $\ChiMinGlob$ by 
	means of a Monte Carlo Simulation. 
        It is the upper bound of the confidence levels
	one may set on the SM,
	which corresponds to 
        the best possible set of $\yQCD$ parameters.	

\end{itemize}

\noindent The \ckmfit\ scheme suffers from two drawbacks:

\begin{itemize}

\item{} It relies on {\it a priori} allowed ranges 
        for the $\yQCD$ parameters.
	
\item{} In the hopeful case where data are such that the method
	is lead to "rule out" the SM,
	it provides no indication as to which $\yQCD$ parameter(s) 
	should 
	be allowed to trespass its allowed range, 
	and by how much, to rescue the SM.

\end{itemize}
 
\subsection{The 95\%~Scan Method}
\label{sec_scan}

%\subsubsection{Brief Overview}

The 95\%~CL Scan method~\cite{schune} 
does not incorporate the theoretical component $\Likthe$
except to define allowed ranges for the $\yQCD$ values:
in effect, 
this is equivalent to the \ckmfit\ scheme.
These $\yQCD$ values are equidistantly scanned within their 
allowed ranges. For each set of $\yQCD$ values 
(denoted {\it model} in the 95\%~CL Scan method terminology)
three operations are performed:

%\subsubsection{Model Selection}
%\label{sec_ModelSelection}
\begin{enumerate}

\item{} One determines
\begin{equation}
\chi^2_{\min ; \ythe}(\yQCD)~,
\end{equation}
the minimal value of the $\chi^2$ function of 
Eq.~(\ref{eq_chi2Function}),
from $\Likexp$ only, for a fixed set of $\yQCD$ values,
when letting all $\ythe$ parameters free to vary.
One then computes the confidence level
\beq
\label{eq_conLevScanI}
\Prob(\yQCD)=\ProbCERN(\chi^2_{\min ; \ythe}(\yQCD),\Ndof)~,
\eeq
where $\Ndof=\Neff-\Nthe$.
If $\Prob(\yQCD)$ is above a threshold $\CLcut$
(usually $\CLcut=0.05$) 
the {\it model} is considered as acceptable,
and selected.
The SM is ruled out if no {\it model} is selected.

%\subsubsection{$95\%~\CL$ Contours}
%\label{CountourDrawing}

\item{} 
Among the $\ythe$ values,
a subset $\a$ is retained as central values to be displayed 
for the current {\it model} (if selected) and
CLs in the $\a$ space are derived using
\begin{equation}
\Delta\chi^2(\a,\yQCD)=\chi^2_{\min ; \ythe\neq\a}
(\a,\yQCD)-\chi^2_{\min ; \ythe}(\yQCD)~,
\end{equation}
and
\beq
\label{eq_conLevSacnII}
\Prob(\a,\yQCD)=\ProbCERN(\Delta\chi^2(\a,\yQCD),\Ndof)~,
\eeq
where $\Ndof=\Min(\Neff-\Nthe+\Na,\Na)$.

%\subsubsection{Graphical Display}
%\label{GraphicalDisplay}

\item{}
The method concludes by a graphical display,
for all selected models,  
of the contours in the $\a$ space defined by 
$\Prob(\a,\yQCD)=\CLcont$ (with $\CLcont=0.05$).
%each possibly completed by its central $\a$ values.

%\subsubsection{Probing the Standard Model}

\end{enumerate}

\subsubsection{Comparison with \ckmfit}

Although the outcome of the 95\%~CL Scan method,
the graphical display,
is quite different from the \ckmfit\ one,
both schemes are close in nature:
they are frequentist approaches,
flawed by the same double drawback mentioned 
in the previous section. 
In addition, 
while it is built on rather firm ground,
the 95\%~CL Scan method presents several unwelcome features which are reviewed 
in Appendix~\ref{DrawbacksoftheSCANmethod}.
The main differences between the two methods are:
\begin{itemize}

\item{} whereas \ckmfit\ seeks for a statistical statement 
	pertaining to $\a$, and only to $\a$,
	the 95\%~CL Scan method leads to statements on $\a$, for given values of $\Mu$
	which take the form of the $95\%~\CL$ contours.

\item{} to correct for this, the 95\%~CL Scan method may conclude by 
	an envelope, which delimits an allowed region in 
	the $\a$ space with at least $95\%~\CL$. 

\item{} \ckmfit\ draws a single smooth \CL\ surface. 
	From this surface one can read off the $95\%~\CL$ contour, 
	or define a family of contours,
	each corresponding to a given \CL.
	These contours encircle a domain,
	the plateau of the \CL\ surface,
	where the \CL\ is essentially equal to unity.

\item{} \ckmfit\ is flexible.
	The default treatment can be extended to atone for the 
	second fundamental drawbacks, and to accommodate for a smooth 
	transition toward the Bayesian treatment, while,
	nevertheless, keeping part of the virtues of the 95\%~CL Scan method and 
	of the \ckmfit\ scheme. This is discussed in the next section.
	
\end{itemize}

\subsection{The Extended Conservative Method (\ckmfoot)}
\label{TheExtendedConservativeMethod}

The \ckmfit\ scheme uses $\Likthe(i)$
functions which trivially take only two values:
either 1 within the allowed range,
or 0 outside,
thereby strictly forbidding any $\yQCD$ to trespass $[\yQCD]$.
Instead,
the extended \ckmfoot\ scheme
uses for $\Likthe(i)$
functions which take values between 1 and 0.
They are equal to 1 within $[\yQCD]$
(there, they do not contribute at all to the full $\chi^2$,
and one recovers the \ckmfit\ scheme)
and drop smoothly to 0 outside.
%(there, they start to enforce the No Trespassing \ckmfit\ law,
%but in a civilised manner).
These functions are not PDFs: they are not combined the ones with 
the others through convolutions, and hence (see next Section) the 
\ckmfoot\ scheme is not a Bayesian scheme.
\vs
The precise way the functions decrease down to zero
is obviously arbitrary: one needs to define a standard.
The proposed expressions for $\Likthe(i)$ are presented 
in Section~\ref{sec_likelihoodsAndSysErrors}.
Their relevant characteristic is the following:
they use two continuously varying parameters,
denoted $\zeta$ and $\kappa$.
The first parameter is a scale factor which
fixes the allowed range where \hbox{$\Likthe(i)=1$.}
The second parameter determines the transition to zero.
The parameter values permit to cover a large spectrum of schemes,
ranging from \ckmfit\ ($\zeta=1$,\, $\kappa=0$),
to a Gaussian scheme ($\zeta=0$,\, $\kappa=1$)
% Big Discovery : This is not Bayesian identical !!!
and defining a standard\footnote
{
	Obviously, it would be better if theorists, and not 
	experimentalists, choose for these two parameters the 
	values which appear the most adequate for each of 
	their predictions.
}, 
denoted \ckmfoot,
for which $\zeta=1$, and $\kappa\simeq 0.8$.
\vs
Because \ckmfoot\ acknowledges the fact that the allowed 
ranges should not be taken literally,
it offers two advantages over \ckmfit:

\begin{itemize}

\item{}
\ckmfoot\ 
%(like any extended scheme with $\zeta\ge 1$)
is more con\-ser\-va\-ti\-ve
%\footnote
%{
%	It was stated above that the 95\%~CL Scan method is 
%	over-conservative with respect to \ckmfit,
%	and this fact was viewed as an unwelcome feature.
%	Since apparently the same fact is here presented as 
%	a virtue in favor of \ckmfoot, 
%	a comment is in order: the 95\%~CL Scan method is 
%	over-conservative when dealing with statistical errors,
%	whereas here, \ckmfoot\ is more conservative when 
%	dealing with systematic errors.
%}
than \ckmfit: by construction,
a \ckmfoot\ CL is always larger than the corresponding 
\ckmfit\ one, and in the $\rhoeta$ plane 
its \CL\ surface exhibits the same plateau 
at \hbox{$\CL=1$ 
(i.e., 
the \ckmfit\ and \ckmfoot\ $\ymodopt$ spaces are identical).}

\item{}
In case the SM tends to be ruled out by \ckmfit,
the \ckmfoot\ scheme is able to detect the eventual $\yQCD$ parameter(s) which,
if allowed to trespass its allowed range, 
would restore an acceptable agreement between data and theory,
and which value it should take.

\end{itemize}
\noindent{\em Despite the two above arguments in favor of }\ckmfoot,
\ckmfit\ {\em is chosen as the scheme advocated in this paper
rather than} \ckmfoot:
because it uses a simpler and unique presciption to incorporate 
theoretical systematics, it is less prone to be confused with a 
Bayesian treatment. Moreover, \ckmfoot\  does not provide a 
clear-cut distinction between statistical and theoretical 
systematic errors in the fit. Finally, in cases where one determines 
theoretical parameters \via\ the fit, as it is the case, \eg, 
for the quantity $\fbdbd$, \ckmfit\ is the natural choice. But, 
obviously, if \ckmfit\ concludes to a SM failure, then \ckmfoot\ 
should be used.

\subsection{The Bayesian Treatment}
\label{sec_TheBayesianTreatment}

%\subsubsection{The General Approach}
%\label{sec_TheGeneralApproach}
The Bayesian treatment~\cite{Achille1,Achille2} 
considers $\Lik$ as a PDF, from which is defined 
$\F(\a)$, the PDF of $\a$, through the convolution
\beq
\label{BayesianII}
\F(\a)=C \int \Lik(\ymod)\ \delta(\a-\a(\ymod))\ d\ymod~,
\eeq
where the constant $C$ is computed {\it a posteriori}
to ensure the normalization to unity of $\F(\a)$.
In practice, the integral can be obtained very conveniently by 
Monte Carlo techniques. For each point in the $\a$ space one 
sets a confidence level $\CL(\a)$, for example, according to
\beq
\label{eq_CLBayesian}
\CL(\a)\;=\hmm\hmm
	\intl_{\F(\a^\prime)\le\F(\a)}\hmm\hmm \F(\a^\prime) 
	\ d\a^\prime~,
\eeq 
but another definition for the domain of integration can de chosen.
The method concludes by a graphical display of \CL.
In particular,
the $95\%~\CL$ contour can be read-off among others.
New physics is not meant to be detected by the Bayesian treatment:
it is aimed at metrology mostly.

\subsubsection{Comparison with \ckmfit}

Although their graphical display appear similar, 
the Bayesian treatment and the \ckmfit\ scheme
depart significantly:
the meaning attached to a given \CL\ value are not the same.
For the Bayesian treatment, the \CL\ is a quantity {\it defined},
using Eq.~(\ref{BayesianII}), for example by Eq.~(\ref{eq_CLBayesian}).
The justification of this definition lies in the understanding 
that a \CL\ value is meant to provide a quantitative measure 
of our qualitative {\it  degree of belief}.
Whereas one understands qualitatively well what is meant by
{\it  degree of belief},
because of its lack of formal definition,
one cannot check that it is indeed well measured by the \CL:
the argument is thus circular.
One is left with the sheer definition of Eq.~(\ref{eq_CLBayesian}),
which, being just a definition, suffers no discussion.
\vs
The key point in the Bayesian treatment is the use of Eq.~(\ref{BayesianII}),
even though the likelihood contains theoretical components.
This implies
that the $\yQCD$ parameters,
which stem from theorist computations,
are to be considered as random realizations ({\it sic})
of their true values. 
The PDFs of these 'random' numbers are then drawned from guess-work 
(The $[\yQCD]$ ranges do not fare better with respect to that.).
For self-consistency,
if one assumes that a large number of theorists perform the same
$\yQCD$ computation,
the distribution of their results should then be interpreted
as a determination of the $\yQCD$ PDF.
Once injected in Eq.~(\ref{BayesianII}),
this PDF,
the shape of which contains no information on nature, 
but information on the way physicist mind work,
will be transformed into information pertaining to nature.
This entails to a surprising confusion between 
what is an experimental result and what is a thinking result.
As illustrated in Appendix~\ref{TheBayesianMethodcaughtunder-conservative} 
and in Section~\ref{IndirectEvidenceforCPViolation},
it is less the {\it ad hoc} shapes of the PDFs which are 
at stake than the implication of using Eq.~(\ref{BayesianII}).

%yield very similar results under Gaussian assumptions.
%However, 
%they depart significantly when using likelihood 
%components of the type presented in 
%Section~\ref{sec_likelihoodsAndSysErrors},
%whereas the Bayesian treatment is rather insensitive to 
%the detail of $\Likthe$, 
%\ckmfit\ results 
%depend strongly on $\Likthe$. Indeed, as shown in the 
%Appendix A, the Bayesian approach introduces an obvious
%bias into the final probabilities which originates from 
%the unmotivated assumption that the shape of the theoretical 
%p.d.f be known and/or the theoretical parameter constitutes
%a statistically distributed quantity.

%
% ----------------- Likelihoods and Systematic Errors ------------
%
\section{Likelihoods and Systematic Errors}
\label{sec_likelihoodsAndSysErrors}

In Section~\ref{sec_statFrame} we have defined the basic
formalism of the \ckmfit\ scheme.
%We have introduced the theoretical likelihoods $\Likthe$ without
%defining their meaning and contents. 
The treatment of
experimental and theoretical systematics is the subject of 
this section.
\vs
Let $\xo$ be a quantity, which is not a random variable, 
but which is not perfectly known. 
We will consider in turn two quantities of this type:
\begin{itemize}
\item  	A theoretical parameter which is not well determined
	(\eg, $\xo=f_{B_d}$): the theoretical prediction of an 
	observable depends on $\xo$ (\eg, $\Delta M_{B_d}$).
\item  	An experimental bias due to detector/analysis defects:
	the measurement should be corrected for this bias.
\end{itemize}
It is the purpose of this section to suggest a prescription 
of how to incorporate the limited knowledge of such 
quantities into the analysis.
%The experimental measurements are denoted $\xexp$ 
%(with $\sigexp$ being the statistical experimental error),
%and their corresponding theoretical predictions are 
%denoted $\xthe$.
The standard treatment of this problem relies on a $\chi^2$
analysis
\beq
\label{eq_chi2expression}
\chi^2
	=\left({\xexp-\xthe\over\sigexp}\right)^2
  	+ \left({\xo-\xobar\over\sigxo}\right)^2~,
\eeq
where 
\begin{itemize}
\item   $\xthe$ (resp. $\xexp$) depends on $\xo$, 
if the latter is 
	a theoretical parameter (resp. experimental systematics),
\item   $\xobar$ is the expected central value of $\xo$,
\item   $\sigxo$ is the {\em uncertainty} on $\xo$.
\end{itemize}
This standard treatment is satisfactory as long as the degree of
belief we put on the knowledge of the value of $\xo$ is peaked 
at $\xobar$ and distributed like a Gaussian.
This is usually summarized by
\beq
\label{eq_defsigxo}
\xo=\xobar\pm\sigxo~.
\eeq
However, this is not necessarily what is intended to be meant by 
Eq.~(\ref{eq_defsigxo}). 
Rather, the theorist (resp. the 
experimentalist) {\it may} mean that the prediction (resp. the 
measurement) can take any value obtained by varying $\xo$ 
at will within the range $[\xobar-\zeta\sigxo,\xobar+\zeta\sigxo]$ 
(denoted {\em the allowed range} below, where $\zeta$ is a constant 
scale factor of order unity), but that it is unlikely that 
$\xo$ takes its true value outside the allowed range. 
This does {\em not} imply that the possible values are 
equally distributed within the allowed range:
they are not distributed {\em at all}\footnote
{
	In some cases (\eg, lattice QCD) 
	statistical fluctuations may enter in the computation.
	In such instances one may reliably define a Gaussian
	likelihood for this component of the  theoretical 
	uncertainty.
}. 
If Eq.~(\ref{eq_defsigxo}) is given 
such a meaning, then the statistical analysis should treat all 
$\xo$ values within the allowed range on the same footing
(which again does not imply with equal {\em probability}):
this corresponds to the \ckmfit\ scheme (with $\zeta=1$).
%But,
\vs
This is not the case for the $\chi^2$ expression of 
Eq.~(\ref{eq_chi2expression}) since the farther $\xo$ moves away 
from $\xobar$, 
the larger becomes the related component of $\chi^2$. 
%However, 
%one would like this component not to contribute at all to the $\chi^2$ 
%if $\xo$ remains within 
%the allowed range in order not to introduce a hierarchy
%for the allowed theoretical values. 
%On the other hand,
\vs
On the other hand,
it might also be useful to define specific tails instead of
sharp cuts,
thus allowing the theoretical parameters to 
leave their allowed ranges,
if needed:
this corresponds to the \ckmfoot\ scheme.
%This provides important hints about the influence of the 
%theoretical parameters in particular 
%(\cf, Section~\ref{sec_constrainedFits}).
\vs
The idea is to move from a pure $\chi^2$ analysis to a 
log-likelihood one, redefining the $\chi^2$ to be
\beq
\label{eq_chisqNew}
\chi^2=
	\left({\xexp-\xthe\over\sigexp}\right)^2
	-2\ln\Hatsyst(\xo)~,
\eeq
where $\Hatsyst(\xo)$, hereafter termed the {\em Hat} function,
is a function equal to unity for $\xo$ within the allowed 
range. Its precise definition is given below.

\subsection{The Hat Function}
\label{sec_flattenedGaussian}

The Hat function $\Hatsyst(\xo,\kappa,\zeta)$ is a continuous function 
defined as
\beq
\label{eq_flattenedGaussian}
-2\ln\Hatsyst(\xo,\kappa,\zeta)=
	\left\{
	\begin{array}{ll}
	0~, 
		& \forall\xo\in[\xobar\pm\zeta\sigxo] \\
	\left(\displaystyle\frac{\xo-\xobar}{\kappa\sigxo}\right)^2
	-\left(\displaystyle\frac{\zeta}{\kappa}\right)^2~,~~
	& 	\forall\xo\notin[\xobar\pm\zeta\sigxo]
	\end{array}
	\right.
\eeq
where the constant $\kappa$ determines the behavior of the function
outside the allowed range.
For the \ckmfit \ scheme $\kappa =0$ is used.
To define a standard 
$\kappa$ can be chosen to be a function of $\zeta$ such that the relative 
normalizations of $\Hatsyst(\xo,\kappa,\zeta)$
(briefly viewed here, for the purpose of defining a standard, as a PDF) 
within and outside the allowed 
range equal those of a Gaussian of width $\sigxo$
\beq
\intl_{-\infty}^{+\infty}\Hatsyst(\xo,\kappa,\zeta)\ d\xo
		\cdot\intl_0^{\zeta/\sqrt{2}}e^{-t^2}\ dt
	= \sqrt{\pi}\,\zeta\sigma_0~.
\eeq
%With this definition,
%the Gaussian function is recovered by setting $\zeta=0$.

\begin{figure}[t]
  \epsfxsize12cm
  \centerline{\epsffile{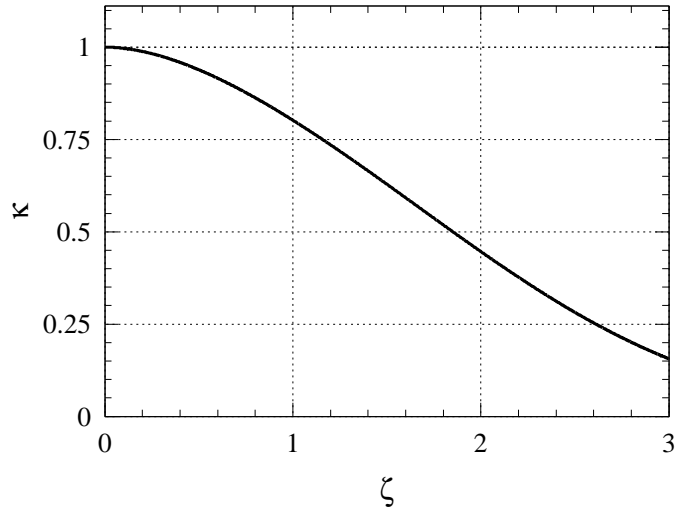}}
  \vspace{-0.0cm}
  \caption[.]{\label{fig_kappa_vs_zeta}\em
              The $\kappa$ parameter as a function 
	      of $\zeta$ (see text).}
\end{figure}

\noindent
The parameter $\kappa$ is numerically computed as a function of $\zeta$. 
The result is shown in Fig.~\ref{fig_kappa_vs_zeta},
in the range of interest $0\le\zeta\le 3$. 
%The $\kappa$ 
%value steadily decreases while $\zeta$ increases. 
For the 
limit $\zeta\rightarrow0$ one obtains $\kappa\rightarrow 1$,
and the Hat becomes a pure Gaussian.
The \ckmfoot\ scheme is defined by $\zeta=1$,
for which one obtains $\kappa\simeq 0.8$.
\vs
Examples of Hat functions with $\xobar=0$ and $\sigxo=1$
are shown on the left plot of 
Fig.~\ref{fig_loglik}. 
Being a function, 
and not a PDF,
$\Hatsyst(\xo)$ needs not be normalized to unity. 

\subsection{Combining Statistical and Systematic Uncertainties}
\label{CombiningStatisticalandSystematicUncertainties}
Having defined $\Hatsyst(\xo)$, 
and following Eq.~(\ref{digest}), 
one proceeds with the 
minimization of the $\chi^2$ of Eq.~(\ref{eq_chisqNew})
by allowing $\xo$ to vary freely.
\vs
For theoretical systematics,
the result depends on the way $\xo$ enters $\xthe$, 
and not much more can be said in generality. 
\vs
For experimental and theoretical systematics
where $\xo$ can be assumed to be an unknown offset\footnote
{
	If systematics take the form of an unknown 
	multiplicative factor, and this is often the case 
	for theoretical uncertainties, a treatment similar 
	to the one discussed here applies.
}: 
the quantity to be confronted to the theoretical prediction $\xthe$ 
is simply $\xexp+\xo$.
Omitting the details of 
straightforward calculations, 
and assuming that $\xobar=0$ 
(otherwise $\xexp$ should be corrected for it), one obtains,
after minimization of the $\chi^2$ with respect to $\xo$:
%and defining the effective likelihood $\expHatsyst$ by
%$\chi^2_{{\min};\,\xo}=-2\ln\expHatsyst(\xexp-\xthe)$
\begin{figure}[t]
  \epsfxsize12cm
  \centerline{\epsffile{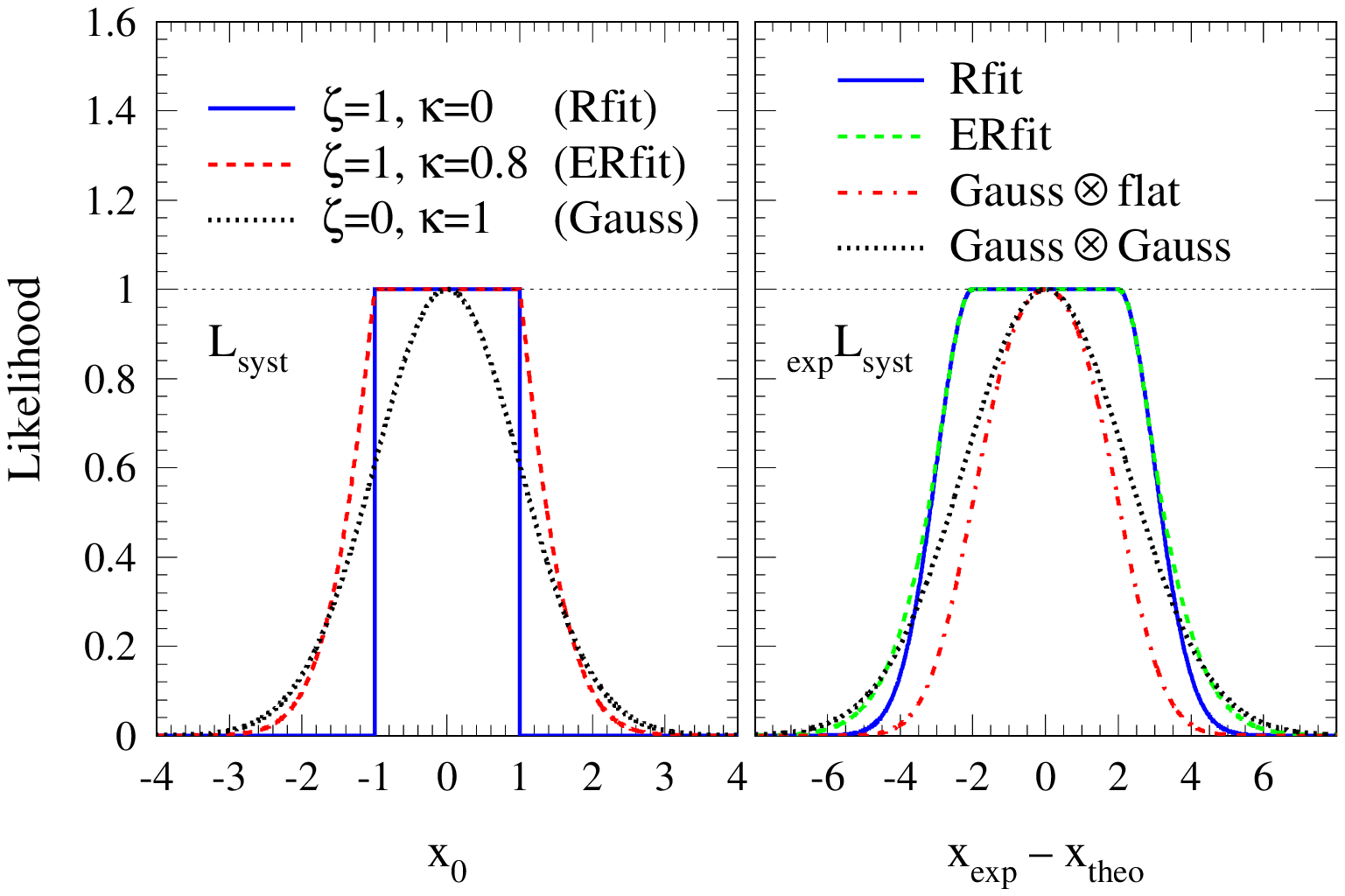}}
  \vspace{-0.0cm}
  \caption[.]{\label{fig_loglik}\em
	The left hand plot shows the Hat functions
        ($\xobar=0$ and $\sigxo=1$) used for the {\rm \ckmfit} 
	scheme, the {\rm \ckmfoot} scheme and the Gaussian treatment.
	The right hand plot shows the combined likelihood
	$\expHatsyst$ (with $\xobar=0$ and $\sigexp=\sigxo=1$)
	obtained from Eq.~(\ref{digest}) for the {\rm \ckmfit} scheme,
	the {\rm \ckmfoot} scheme,
	a convolution of a Gaussian with a uniform distribution
	(hence taken as a PDF, following the Bayesian approach)
	and a convolution of two Gaussians 
	(see Appendix~\ref{TheBayesianMethodcaughtunder-conservative}).}
\end{figure}

\begin{itemize}

\item   $\mid\xexp-\xthe\mid\le\zeta\sigxo$~:\hskip .5truecm
$\chi^2_{\min ;\,\xo}=0~.$
%	\beq
%	\chi^2_{\min ;\,\xo}=0~.
%	\eeq

\item   $\zeta\sigxo\le\mid\xexp-\xthe\mid\le\zeta\sigxo
	(1+({\sigexp\over\kappa\sigxo})^2)$~:\hskip .5truecm
	$\chi^2_{\min ;\,\xo}
		=\left({\mid\xexp-\xthe\mid
		-\zeta\sigxo\over\sigexp}\right)^2~.$
%	\beq
% 	\chi^2_{\min ;\,\xo}
%		=\left({\mid\xexp-\xthe\mid
%		-\zeta\sigxo\over\sigexp}\right)^2~.
%	\eeq

\item   $\mid\xexp-\xthe\mid\ge\zeta\sigxo
	(1+({\sigexp\over\kappa\sigxo})^2)$~:\hskip .5truecm
	$\chi^2_{\min ;\,\xo}
		={(\xexp-\xthe)^2\over\sigexp^2+(\kappa\sigxo)^2}
		-\left({\zeta\over\kappa}\right)^2~.$
%	\beq
%	\chi^2_{\min ;\,\xo}
%		={(\xexp-\xthe)^2\over\sigexp^2+(\kappa\sigxo)^2}
%		-\left({\zeta\over\kappa}\right)^2~.
%	\eeq

\end{itemize}
In the limit $\zeta\rightarrow 0$ 
(and hence, $\kappa\rightarrow 1$) only the third instance 
is met, and one recovers the usual rule of adding in 
quadrature the statistical and the systematic uncertainties.
Otherwise, the result is non trivial. 
An example of the effective likelihood
$\expHatsyst(\xexp-\xthe)\equiv{-{1\over 2}\chi^2_{\min ;\,\xo}}$
(with $\xobar=0$ and $\sigexp=\sigxo=1$)
is shown in the right hand plot of Fig.~\ref{fig_loglik} for 
the \ckmfit\ scheme,
the \ckmfoot\ scheme,
a convolution of a Gaussian with a uniform distribution
(hence taken as a PDF, following the Bayesian approach)
and a convolution of two Gaussians.

%
% --------------------- Ingredients -----------------------------
%
\section{Fit Ingredients}
\label{sec_fitIngredients}

This section provides a compendium of the measurements and 
SM predictions entering the overall constrained CKM fit.
In some cases, we pre-combine compatible measurements by 
means of a simple weighted mean in order to speed up the 
fit by reducing the effective number of degrees of freedom 
(see the discussion in Section~\ref{sec_TheExperimentalLikelihood}). 
Since the solution of a $\chi^2$ minimization of Gaussian distributed 
measurements corresponds to the weighted mean, this entails no loss 
of information. Below, we give a status of the input quantities used. 
The corresponding numerical values that enter the fit and the
treatment of their uncertainties within \ckmfit\
are summarized in Table~\ref{tab_ckmValues}.

\subsection{The CKM Matrix Elements}
\label{TheCKMMatrixElements}

\begin{itemize}
\item   {\boldmath$|V_{ud}|:$}
        The matrix element $|V_{ud}|$ has been extracted by 
	means of three different methods: 
	super-allowed nuclear $\beta$-decays, neutron $\beta$-decay and 
        pion $\beta$-decay.
	\vs
        Using the lifetime measurements of super-allowed nuclear 
        $\beta$-decays with pure Fermi-transitions ($0^+\rightarrow 0^+$), 
	$|V_{ud}|$ can be extracted with high precision. 
        Averaging the values of nine different 
        super-allowed nuclear $\beta$-decays~\cite{superAllow} 
        the result including nucleus-independent 
        and nucleus-dependent radiative corrections as well as 
	charge-dependent corrections 
	(see Ref.~\cite{heraB} for a compendium of references) is: 
	$|V_{ud}|=0.97400\pm0.00014_{\rm exp}
	\pm0.00048_{\rm theo}$~\cite{heraB,superAllow}. 
        The precision of $|V_{ud}|$ from nuclear 
	$\beta$-decays is often questioned in light of the 
	observed '$2\sigma$' deviation from the unitarity condition 
	when combining this value with the best knowledge 
	for $|V_{us}|$ and $|V_{ub}|$. 
        A possible enhancement of $|V_{ud}|$ is predicted
	by a quark-meson coupling model due to a change 
	of charge symmetry violation for quarks inside bound 
        nucleons compared to unbound nucleons~\cite{vudsaito}. 
        Since the status here is unclear, the error has been 
        enlarged by the amount of the possible correction using 
        the PDG rescaling~\cite{pdg2000}: 
	$|V_{ud}| = 0.9740 \pm 0.0010$.\footnote
	{
		It should be kept in mind that such a procedure might 
		hide a possible violation of the unitarity condition 
		in the first family.
	} 
	\vs
        For the neutron $\beta$-decay no nuclear structure effects 
	play any role. 
        However, $|V_{ud}|$ has to be extracted from two quantities, 
	the neutron lifetime and the ratio $g_{A}/g_{V}$. 
	In contrast to nuclear $\beta$-decays 
        these measurements are not dominated by theoretical 
	uncertainties. The weighted mean for the neutron lifetime 
        measurements is
        $\tau_{n} = (885.7 \pm 1.0)~{\rm s}$~\cite{vudarzu}
	and the average value for 
	$g_{A}/g_{V}$ reads $-1.2699 \pm 0.0029$~\cite{vudabele} 
        where the error was rescaled by a factor of two 
        due to inconsistencies in the data set. 
        Combining these numbers we get:
        $|V_{ud}| = 0.9738\pm0.0020_{g_{A}/g_{V},\tau_{\rm n}}\pm0.0004_{\rm rad}$,
	where the error is dominated by the experimental error on 
	$g_{A}/g_{V}$.
	\vs
        The pion $\beta$-decay $\pi^{+} \rightarrow \pi^{0} e^{+} \nu_{e}$
        is a very attractive candidate to extract $|V_{ud}|$ 
        from the branching ratio and the pion lifetime, since 
	it is mediated by a pure vector transition and 
	does not suffer from nuclear structure effects. 
	However, due to the small branching ratio,
	${\rm BR} = (1.025 \pm 0.034) \times 10^{-8}$~\cite{pdg2000}, 
	the present statistical accuracy 
	is not competitive with the other methods: 
        $|V_{ud}|=0.967 \pm 0.016_{\rm BR} \pm 0.0009_{\rm theo}$.
	\vs
	Assuming all three measurements to be Gaussian distributed\footnote
	{
	        Being consequent, also in this case, one should treat 
		the theoretical errors as ranges.
        	However, as long as the relative uncertainty on $\lambda$ 
		from $|V_{ud}|$ and $|V_{us}|$ is much smaller than 
		what one obtains for $\rhoeta$ from constraints like 
		$|V_{ub/V_{cb}}|$, $|\epsk|$, $\dmd$ and
	        $\dms/\dmd$, the procedure used is certainly not a
		critical issue.
	} 
	we combine them \via\ a weighted mean, yielding:
	$|V_{ud}|=0.97394 \pm 0.00089$.
\item   {\boldmath$|V_{us}|:$ }
        The analyses of kaon and hyperon semileptonic decays 
	provide the best determination of $|V_{us}|$ 
	directly related to $\lambda$ in the Wolfenstein parameterization.
        However, due to theoretical uncertainties from the 
	breakdown of SU(3) flavor symmetry, the hyperon decay data 
        are less reliable~\cite{vusdhk,vusflores}. As a consequence, 
	we use only the value obtained from the vector transitions 
        $K^{+} \rightarrow \pi^{0} e^{+} \nu_{e}$ and 
        $K_{L} \rightarrow \pi^{-} e^{+} \nu_{e}$~\cite{pdg2000}. 
        Owing to the small electron mass, only one form factor 
        plays a role in $K_{\rm e3}$ decays the functional dependence 
	of which can be extracted from data. 
        The form factor value at zero recoil, $f_{1}(0)$, 
	is calculated within the framework of chiral perturbation theory 
	and is found to be:
	$f_{1}^{K^{0}\pi^{-}}(0) = 0.961 \pm 0.008$~\cite{vusleutw}.
	The error estimate for this value was criticized in 
	Ref.~\cite{vuspaschos}.
	However, a relativistic constituent quark model, successful in
	the description of electroweak properties of light mesons
	gives a consistent result:
	$f_{1}^{K^{0}\pi^{-}}(0) = 0.963 \pm 0.004$~\cite{vusjaus}. 
	Channel-independent and channel-dependent 
        radiative corrections~\cite{vusmarciano,vuswilliams}
	as well as charge symmetry ($K^{+}/K_{L}$) 
        and charge independence ($\pi^{-}/\pi^{0}$) breaking 
        corrections~\cite{vusleutw} are applied 
	in order to compare the results from both channels:
	$f_{1}^{K^{0}\pi^{-}}(0) |V_{us}|
         =0.2134\pm 0.0015_{\rm exp}\pm 0.0001_{\rm rad}$ 
	($K^{+} \rightarrow \pi^{0} e^{+} \nu_{e}$) and
	$f_{1}^{K^{0}\pi^{-}}(0) |V_{us}| 
        = 0.2101 \pm 0.0013_{\rm exp} \pm 0.0001_{\rm rad}$
	($K_{L} \rightarrow \pi^{-} e^{+} \nu_{e}$).
	The weighted average is: 
	$f_{1}^{K^{0}\pi^{-}}(0) |V_{us}| = 0.2114 \pm 0.0016$
	where the error was rescaled by a factor of 1.6 
	to account for inconsistencies.
	\vs
	The result then reads
	$|V_{us}| = 0.2200 \pm 0.0017_{\rm exp} \pm 0.0018_{\rm theo} = 
                    0.2200 \pm 0.0025$.
        As in the case for $V_{ud}$ all uncertainties were considered
        as Gaussian errors.

%\item   {\boldmath$|V_{ub}/V_{cb}|:$}
%	An excess of leptons in the momentum region 
%	$2.3~{\rm GeV}\le p_\ell^*\le2.6~{\rm GeV}$ stems from
%	charmless $b\rightarrow {\rm u}\,\ell^-{\bar\nu}_\ell$ 
%	transitions rather than $b\rightarrow {\rm c}\,\ell^-{\bar\nu}_\ell$,
%	as the momenta are beyond the phase space 
%	of the latter~\cite{vub_cbCleo1,vub_cbCleo2,vub_cbArgus}.
%	The small phase space window introduces theoretical 
%	uncertainties from the phenomenological models
%	used for the analysis~\cite{modACCMM,modISGW,modKS,modWSB}.
%	In addition, possible violations of quark-hadron duality
%	might be enhanced in a small part of the phase space. 
%	A new measurement of $|V_{ub}/V_{cb}|$ has been performed 
%        by the DELPHI \CoL\ on the basis of the shape of the invariant 
%        mass distribution of the recoil tracks with respect to the 
%        lepton. No cut on the lepton momentum has been applied which 
%        considerably reduces the model dependence of the result which 
%        is used in the combined LEP result on $|V_{ub}|$~\cite{combB}.
\item   {\boldmath$|V_{ub}|:$	}
        Both, inclusive B-decays ($b\rightarrow X_u\ell^-{\bar\nu}_\ell$), 
        measured at LEP~\cite{vubAleph,vubL3,vub_cbDelphi}, and 
	exclusive B-decays ($B^0\rightarrow\pi^-\ell^+\nu_\ell$, 
	$B^0\rightarrow\rho^-\ell^+\nu_\ell$),
        measured by CLEO~\cite{vubCleo}, allow an extraction of the
	third column element $|V_{ub}|$\footnote
	{
		The determination of $|V_{ub}|$ from the lepton endpoint 
		spectrum, obtained by ARGUS~\cite{vub_cbArgus} and 
		CLEO~\cite{vub_cbCleo1,vub_cbCleo2}, 
		suffers from large model 
		dependencies~\cite{modACCMM,modISGW,modKS,modWSB}.
		In addition, possible violations of quark-hadron duality might 
		be enhanced in this small part of the phase space. Hence, 
		these results were not taken into account for this analysis. 
	}.
	\vs
	The exclusive measurements are dominated by the theoretical 
	uncertainty due to the model dependence in the determination 
	of the form factor. The exclusive CLEO measurements give: 
        $|V_{ub}|=(3.25 \pm 0.14_{\rm stat}{^{+0.21}_{-0.29}}_{\rm sys} 
                         \pm 0.55_{\rm theo}) \times10^{-3}$,
        where the error is dominated by the theoretical uncertainties.
	We add the statistical and experimental systematic error
	in quadrature and consider the theoretical error as a range: 
	$|V_{ub}|=(3.25 \pm 0.29_{\rm exp}\pm0.55_{\rm theo})\times10^{-3}$.
	There is some hope that exclusive measurements in the future may 
	take advantage of unquenched lattice QCD calculations and thus
	reduce the model dependent error.
	\vs
        The three inclusive LEP measurements rely on different techniques 
	and are combined in Ref.~\cite{combB}, 
	taking into account all uncorrelated and correlated errors:\\
        $|V_{ub}| = (4.04^{+0.41}_{-0.46}  ({\rm stat+det})^{+0.41}_{-0.46}
	(b \rightarrow c) ^{+0.24}_{-0.25}(b \rightarrow u) 
	\pm 0.02(\tau_{b}) \pm 0.19({\rm HQE})) \times 10^{-3}$.
        The theoretical uncertainty from the Heavy Quark Expansion (HQE)
	is a combination of three sources: 
        the neglect of higher terms including $1/m_b^3$, the 
	uncertainty in the b-quark mass $m_b$ and from 
	perturbative corrections.
	The extraction of $|V_{ub}|$ from the inclusive semileptonic 
	${\rm BR}(b \rightarrow {\rm u} \,\ell^{-} \bar\nu_{\ell})$ 
	relies on the validity of quark-hadron duality.
	Although quark-hadron duality can not be expected to be exact, there 
	are good reasons that inclusive semileptonic decays of beauty hadrons 
	are described quite accurately by HQE~\cite{bigiRioManifesto}.
	However, since the analyses have to apply cuts in order to suppress 
	the background from $b \rightarrow c$ transitions only a part of
	the total semileptonic rate is measured which could lead to sizable 
	effects from quark-hadron duality violation which are difficult
	to quantify. 
	In the case of the inclusive measurements we combine the theoretical 
	uncertainties from HQE and the large systematic uncertainties due to 
	$b \rightarrow c$ and $b \rightarrow u$ transitions
	by adding them in quadrature and obtain:
        $|V_{ub}| = (4.04 \pm 0.44_{\rm exp} \pm 0.54_{\rm model}) \times 10^{-3}$.
	\vs	
	For the combined result of inclusive and exclusive measurements 
	we obtain:
        $|V_{ub}| = (3.49 \pm 0.24 \pm 0.55_{\rm theo}) \times 10^{-3}$ 
	where only the first error was used for the weighted mean. 
	The maximum of both single ranges was assigned as the final 
	systematic theoretical uncertainty. 
%	\vs
%	In an alternative treatment for the inclusive result only the combination
%	of the $b \rightarrow u$ and the HQE uncertainties can be combined.
%	In this case, the inclusive result reads: 
%	$|V_{ub}| = (4.04 \pm 0.63_{\rm exp} \pm 0.31_{\rm model}) \times 10^{-3}$
%	and the combined result: 
%	$|V_{ub}| = (3.39 \pm 0.26 \pm 0.55_{\rm theo}) \times 10^{-3}$.
%	Since the exclusive result dominates the final result we always attribute
%	the exclusive model uncertainty to the final result.
\item   {\boldmath$|V_{cd}|:$}
        Both matrix elements, $|V_{cd}|$ and $|V_{cs}|$, can be determined 
        from di-muon production in deep inelastic scattering (DIS) of 
        neutrinos and anti-neutrinos on nucleons. In an analysis performed 
        by CDHS~\cite{vcdCdhs}, $|V_{cd}|$ and $|V_{cs}|$ are extracted by 
	fitting the data of three experiments, CDHS~\cite{vcdCdhs}, 
	CCFR~\cite{vcdCcfr} and CHARM II~\cite{vcdCharmii}, giving: 
        $|V_{cd}|^2 \times B_{c} = (4.63 \pm 0.34) \times 10^{-3}$, where
        $B_{c} = 0.0919 \pm 0.0094$~\cite{bolton,ushida,kubota} is the 
	weighted average of semileptonic 
        branching ratios of charmed hadrons produced in neutrino-nucleon DIS.
	This results in: $|V_{cd}| = 0.224 \pm 0.014$~\cite{heraB}.
\item   {\boldmath$|V_{cs}|:$}
        Besides DIS, the matrix element $|V_{cs}|$ can be obtained from 
        $D_{\rm e3}$ decays, charm tagged $W$ decays and hadronic $W$ decays. 
	\vs
        The average DIS result from CDHS, CCFR and CHARM II is
        $\kappa |V_{cs}|^{2} B_{c} = (4.53 \pm 0.37) \times10^{-2}$ where 
        $\kappa = 0.453 \pm 0.106^{+0.028}_{-0.096}$ is the relative size 
        of strange quarks in the sea compared to $\bar{u}$ and $\bar{d}$
        resulting in $|V_{cs}| = 1.04 \pm 0.16$~\cite{pdg2000}.
	\vs
        Similar to $K_{\rm e3}$, decays $|V_{cs}|$ can be also extracted 
	from $D_{\rm e3}$ decays. 
	However, the theoretical uncertainty in the form factor calculation
        $f_{1}(0) = 0.7 \pm 0.1$~\cite{vcsMontanet} limits the precision: 
	$|V_{cs}| = 1.04 \pm 0.16$~\cite{heraB}, 
        in perfect agreement with $|V_{cs}|$ from DIS.
	\vs
        Under the assumption that unitarity holds for three families, 
	the ratio $R_{c} = \Gamma(W^{+} \rightarrow c\bar{q})/ \Gamma(W^{+} 
	\rightarrow {\rm hadrons}) 
        = \sum_{i=d,s,b}|V_{ci}|^{2}/ (\sum_{i=d,s,b;j=u,c}|V_{ji}|^{2})$ 
        for $W$ decays is expected to be 1/2. The results of all 
	LEP2 experiments are consistent with this expectation and give
        $|V_{cs}| = 0.97\pm 0.09_{\rm stat}\pm 0.07_{\rm sys}$
	~\cite{pdg2000,vcsAleph,vcsDelphi,vcsL3,vcsOpal}.
        The ratio of hadronic and leptonic $W$ decays measured 
	by LEP2 provides the tightest bound on $|V_{cs}|$ 
	if unitarity for three families is assumed: 
        $R_{c} = \Gamma(W^{+} \rightarrow {\rm hadrons}/ 
	\Gamma(W^{+} \rightarrow {\rm leptons}) 
               = \sum_{i=d,s,b;j=u,c}|V_{ji}|^{2} 
	\times (1+\alpha_s(m_{W})/\pi)$.
        From the four LEP experiments 
	$|V_{cs}| = 0.989 \pm 0.016$~\cite{pdg2000} 
	is found where the errors on the single measurements are dominated 
	by statistical errors\footnote
	{
		The measurement of $R_{c}$ should be used in 
		the fit, rather than the quoted $|V_{cs}|$ 
		determination which is derived from it.
		This piece of information is not used here,
		for the sake of simplicity.
	}.
	\vs
	Very recently, the OPAL collaboration has presented a new direct
	determination of $|V_{cs}|$ from $W \rightarrow X_{c} X$ resulting in
	$|V_{cs}| = 0.969 \pm 0.058$~\cite{vcsOpal2}, which we use 
	in the fit.
\item   {\boldmath$|V_{cb}|:$}
	In the Wolfenstein parameterization, $|V_{cb}|$
	determines the parameter $A$ the precision of which plays 
	an important part for the constraints 
	$|V_{ub}/V_{cb}|$, $|\epsk|$ and $\dmd$. 
	It is obtained from exclusive $B\rightarrow D^{(*)}\ell{\bar\nu}_\ell$
	and inclusive semileptonic b decays to charm, 
        $b \rightarrow \ell^{-} \bar{\nu_{\ell}} X_{c}$, 
        both measured by CLEO and the LEP experiments. 
        The theoretical framework for extracting numerical values for
	$|V_{cb}|$ from the measured decay rates is 
        Heavy Quark Effective Theory (HQET)~\cite{hqet} 
        for exclusive measurements
	and HQE~\cite{hqt} for inclusive measurements.
	\vs
	The exclusive results are given in the form
	${\cal F}_{D^*}(1)|V_{cb}|$, so that they must be divided 
	by the value of the Isgur-Wise function at zero-recoil.
        The form factor at zero-recoil, ${\cal F}_{D^*}(1)$, is 1 in
        the heavy quark limit. For finite quark masses the corrections 
        can be calculated in HQET. 
	There are open discussions in the literature concerning the 
	calculation of the $1/m_{Q}^2$ corrections for ${\cal F}_{D^*}(1)$.
	Here we use the value
        ${\cal F}_{D^*}(1) = 0.913 \pm 0.042$~\cite{schune} for which
	several references have been taken into account. 
	In the future, the most accurate determinations 
	of ${\cal F}_{D^*}(1)$ are expected to come from lattice QCD. 
	A first result with a quite small error reads 
	${\cal F}_{D^*}(1) = 0.935 \pm 0.030$~\cite{formFaSimone}.
        The results from LEP and CLEO read
	\beqns
	  \begin{array}{rclll}
	   {\cal F}_{D^*}(1)|V_{cb}|_{\rm LEP} &=& 
                           0.0350 \pm 0.0007_{\rm stat} 
                                  \pm 0.0015_{\rm sys}&\cite{combB,newcombB}\\
	   {\cal F}_{D^*}(1)|V_{cb}|_{\rm CLEO} &=& 
			   0.0424 \pm 0.0018_{\rm stat} 
				  \pm 0.0019_{\rm sys}&\cite{CLEO2000VCB}\\
	   {\cal F}_{D^*}(1)|V_{cb}| &=& 
			   0.0373 \pm 0.0013\\
	    |V_{cb}| &=& 
			   0.0409 \pm 0.0014 \pm 0.0019_{\rm theo}.
	  \end{array}
	\eeqns
	The combined fit of the CLEO and LEP numbers results in a 
	confidence level of $7\%$.
	\vs
        The theoretical error for the inclusive measurement contains the 
	uncertainty in the kinetic energy $\mu_{\pi}^2$ of the b-quark 
	inside the b-hadron and uncertainties from perturbative corrections, 
	the b-quark mass and the neglect of higher order terms in the 
	$1/m_b$ expansion including $1/m_b^3$ terms~\cite{bibivubvcb}. 
	As in the case of the inclusive determination of $V_{ub}$ possible 
	violations of quark-hadron duality could imply sizable effects.
	Future experimental investigations should aim to shed more light
	on this topic. The most recent inclusive results read	
	\beqns
	  \begin{array}{rclll}

	   |V_{cb}|_{\rm LEP} &=& 
                   0.04076 \pm 0.00050_{\rm exp} 
                                   \pm 0.00204_{\rm theo}&\cite{combB,newcombB}\\
	   |V_{cb}|_{\rm CLEO} &=& 
                   0.041   \pm 0.0010_{\rm stat} \pm 0.0020_{\rm sys}
                           \pm 0.00205_{\rm theo}&\cite{vcbInclCleo,vcbMarina}\\
	  \end{array}
	\eeqns
	The weighted mean of exclusive and inclusive results is
	$|V_{cb}| = (40.76 \pm 0.50 \pm 2.0_{\rm theo})\times10^{-3}$, 
	and is dominated by the inclusive measurement.
	In light of the controversial experimental and theoretical 
        situation in the exclusive sector and possible violations of 
	quark-hadron duality the theoretical uncertainty was not further 
	reduced.
\item   {\boldmath$|V_{tb}|:$	}
	Assuming unitarity for three families, one obtains $|V_{tb}|$ 
        from the ratio of the bottom quark production in top decays to 
        the total top decay width: $|V_{tb}| = 0.99\pm 0.15$~\cite{vtbCdf}. 
        The unitarity assumption, here explicitly used, could be removed.
	However, owing to the poor precision presently achieved, 
	we do not use this measurement in the global CKM fit.
\item   {\boldmath$|V_{ts}V_{tb}/V_{cb}|:$	}
	The inclusive ratio of $b\rightarrow {\rm s}\,\gamma$ to
	$b\rightarrow {\rm c}\,\ell^-{\bar\nu}_\ell$ production 
	provides a measure of the third row CKM elements. Present 
        accuracy is only fair:  
	$|V_{ts}V_{tb}/V_{cb}| 
	= 0.93\pm0.14\pm0.08$~\cite{refHeraB,vtstbAleph,vtstbCleo}.
        Hence, the value is not used in the fitting procedure.
\end{itemize}

\subsection{CP Observables and Mixing}
\label{CPObservablesandMixing}

Constraints on the CKM phase are obtained by the CP-violating 
observables in the $\kdo$ and $\bdo$ systems, and by $\bdo$
and $\bso$ mixing.
\begin{itemize}
\item   {\boldmath$|\epsk|:$}
	Indirect CP violation in the $\kdo$ system is measured by 
	\beq
	   \epsk = \frac{2}{3}\eta_{+-} + \frac{1}{3}\eta_{00}~,
	\eeq
	with $\eta_{+-}$ ($\eta_{00}$) being the ratio of the
	amplitudes of the long-lived and short-lived neutral 
	kaons decaying into two charged (neutral) pions. They have 
        been measured to an accuracy of 1\%~\cite{gew1,apo1,ang1,chr1}. 
        The averaged value of Ref.~\cite{pdg2000} is used in this 
	analysis. Within the SM, CP violation is induced by 
	$\Delta S=2$ transitions owing to box diagrams. 
        It can be related to the CKM-matrix elements by means of the 
        vacuum insertion approximation, used to determine the
	hadronic matrix element
        \beq
        \langle 
	       \bar{K^0}|(\bar{s}\gamma^{\mu}(1-\gamma^{5})d)^2|K^0
	\rangle 
         = \frac{8}{3} m_{K}^2 f_{K}^2 B_{K}~.
	\eeq
	Neglecting the real part of the non-diagonal element of 
	neutral kaon mass matrix $M_{12}$, one obtains
        \beqn
	\label{eq_epsk}
 	|\epsilon_{K}|
		&=& \frac{G_{F}^2 m_{W}^2 m_{K} f_{K}^2}
                       {12 \sqrt{2} \pi^2 \Delta m_{K}}
                  B_{K}\bigg(  
                      \eta_{cc} S(x_{c},x_{c}) 
                         {\rm Im}\big[(V_{cs} V_{cd}^*)^2\big]
                      + \eta_{tt} S(x_{t},x_{t}) 
                         {\rm Im}\big[(V_{ts} V_{td}^*)^2\big]\nonumber\\
                & &\hspace{3.5cm}
                      +\; 2\eta_{ct} S(x_{c},x_{t}) 
                         {\rm Im}\big[V_{cs} V_{cd}^* V_{ts} V_{td}^*\big]
                        \bigg)
	\eeqn
	Here, the $S(x_{i},x_{j})$ are the Inami-Lim 
        functions~\cite{inamilim}
   	\beqn
	       S(x)\equiv S(x_i,x_j)_{i=j} 
			&=& x\left(\frac{1}{4} + \frac{9}{4(1 - x)}
                               - \frac{3}{2(1 - x)^2}\right)
                          - \frac{3}{2}\left(\frac{x}{1 - x}\right)^{\!3}
                            {\rm ln(x)}~, \nonumber \\
	\label{eq_inamilim}
               S(x_i,x_j)_{i\ne j}  &=&  x_i x_j \Bigg[
                                \left(\frac{1}{4} + \frac{3}{2(1 - x_i)}
                                       - \frac{3}{4(1 - x_i)^2}\right)
                                \frac{1}{x_i - x_j}{\rm ln(x_i)}\nonumber \\
                           & &  \hspace{1cm}+\; 
				(x_i\leftrightarrow x_j)
                                - \frac{3}{4}\frac{1}{(1 - x_i)(1 - x_j)}
                                      \Bigg]~,
        \eeqn
	depending on the masses of the virtual charm and top quarks in the 
	box diagrams ($x_{i} = m_{i}^2/m_{W}^2$).         
        The QCD corrections to the Inami-Lim functions have been calculated
        to next-to-leading order: $\eta_{cc} =1.38 \pm 0.53$~\cite{hn1}, 
	$\eta_{tt} =0.574 \pm 0.004$~\cite{bjw} and 
	$\eta_{ct} = 0.47 \pm 0.04$~\cite{hn1} 
	(for a compendium see also~\cite{bbl}).
	The kaon decay constant has been extracted from the 
	leptonic decay rate: 
	$f_{K}=(159.8\pm1.4_{|V_{us}|}\pm 0.44_{\rm theo})~{\rm MeV}$~\cite{pdg2000}.
        The $K_{S}-K_{L}$ mass difference is known with 
	excellent accuracy, 
        $\Delta m_{K}=(3.4885 \pm 0.0008)\times 10^{-15}~{\rm GeV}$~\cite{pdg2000}.
	\vs
        The main uncertainty in Eq.~(\ref{eq_epsk}) originates 
	from the bag parameter $B_{K}$ which cannot be measured 
	but has to be predicted by theory. The most reliable 
	calculations of $B_{K}$ are supposed to come 
	from lattice QCD. Currently, these calculations are 
	performed only under the assumption of $SU(3)$ symmetry 
	using the quenched approximation, \ie, using quarks 
	with infinite masses and neglecting the
	contribution of sea-quarks in closed loops, which leads 
	to a substantial reduction in computing time. The world 
	average reads: 
        $B_{K} = 0.87 \pm 0.06 \pm 0.14_{\rm quench}$~\cite{lellouch}, 
	where the first error combines statistical and accountable 
	systematic uncertainties while the second stands for an 
	estimate of the error introduced by the quenched approximation 
	and $SU(3)$ breaking effects.

\item   {\boldmath$\epe:$}
	In terms of the neutral kaon amplitude ratios $\eta_{+-}$
	and $\eta_{00}$, one finds to a very good approximation
	\beq
	   {\rm Re}(\epe) = \frac{1}{6}(1 - |\eta_{00}/\eta_{+-}|^2)~.
	\eeq
	The first evidence of direct CP violation in the neutral kaon system 
	has been found by the NA31 Collaboration at CERN~\cite{na31} which was 
	not confirmed by the E731 Collaboration~\cite{e731}.
	Since then, measurements at KTeV~\cite{ktev} and NA48~\cite{na48} 
	have obtained significant positive results while some inconsistencies
	about the central value remain. The experimental values are
	\beq
	{\rm Re}(\epe)\times10^4 = \left\{ 
	 \begin{array}{rclll}
	   23.0 & \pm & 6.5 & ({\rm NA31}&\hmm\cite{na31}) \\
	    7.4 & \pm & 5.9 & ({\rm E731}&\hmm\cite{e731}) \\
	   28.0 & \pm & 4.1 & ({\rm KTeV}&\hmm\cite{ktev}) \\
	   14.0 & \pm & 4.3 & ({\rm NA48}&\hmm\cite{na48})
	 \end{array}
	 \right.
	\eeq
	for which the weighted mean of 
	$\langle{\rm Re}(\epe)\rangle=(19.2\pm2.5)\times10^{-4}$ 
 	has a consistency of $\chi^2/3 =3.5$. The experimental
	situation being somewhat inconsistent, the theoretical
	prediction in the framework of the SM is still
	under strong investigations. The basic expression is of the
	form~\cite{ciu_mart}
	\beq
  	  {\rm Re}(\epe) = {\rm Im}\left[V_{ts}^*V_{td}V_{us}^*V_{ud}\right]
	         \Sigma\left(B_6, B_8, m_s\langle s\bar s\rangle\right)
	\eeq
	with $\Sigma(\dots)$ being a function of the hadronic
	matrix elements, $B_6$ and $B_8$, of the dimension-6 and
	dimension-8 non-perturbative power corrections
	which contribute to the effective Hamiltonian. The 
	quantitative size of these operators, in particular $B_6$,
	and the exact ingredients of $\Sigma(\dots)$ are still 
	under consideration~\cite{ciu_mart}. 

	{\em As a consequence, we shall not use $\epe$ in the current
	analysis.}
\item   {\boldmath$\dmd:$}
        The frequency of $\bdo$ oscillation is given by the mass 
	difference, $\dmd$, between the two $B^0_d$ mass eigenstates, 
	$B_{\rm H}$ and $B_{\rm L}$. It has been measured to an 
	accuracy of $3\%$~\footnote
	{
		The consistent, though preliminary measurements of 
		\babar~\cite{dmdBabar} and Belle~\cite{dmdBelle}, 
        	were not considered in the average quoted in 
		table~\ref{tab_ckmValues}.
	}
	(see table~\ref{tab_ckmValues}).        
	In analogy to $|\epsk|$, $\bdo$ oscillation in the SM 
	is driven by effective flavor-changing neutral current (FCNC) 
        processes through $\Delta B=2$ box diagrams. 
        In contrary to $|\epsk|$, where the large hierarchy
        in the Inami-Lim functions is partly compensated by
	the CKM matrix elements, the $\Delta B=2$ box diagrams 
	are dominated by top quark exchange between the virtual
	$W^\pm$ boson lines. This simplifies the theoretical 
	prediction which then reads
	\beq
	\label{eqof-dmd}
	   \dmd = \frac{\GF^2}{6 \pi^2}\eta_B m_{B_d}f_{B_d}^2B_d
		  m_W^2 S(x_t) \left|V_{td} V_{tb}^*\right|^2~,
	\eeq
	where $\eta_B = 0.55 \pm 0.01$ (for a review see~\cite{bbl}) 
	is a correction to the Inami-Lim function $S(x_t)$ 
	(see Eq.~(\ref{eq_inamilim})) from perturbative QCD.
        The leptonic decay constant $\fbd$ has not been measured
        so far and, like the bag parameter $B_d$, has to be determined
        by theory, in particular lattice QCD.
        Up to now, calculations are mainly performed in the quenched 
	approximation where the different groups find consistent results. 
	The most recent world averages for the decay constant is 
	$\fbd = (175\pm20)~{\rm MeV}$~\cite{latt2000}
        where the error includes statistical and accountable systematic 
        uncertainties.
        The unquenched result is estimated to be about $10\%$ 
        higher than the quenched value,
        $\fbd = (200\pm23^{+27}_{-17})~{\rm MeV}$~\cite{latt2000,bernard}.
	Recently, first (partly) unquenched calculations with
        two degenerate sea quarks were published and are, 
	within the given uncertainties, in agreement with 
	the expected increase~\cite{alikhan,bernard1}.
        The world average for the bag parameter in the quenched
        approximation is $B_d = 1.30\pm0.12\pm0.13$, where the
        second error is the estimated uncertainty due to 
	the quenched approximation~\cite{latt2000}.
	In the present work we use 
        $\fbd \sqrt{B_d} = (230\pm28\pm28)~{\rm MeV}$~\cite{latt2000}, 
	where the second error has been symmetrized.
\item   {\boldmath$\dms:$}
	Although $\dms$ itself has only a weak dependence on the 
	CKM phase the ratio $\dms/\dmd$ introduces a strong constraint since
	the dependence of the SM prediction on the parameters $\eta_{B}$ and 
	$m_{t}$ cancel in the ratio. Furthermore, the ratio 
	$\xi = \fbs \sqrt{B_s}/\fbd \sqrt{B_d}$ can be calculated more 
	reliably from lattice QCD than $\fbd \sqrt{B_d}$ alone since 
	most of the systematics cancel. For $\xi$, we are combining the 
	average values for quenched calculations from 
	Refs.~\cite{latt2000,Achille1} and choose: 
	$\xi = 1.16 \pm 0.03_{\rm stat,sys} \pm 0.05_{\rm quench}$.
	\vs
        Limits on $\bso$ oscillation governed by the mass difference $\dms$ 
	have been obtained by ALEPH~\cite{alephbs}, DELPHI~\cite{delphibs}, 
	OPAL~\cite{opalbs}, SLD~\cite{sldbs} and CDF~\cite{cdfbs}. 
	A convenient approach to average various results on $\dms$ is the 
	Amplitude Method~\cite{amplit} (see also the exhaustive study in 
	Ref.~\cite{toy}), which consists of a likelihood fit to the measured 
	proper time distribution with the amplitude of the oscillating term 
	being the free parameter at given frequency
\begin{figure}[t]
  \epsfxsize\largefig
  \centerline{\epsffile{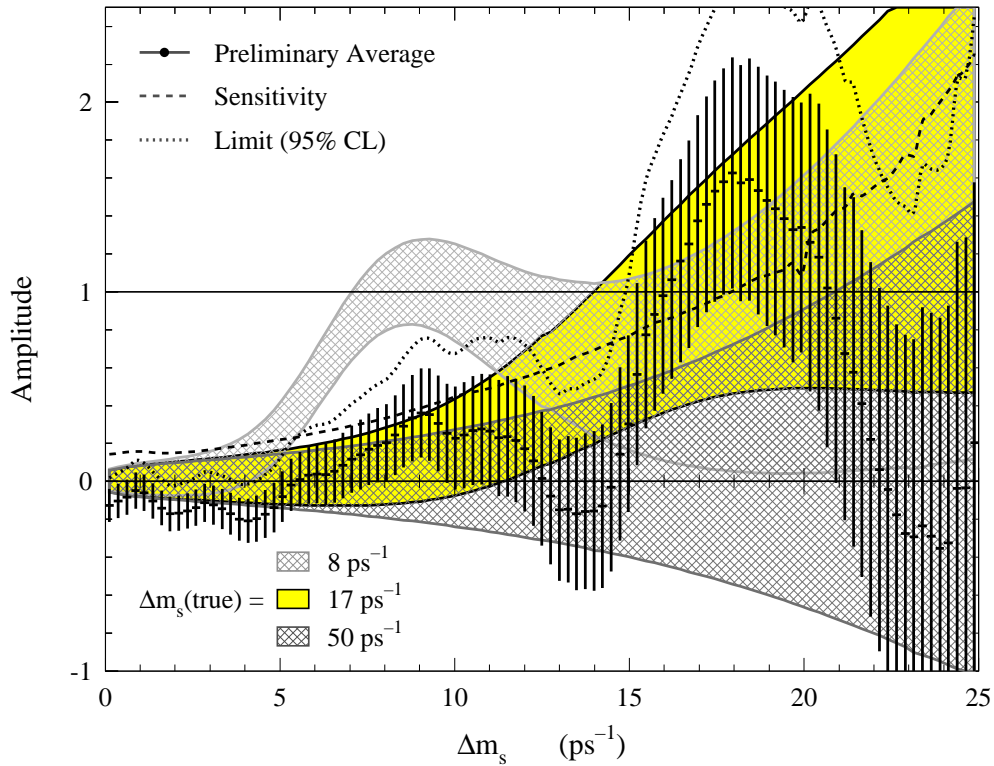}}
  \caption[.]{\label{fig_bs}\em
	Preliminary average of the experimental amplitude 
	spectrum using results
	from ALEPH, DELPHI, OPAL, SLD and CDF~\cite{newOscillB}. 
	The dashed and dotted lines give the sensitivity 
	($1.645\times\sigma_A$) and
	the $95\%$ limit $A+1.645 \times \sigma_A$), respectively.
	Shown in addition are the distributions obtained from
	a toy Monte Carlo simulation using as true values
	$\dms=8,\,17,\,50~{\rm ps}^{-1}$.}
\end{figure}
	$\dms$. Fig.~\ref{fig_bs} shows the average of the 
	measured amplitude spectrum~\cite{newOscillB} with 
	the expected spectra for different true $\dms$ superimposed. 
	The latter have been obtained, following the prescription of
	Ref.~\cite{toy}, from a toy simulation in which the decay length 
	and momentum resolutions are tuned to reproduce the measured errors 
	on the amplitudes (the RMS of the relative difference
	between measured and simulated errors is smaller
	than $2\%$ in the relevant sensitive region of $\dms$).
	Shown in addition are the experimental sensitivity	
	defined as $1.645 \times \sigma_A$ for a given $\dms$
	and the $95\%$~\CL~limit which is given by the sum
	of the sensitivity and the measured central amplitude.
	A sensitivity of $18.0~{\rm ps}^{-1}$ and
	a lower limit for $\dms$ of $14.9~{\rm ps}^{-1}$ at
	$95\%$~\CL~is obtained~\cite{newOscillB}.
	\vs
	The information from $\bso$ oscillations is usually
	implemented into $\chi^2$ fits using~\cite{amplit}
	$\chi^2_{|1-A|} = ((1-A)/\sigma_{A})^2$ and  
	${\rm CL}(\chi^2_{|1-A|})
	= {\rm Erfc}(|1-A|/\sigma_{A}/\sqrt{2}))$.
	However, this procedure does not properly interpret
	the information of the amplitude spectrum. 
	For instance, two measured amplitudes $A_{1}$ and $A_{2}$,
	where $A_{1} > 1$ and $A_{2} < 1$ but $A_{1}-1 = 1-A_{2}$,
	result in the same confidence level in this approach although
	it would be natural to assign a larger likelihood for an 
        oscillation to $A_{1}$ than to $A_{2}$.
	We propose an alternative procedure which exploits the information
	from the sign of $1-A$ by suppressing the module in the above 
	definition of $\chi^2_{|1-A|}$:
	\beq
	\chi^2_{1-A} = 2\cdot
	\left[{\rm Erfc}^{-1}
	\left(
		\frac{1}{2}\,{\rm Erfc}
		\left(\frac{1-A}{\sqrt{2}\sigma_A}\right)
	\right)
	\right]^{2}~.
	\eeq	
	It has been pointed out that the maximum information from the fit 
	to the proper time distributions of mixed and unmixed 
	$B^0_s(\bar{B_s^0})$ decays is obtained from the ratio of the 
        likelihood at given frequency $\dms$, ${\cal L}(\Delta m_s)$,
	to the likelihood at infinity, 
	${\cal L}(\Delta m_s = \infty)$~\cite{amplit,checchiaLogLik,Achille1}. 
	The logarithm of this ratio reads
	\beq
	\label{eq_likratio}
		2\Delta {\rm ln} {\cal L}^{\infty}(\Delta m_s) 
		= \frac{(1-A)^2}{\sigma_A^2} - \frac{A^2}{\sigma_A^2}~,
	\eeq
	which is assigned \via\ Eq.~(\ref{eq_chi2Function}) to $\chi^2_\infty$.
	The behaviour of the above defined $\chi^2$ and likelihood functions
	versus $\dms$ for the measured amplitude spectrum is plotted 
	in Fig.~\ref{fig_dmslik}. The dashed line shows the drawback
	of using $\chi^2_{|1-A|}$: the strongest signal yield is obtained
	at the crossings $A=1$ and not at the maximum amplitude
	situated around $17~{\rm ps}^{-1}$. This drawback is cured
	for $\chi^2_{1-A}$ (solid line). The dotted line shows
	the ratio $\chi^2_\infty$, providing a significantly 
	stronger constraint. For the current analysis we decided not to 
	use the likelihood ratio since the validity of the normalisation 
        of the likelihood which allows to identify ${\cal L}^{\infty}$ 
	with a probability density function is questionable. 
	This problem could be circumvented by means of a realistic 
	Monte Carlo simulation which permits the conversion of 
	likelihoods to confidence levels which however is currently
	not available.
\begin{figure}[t]
  \epsfxsize\smallfig
  \centerline{\epsffile{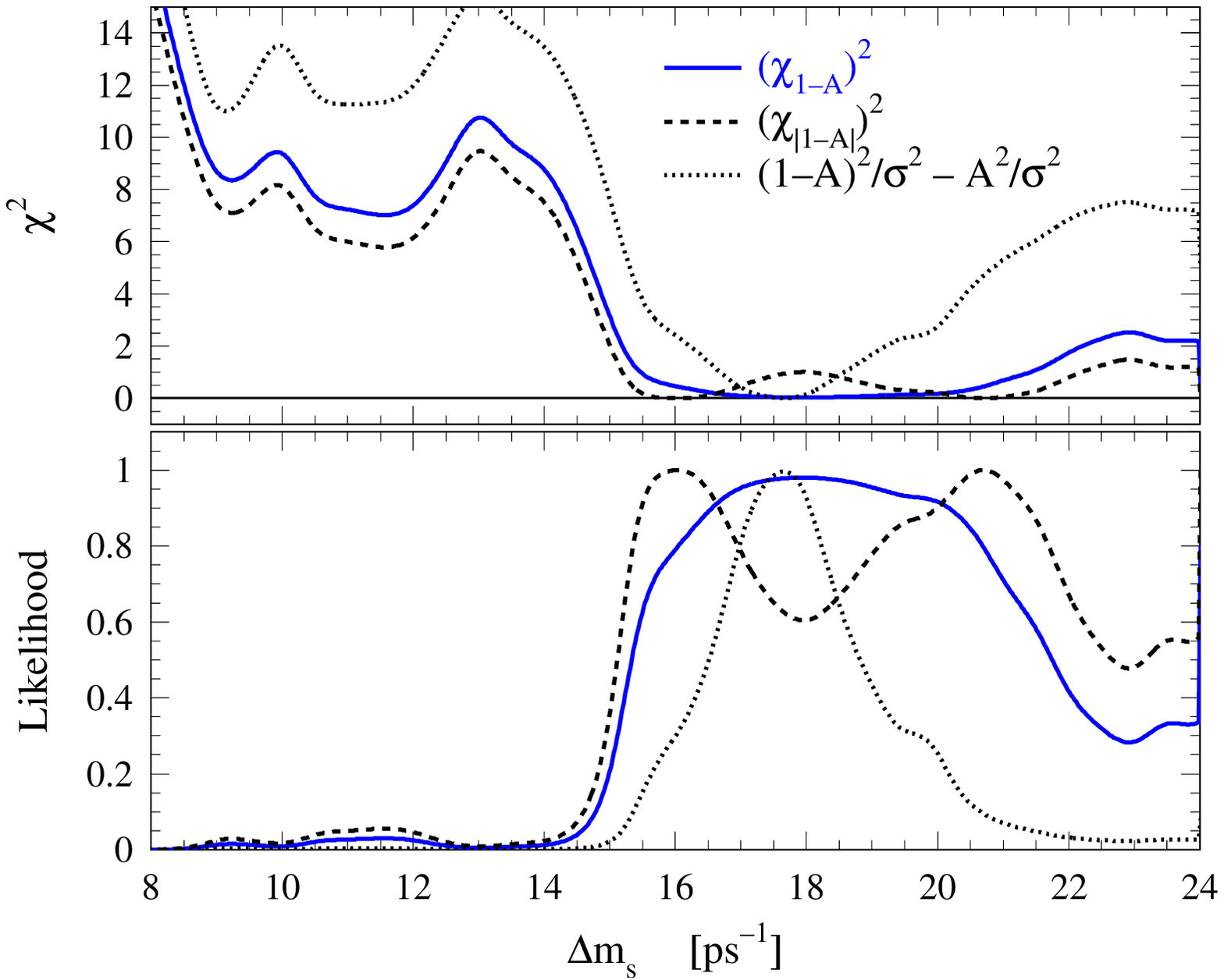}}
  \caption[.]{\label{fig_dmslik}\em
	The measured $\chi^2$ (upper plot) and likelihood (lower plot)
	functions (defined in the text) versus the frequency 
	of $\bso$ oscillation, $\dms$.}
\end{figure}
\begin{table}[p]
\begin{center}
\setlength{\tabcolsep}{0.35pc}
{\normalsize
\begin{tabular}{lcccccc}\hline
	&	&	&	\mc{3}{c}{Error treatment in \ckmfit:}&	\\
  \rs{Parameter}	& \rs{Value $\pm$ Error(s)}
		& \rs{Reference(s)}
		& Gauss.
		& Theo.
		& Prop.
		& \rs{Float.} \\ 
\hline
&&&& \\
  \rs{$|V_{ud}|$} 	& \rs{$0.97394\pm0.00089$}
		& \rs{\footnotesize see text}
		& \rs{*} & \rs{-} & \rs{-} & \rs{-} \\
  \rs{$|V_{us}|$} 	& \rs{$0.2200\pm0.0025$}
		& \rs{\footnotesize see text}
		& \rs{*} & \rs{-} & \rs{-} & \rs{-} \\
  \rs{$|V_{ub}|$} 	& \rs{$(3.49 \pm 0.23 \pm 0.55) \times 10^{-3}$}
		& \rs{\footnotesize see text}
		& \rs{*} & \rs{*} & \rs{-} & \rs{*} \\
  \rs{$|V_{cd}|$} 	& \rs{$0.224\pm0.014$}
		& \rs{\footnotesize see text}
		& \rs{*} & \rs{-} & \rs{-} & \rs{-} \\
  \rs{$|V_{cs}|$}	& \rs{$0.969 \pm 0.058$}
		& \rs{\footnotesize see text}
		& \rs{*} & \rs{-} & \rs{-} & \rs{-} \\
  \rs{$|V_{cb}|$}	& \rs{$(40.76\pm 0.50 \pm 2.0)\times10^{-3}$}
		& \rs{\footnotesize see text}
		& \rs{*} & \rs{*} & \rs{-} & \rs{*} \\
\hline
&&&& \\
  \rs{$|\epsk|$}     	& \rs{$(2.271\pm0.017)\times10^{-03}$}
		& \rs{\footnotesize see text}
		& \rs{*} & \rs{-} & \rs{-}  & \rs{-}\\
  \rs{$\dmd$}      	& \rs{$(0.487 \pm 0.014)~{\rm ps}^{-1}$}
		& \rs{\footnotesize see text}
		& \rs{*} & \rs{-} & \rs{-}  & \rs{-}\\
  \rs{$\dms$}           
		&  \rs{Amplitude spectrum}
		& \rs{\cite{newOscillB}, \footnotesize see text}
		& \rs{*} & \rs{-} & \rs{-}  & \rs{-}\\
  \rs{$\stbwa$} & \rs{$0.48 \pm 0.16$}
		& \rs{\footnotesize see text}
		& \rs{*} & \rs{-} & \rs{-}  & \rs{-}\\
\hline
&&&& \\
  \rs{$m_c$}	& \rs{$(1.3\pm0.1)~{\rm GeV}$}
		& \rs{\cite{pdg2000}}
		& \rs{-} 
		& \rs{*}
		& \rs{-}  & \rs{*}\\
  \rs{$m_t(\MSbar)$}	& \rs{$(166.0\pm5.0)~{\rm GeV}$}
		& \rs{\cite{pdg2000}}
		& \rs{*} 
		& \rs{-}
		& \rs{-}  & \rs{*}\\
  \rs{$m_K$}		& \rs{$(493.677\pm0.016)~{\rm MeV}$}
		& \rs{\cite{pdg2000}}
		& \rs{-}
		& \rs{-}
		& \rs{*}   & \rs{-}\\
  \rs{$\Delta m_K$}	& \rs{$(3.4885 \pm 0.0008)\times 10^{-15}~{\rm GeV}$}
		& \rs{\cite{pdg2000}}
		& \rs{-}
		& \rs{-}
		& \rs{*}   & \rs{-}\\
  \rs{$m_{B_d}$}	& \rs{$(5.2794\pm0.0005)~{\rm GeV}$}
		& \rs{\cite{pdg2000}}
		& \rs{-}
		& \rs{-}
		& \rs{*} & \rs{-}\\
  \rs{$m_{B_s}$}	& \rs{$(5.3696\pm0.0024)~{\rm GeV}$}
		& \rs{\cite{pdg2000}}
		& \rs{-}
		& \rs{-}
		& \rs{*} & \rs{-}\\
  \rs{$m_W$}		& \rs{$(80.419\pm0.056)~{\rm GeV}$}
		& \rs{\cite{pdg2000}}
		& \rs{-}
		& \rs{-}
		& \rs{*}  & \rs{-} \\
  \rs{$G_F$}		& \rs{$(1.16639\pm0.00001)\times 10^{-5}~{\rm GeV^{-2}}$}
		& \rs{\cite{pdg2000}}
		& \rs{-}
		& \rs{-}
		& \rs{-}  & \rs{-} \\
  \rs{$f_K$}		& \rs{$(159.8\pm1.5)~{\rm MeV}$}
		& \rs{\cite{pdg2000}}
		& \rs{-}
		& \rs{-}
		& \rs{*}  & \rs{-} \\
\hline
&&&& \\
  \rs{$B_K$}		& \rs{$0.87\pm0.06\pm0.13$}
		& \rs{\footnotesize see text}
		& \rs{*} 
		& \rs{*}
		& \rs{-} & \rs{*} \\
  \rs{$\eta_{cc}$}	& \rs{$1.38\pm0.53$}
		& \rs{\footnotesize see text}
		& \rs{-} 
		& \rs{*}
		& \rs{-} & \rs{*} \\
  \rs{$\eta_{ct}$}	& \rs{$0.47\pm0.04$}
		& \rs{\footnotesize see text}
		& \rs{-}
		& \rs{-}
		& \rs{*} & \rs{-} \\
  \rs{$\eta_{tt}$}	& \rs{$0.574\pm0.004$}
		& \rs{\footnotesize see text}
		& \rs{-}
		& \rs{-}
		& \rs{*}  & \rs{-}\\
  \rs{$\eta_B(\MSbar)$} & \rs{$0.55\pm0.01$ }
		& \rs{\footnotesize see text}
		& \rs{-}
		& \rs{*}
		& \rs{-} & \rs{*}\\
  \rs{$\fbd\sqrt{B_d}$} & \rs{$(230\pm28\pm28)$~{\rm MeV}}
		& \rs{\footnotesize see text}
		& \rs{*} 
 		& \rs{*}
		& \rs{-} & \rs{*}\\
  \rs{$\xi$}		& \rs{$1.16 \pm 0.03 \pm 0.05$}
		& \rs{\footnotesize see text}
		& \rs{*} 
		& \rs{*}
		& \rs{-} & \rs{*}\\		
\hline
\end{tabular}
}
\caption{\label{tab_ckmValues} \em
	Inputs to the global CKM fit. 
	If not stated otherwise: for two errors given, the 
	first is statistical and accountable systematic and the 
	second stands for systematic theoretical uncertainties.
	The fourth, fifth and sixth columns indicate
	the treatment of the parameters within {\rm \ckmfit}: measurements
	dominated by experimental errors (or statistical components
	of theoretical parameters) are marked as ``Gauss.'' by 
	an asterisk; parameters dominated by systematic theoretical
	uncertainties, treated as ranges in {\rm \ckmfit}, are marked 
	as ``Theo.''; for parameters
	that have experimental and systematic theoretical errors,
	treated in the fit according to Eq.~(\ref{eq_flattenedGaussian}),
	both fields, ``Exp.'' and ``Theo.'', are marked; parameters
	with small errors, marked as ``Prop.'', have
	their uncertainties propagated to the 
	corresponding measurements to whose errors they are added 
	in quadrature. The last column indicates whether
	or not the parameter is floating in the fit. In general,
	measurements with non-vanishing systematic theoretical
	errors have a floating theoretical component. Theoretical
	parameters with significant errors are necessarily floating.
	\underline{Upper part:} 
	experimental determinations of the CKM matrix elements.
	\underline{Middle upper part:} 
	CP-violating and mixing observables.
	\underline{Middle lower part:} 
	parameters of the SM predictions obtained from experimental data.
	\underline{Lower part:} 
	parameters of the SM predictions obtained from 
	theory.
}
\end{center}
\end{table}
\item   {\boldmath$\stb$}
	The first measurements for $\stb$ in $B$ decays to CP 
	eigenstates containing charmonium from the 
	$B$ factories, CDF and LEP give results
	which are compatible with both, the SM expectation and zero:
	{\small
	\beq
	{\normalsize \stb = }\left\{ 
	\begin{array}{rclll}
          0.34 & \pm & 0.21 & ({\rm \babar}) &\hmm\cite{sin2betaBabar})\\
          0.58 & \pm & 0.34 & ({\rm Belle}) &\hmm\cite{sin2betaBelle})\\
	  0.79 & \pm & 0.43 & ({\rm CDF})   &\hmm\cite{sin2betaCdf})\\
          0.84 & \pm & 0.93 & ({\rm ALEPH}) &\hmm\cite{sin2betaAleph})\\
          3.2  & \pm & 2.0  & ({\rm OPAL})  &\hmm\cite{sin2betaOpal})\\
	\end{array}
	\right.
	\eeq
	}
	From these measurements (asymmetric errors have been averaged)
	we obtain the weighted mean
	$\stbwa=0.48 \pm 0.16$, where the (small) effect from using 
	different $\dmd$ values in the single analyses has not 
	been taken into account.
\end{itemize}

\subsection{Future Prospects: Rare Decays of $K$ and $B$-Mesons}
\label{sec_FutureProspects}

Theoretically clean measurements of CKM matrix elements
are obtained by virtue of rare $K$ and $B$ decays.
The countours in the $\rhoeta$ plane expected from rare $K$ decays are 
drawned on Fig.\ref{fig_rhoetaall}.
\begin{itemize}
\item	The decay {\boldmath$K^0_{\rm L}\rightarrow\pi^0\nu\bar\nu$}
	has not been observed yet. The current upper limit
	reads~\cite{pdg2000}
	\beq
	   {\rm BR}(K^0_{\rm L}\rightarrow\pi^0\nu\bar\nu)
		< 5.9\times10^{-7}~~~(\CL=90\%)~,
	\eeq
	while the expected SM branching ratio is of the order
	of $2\times10^{-11}$.
	The decay proceeds \via\ a loop induced FCNC transition 
	at short distance and is greatly dominated by
	a direct CP-violating amplitude in the SM~\cite{littenberg},
	$A(K^0_{\rm L}\rightarrow\pi^0\nu\bar\nu)\propto
	{\rm Im}[V_{td}V^*_{ts}]\langle\pi^0|(\bar s d)_{V-A}|K^0\rangle$,
	due to the cancellation of the charm contributions. The
	SM prediction for the branching fraction of the decay
	reads~\cite{littenberg,buras2}
	\beq
	\label{eq_klpi0nunu}
	    {\rm BR}(K^0_{\rm L}\rightarrow\pi^0\nu\bar\nu)
		= r_{K_{\rm L}}\frac{\tau_{K_{\rm L}}}{\tau_{K^+}}
	          \frac{3\alpha^2}{2\pi^2}
		  \frac{{\rm BR}(K^+\rightarrow\pi^0 e^+ \nu)}
		       {|V_{ud}V_{us}^2|^2{\rm sin}^4\theta_{\rm W}}
		  \bigg(\eta_XX_0(x_t)
			{\rm Im}\left[V_{ts}^*V_{td}V_{us}^*V_{ud}
				\right]\bigg)^{\!2}~.
	\eeq
	Here, $r_{K_{\rm L}}=0.944$ corrects for isospin breaking effects
	and the different phase space~\cite{marciano} involved in the relation
	between the $K^0_{\rm L}$ and the $K^+$ branching fractions.
	The other parameters in Eq.~(\ref{eq_klpi0nunu})
	are the kaon lifetimes, the QED running
	fine structure constant and the Weinberg angle. The
	Inami-Lim function $X(x_t)$ for $x_t=(m_t/m_W)^2$ is
	defined as
	\beq
	\label{eq_inamilim2}
	   X_0(x) = \frac{x}{8}\left(\frac{x+2}{x-1}
				     + \frac{3x-6}{(x-1)^2}{\rm ln}(x)
			             \right)
	\eeq
	for which $\eta_X=0.994$ (Eq.~(\ref{eq_klpi0nunu}))
	represents the NLO correction~\cite{buras2}.
	In Ref.~\cite{buchalla} the CP conserving contribution 
	to $K^0_{\rm L}\rightarrow\pi^0\nu\bar\nu$ has been 
	found to be suppressed by a factor of $6\times10^{-5}$
	with respect to the CP-violating rate. 
	Expressed in the Wolfenstein parameterization, the SM
	prediction corresponds to
	\beq
	    {\rm BR}(K^0_{\rm L}\rightarrow\pi^0\nu\bar\nu)
		\propto \lambda^8 A^4 \etabar^2~,
	\eeq
	showing that a measurement would provide a pair of horizontal
	lines in the $\rhoeta$ plane.
	Proposed experiments that could measure 
	${\rm BR}(K^0_{\rm L}\rightarrow\pi^0\nu\bar\nu)$ are
	the KOPIO experiment at BNL~\cite{KOPIO} and KAMI at 
	FNAL~\cite{KAMI} expecting 60 and 120 events, respectively.
	The experiment(s) will not start before 2005 and have
	to take data for several years. The expected precision
        on the branching ratio in 2010 is of the order
        $5\%$ to $10\%$, thereby yielding a precision of a 
	few percent on the measurement of $\etabar$.
\item	{\boldmath$K^+\rightarrow\pi^+\nu\bar\nu:$}
	The SM prediction of the rare decay
	$K^+\rightarrow\pi^+\nu\bar\nu$ is given by~\cite{buras2}
	\beq
	\label{eq_kpppnunu}
	    {\rm BR}(K^+\rightarrow\pi^+\nu\bar\nu)
		= r_{K^+}
	          \frac{3\alpha^2}{2\pi^2}
		  \frac{{\rm BR}(K^+\rightarrow\pi^0 e^+ \nu)}
		       {|V_{us}|^2{\rm sin}^4\theta_{\rm W}}
		  \sum_{i=e,\mu,\tau}
		\left|\eta_XX_0(x_t)V_{td}V^*_{ts}
		      + X_{\rm NL}^{(i)}V_{cd}V^*_{cs}
		\right|^2
	\eeq
	where the function $X_0(x_t)$ is given by Eq.~(\ref{eq_inamilim2})
	and where the $X_{\rm NL}^{(\ell)}$ terms account for the 
	charm contributions (not suppressed here) and are calculated in 
	Ref.~\cite{buras4}. The values depend on the QCD scale
	$\Lambda_{\MSbar}^{(4)}$ and the running charm quark
	mass: $X_{\rm NL}^{(e)} = (8-13)\times10^{-4}$
	and $X_{\rm NL}^{(\tau)} = (5-9)\times10^{-4}$. A detailed
	discussion of the theoretical uncertainties connected
	with this and the above SM predictions is provided in 
	Ref.~\cite{buras2}.
	The BNL experiment E787 has observed one event resulting in
	${\rm BR}(K^+\rightarrow\pi^+\nu\bar\nu)
	=(1.5 ^{+3.4}_{-1.2}) \times 10^{-10}$~\cite{E787}.
	About 5 - 10 events are expected to be observed by the successor 
	E949~\cite{E949} whereas the CKM project at FNAL~\cite{CKM},
	starting about 2005, expects to collect about 100 events within
	some years of data taking. A similar precision for the branching
	ratio measurement as in the case of 
	$K^0_{\rm L}\rightarrow\pi^0\nu\bar\nu$ may be achieved
	in the year 2010. A theoretical uncertainty of the order $5\%$ 
	but likely not well below this value might be possible~\cite{buras2}.
\item	The {\boldmath$B^+\rightarrow\tau^+\nu$} decay has not
	been observed yet. The current upper limit for its branching 
	fraction reads~\cite{pdg2000}
	\beq
	   {\rm BR}(B^+\rightarrow\tau^+\nu) 
		< 5.7\times10^{-4}~~~(\CL=90\%)~.
	\eeq
	In the SM the branching ratio is given by
	\beq
	   {\rm BR}(B^+\rightarrow\tau^+\nu) 
		= \frac{G_{\rm F}^2m_{B}m_\tau^2}{8\pi}
		  \left(1 - \frac{m_\tau^2}{m_{B}^2}\right)
	          \fbd^2|V_{ub}|^2\tau_{B}~,
	\eeq
	with the $B$ meson decay constant $\fbd$ (see
	Tab.~\ref{tab_ckmValues}) and the lifetime of the charged $B$,
	$\tau_{B}=1.653\pm0.028~{\rm ps}$~\cite{pdg2000}.
	Depending on the precision of the lattice 
	calculation of $\fbd$, a measurement of 
	${\rm BR}(B^+\rightarrow\tau^+\nu)$ may either yield 
	a direct measurement of $|V_{ub}|$, or may improve
	the prediction of $\dmd$ through the constraint of 
	$\fbd$ which is the more likely way to proceed.
	Additional information may be obtained by measuring
	the radiative decay $B^+ \rightarrow \ell^+ \nu_{\ell} \gamma$
	in which the helicity suppression is circumvented
	due to the emission of the photon from the primary u-quark
	(see, {\rm e.g.}, Ref.~\cite{atwood}).
	Although the calculation of the branching ratio is model
	dependent a measurement possibly provides a useful 
	experimental check of lattice calculations.
\item   CP-violating {\bf Partial Rate Asymmetries} (PRA) of 
	inclusive $b \rightarrow s(d) \gamma$ decays 
        can be calculated in the SM~\cite{soares,chet,kagan,kagan2,kiers}.
	They are defined by the ratio
	\beq
	   A_{\rm CP}^{b\rightarrow s(d)\gamma}
	    = \frac{{\rm BR}({\bar B}\rightarrow X_{s(d)}\gamma)
            - {\rm BR}({\bar B}\rightarrow X_{{\bar s}({\bar d})}\gamma)}
            {{\rm BR}({\bar B}\rightarrow X_{s(d)}\gamma)
            + {\rm BR}({\bar B}\rightarrow X_{{\bar s}({\bar d})}\gamma)}
       	\eeq
	Their theoretical predictions depend on various Wilson 
	coefficients and CKM matrix elements involving the 
	CP-violating phase~\cite{kagan}.
	Lumping all coefficients together, where external parameters
	like the strong coupling constant, the $b$-quark mass,
	the photon infrared cut-off and the 
	renormalization scale have to be fixed, gives the 
	estimate~\cite{kiers}
	\beq
	   A_{\rm CP}^{b\rightarrow s(d)\gamma}
	     \approx 0.33 \times {\rm Im}
	     \left[\frac{V_{ub}V^*_{us(d)}}{V_{tb}V^*_{ts(d)}}
	     \right]~.
	\eeq
	This yields asymmetries of
	$A_{\rm CP}^{b\rightarrow s\gamma}\approx0.6\%$ and 
	$A_{\rm CP}^{b\rightarrow d\gamma}\approx-16\%$ for some
	typical values of the CKM elements.
\item  	{\boldmath$B\rightarrow\pi K:$}
	Due to the work of many authors (see, \eg, 
	Refs.~\cite{fleischer,fleischer2,bbns,sanda,schuneplasz} 
	- this list is far from being complete), it could be 
	shown that the (ratios of) branching fractions of 
	charmless $B_d$ decays into $\pi$ and $K$ final states 
	provide constraints on the UT angle $\gamma$. 
	Most recent branching ratios read
	\beqn
	\label{charmless_br}
         {\rm BR}(B^0_d \rightarrow \pi^+ \pi^-)
		      + {\rm BR}(\bar{B^0_d} \rightarrow \pi^+ \pi^-) 
			&=&  (4.43 \pm 0.89)\times10^{-6}~,\\
          {\rm BR}(B^0_d \rightarrow K^+ \pi^-)
		      + {\rm BR}(\bar{B^0_d} \rightarrow K^- \pi^+)
			&=& (17.25 \pm 1.55)\times10^{-6}~,\\
          {\rm BR}(B^+   \rightarrow K^+ \pi^0)
		      + {\rm BR}(B^-         \rightarrow K^- \pi^0)
			&=& (12.10 \pm 1.70)\times10^{-6}~,\\
          {\rm BR}(B^+   \rightarrow K^0 \pi^+)
		      + {\rm BR}(B^-         \rightarrow \bar{K^0} \pi^-)
			&=& (17.19 \pm 2.54)\times10^{-6}~,\\
          {\rm BR}(B^0_d \rightarrow K^0 \pi^0)
		      + {\rm BR}(\bar{B^0_d} \rightarrow \bar{K^0} \pi^0)
			&=& (10.33 \pm 2.53)\times10^{-6}~,
	\eeqn
	where the values given are the weighted means of the 
	{\em preliminary} results on charmless $B$ decays presented 
	by the \babar, Belle and CLEO 
	collaborations~\cite{babarbk,bellebk,cleobk} 
	(asymmetric errors have been averaged). 
	The authors of Refs.~\cite{bbns,sanda} have obtained
	predictions of relative amplitudes and phases of the 
	tree and penguin diagrams involved in the above decays.
	Very recently, a new theoretical analysis of two-body $B$
	decays to pions and kaons, based on non-leading 
	Factorization Approximation, has been published~\cite{Newbbns}.	
	The authors obtain an allowed region for $|V_{ub}|e^{-i \gamma}$ 
	which is in agreement with the results found in this work.
\vs
	A statistical discussion and formulae for the treatment of 
	ratios of branching fractions or, more precisely, constraints
	from parameters with arbitrary absolute, but known relative 
	normalization is 
	given in Appendix~\ref{sec_RatiosOfBranchingFractions} of
	this paper.
\end{itemize}
\vspace{0.3cm}

\noindent
We have attempted in this section to recall some of the most
striking prospects for future CKM constraints which, however, is 
far from being complete. A more quantitative
and broader selection of CKM sensitive quantities as well
as extrapolations into the future can be found in Ref.~\cite{heikobcp4}.
If $\dms$ is not much larger than suggested by the current
SM constraints it will likely be measured during the forthcoming
Tevatron II run. The precision of the combined $\dmd$ and
$\dms$ constraints on $\rhobar$, $\etabar$ will then be
dominated by the QCD parameter $\xi$. An experimental 
determination of the decay constant ratio $f_{D_{s}}/f_{D_{d}}$ 
at a $\tau$/charm factory would be helpful to
check lattice QCD calculations of $\xi$. 	 
Measurements of time-dependent CP-asymmetries at the $\Upsilon(4S)$ 
and at hadron machines, in particular at the forthcoming 
experiments LHCb and BTeV, aim to extract the UT angle $\alpha$ 
in non-strange, charmless two and three body decays.
The remaining angle $\gamma$ is expected to be determined 
in $B_{d} \rightarrow D K$ and $B_{s} \rightarrow D_{s} K$ decays,
though these measurements require very large data samples of the
corresponding $B_{d(s)}$ mesons. The dedicated experiments
LHCb and BTeV will also measure the most promising 
channel $B_{s} \rightarrow \psi \phi$.

%
% --------------------- Constrained Fits -----------------------------
%
\section{Constrained Fits Within the SM}
\label{sec_constrainedFits}

After the discussions in the preceding sections, we are
prepared to perform the constrained fits of the CKM
parameters and related quantities. 
We place ourselves in the framework of the \ckmfit\ scheme
(see Section~\ref{sec_CkmFitScheme} for an introduction
and Section~\ref{sec_reminderCkmFitSchem} for a summary
of its main features) and hence define the theoretical 
likelihoods of Eq.~(\ref{eq_likThe}) to be one within the allowed 
ranges and zero outside\footnote{
In other words,
we use $\kappa=0$ and $\zeta=1$ for the Hat function $\Hatsyst(\xo)$ of 
Eq.~(\ref{eq_flattenedGaussian}).
}. 
As a consequence, no hierarchy 
is introduced for any permitted set of theoretical parameters, \ie, 
%the likelihood which is maximized in the fit
the $\chi^2$ which is minimized in the fit receives no 
contribution from theoretical systematics,
but theoretical paramaters cannot exit their allowed ranges.
%is independent of the systematic theoretical part provided
%the parameters are within their allowed ranges.
When relevant,
statistical and theoretical uncertainties are combined beforehand,
as presented in Section~\ref{sec_flattenedGaussian} 
(\ie, applying Eq.~(\ref{digest})).
Floating theoretical
parameters are labelled by an asterisk in the ``Float.'' column
of Table~\ref{tab_ckmValues}. 
For parameters with small uncertainties,
errors are propagated through the theoretical predictions, 
and added in quadrature to the experimental error of the corresponding 
measurements\footnote{
This procedure neglects the correlations
occurring when such parameters are used in 
more than one theoretical prediction. 
}:
they are labelled by an asterisk in the ``Prop.'' 
column of Table~\ref{tab_ckmValues}.

\subsection{Two Dimensional Parameter Spaces}
\label{TwoDimensionalParameterSpaces}

It is customary to
present the constraints on the CP-violating phase in
the two-dimensional $\rhoeta$ plane of the Wolfenstein 
parameterization.
%(related \via\ 
%${\rm tan}\delta=\etabar/\rhobar$ to the CP-violating CKM
%phase in the standard parameterization Eq.~(\ref{eq_ckmPdg}),
%or \via\ $J\simeq \lambda^6A^2\eta$ to the Jarlskog parameter).
Other representations involving the UT angles $\alpha$, $\beta$
and $\gamma$ are also 
considered in the analysis. 
For the two-dimensional
graphical displays we define the $\a$ parameter space by 
the coordinates $\a=\{x, y\}$ (\eg, $\a=\arhoeta$)
and the $\Mu$ space by the other CKM parameters $\lambda$ 
and $A$, as well as the 
$\yQCD$ parameters.
%remaining theoretical 
%parameters $\ymod$ with sizeable uncertainties. 

\subsubsection{Metrology in the \boldmath$\rhoeta$ Plane}
\label{sec_constraintsRhoeta}

\begin{figure}[t]
  \epsfxsize\largefig
  \centerline{\epsffile{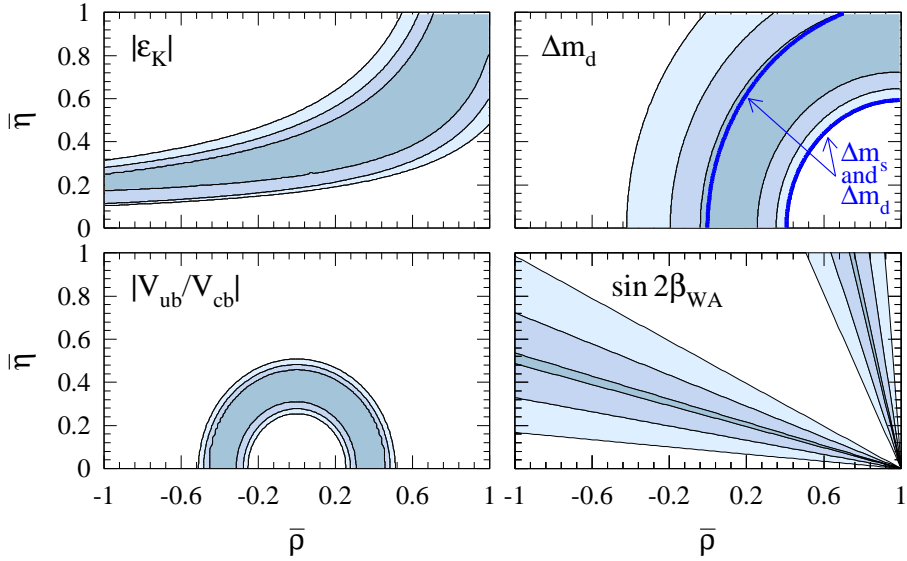}}
  \vspace{-0.0cm}
  \caption[.]{\label{fig_rhoeta_vars}\em
	Confidence levels in the $\rhoeta$ plane for the 
	individual constraints. The upper right hand plot
	shows in addition to $\dmd$ the improved constraint
	from $\dms$ \via\ $\xi$ on $\dmd$. The shaded areas 
	indicate the regions of	$\ge90\%$, $\ge32\%$ and $\ge5\%$ 
	CLs, respectively.}
\end{figure}
\begin{figure}[p]
  \vspace{1.0cm}
  \epsfxsize\largefig
  \centerline{\epsffile{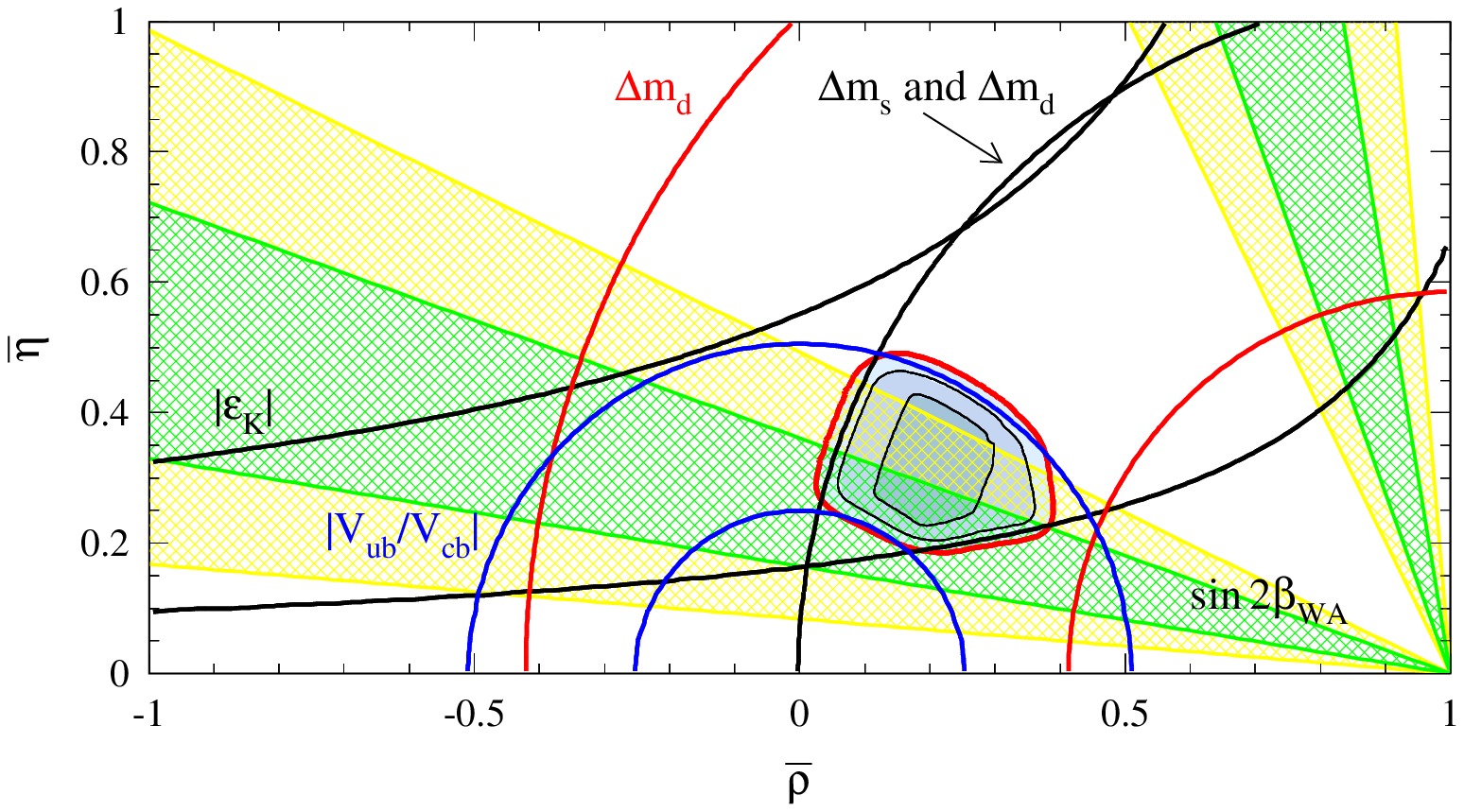}}
  \vspace{0.5cm}
  \centerline{\epsffile{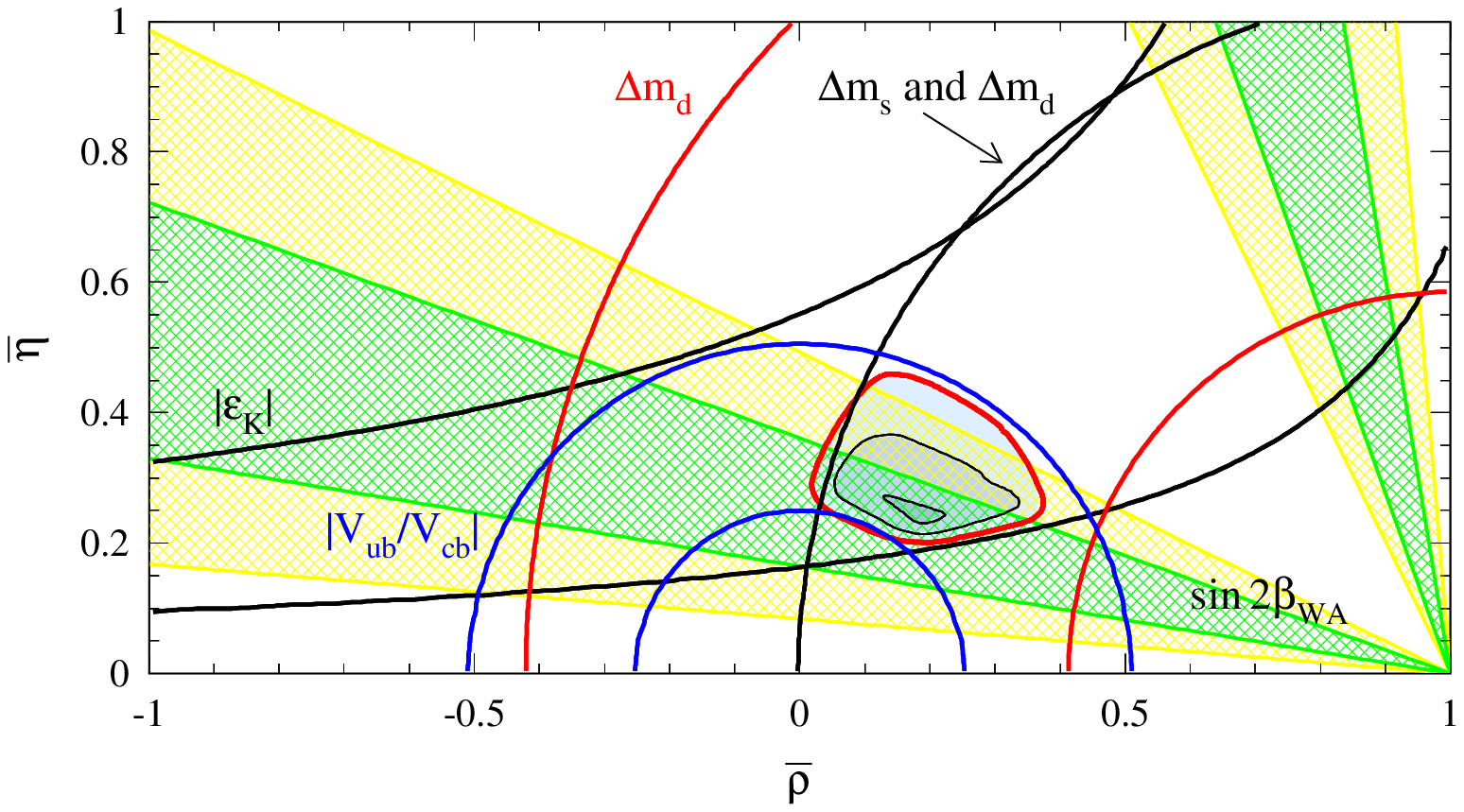}}
  \caption[.]{\label{fig_rhoeta}\em
	\underline{Upper plot:} 
	confidence levels in the $\rhoeta$ plane for the 
	global CKM fit. The shaded areas indicate the regions of
	$\ge90\%$, $\ge32\%$ and $\ge5\%$ CLs, respectively. 
	Also shown are the $5\%$
	CL contours of the individual constraints. The 
	$\ge32\%$ and $\ge5\%$ CL constraints from the 
	world average of the $\stb$ measurements, not entering the 
	combined fit, are depicted by the dashed areas. 
	See Table~\ref{tab_ckmValues} for a compendium of the 
	fit input values.
	\underline{Lower plot:} 
	confidence levels obtained when including
 	the world average of $\stb$ in the combined fit.
}
\end{figure}

The individual constraints, sensitive to $\rhobar$ and $\etabar$,
are drawn in Fig.~\ref{fig_rhoeta_vars}. 
Shown are the CLs of Eq.~(\ref{eq_conLev}) which, according to the frequentist
approach adopted in \ckmfit, have to be
interpreted as {\em upper bounds for the optimal set
of theoretical parameters at a given point in the $\rhoeta$ plane}
(this is implicit in the following when invoking the term CL).
Obviously, 
CLs should not be interpreted as PDFs, 
\ie, inferring equal relative probability density from equal shades. 
Instead, a CL value expresses a probability
which is defined for a given coordinate $\{\rhobar,\etabar\}$:
it is the probability that the agreement between data and the most 
favorable realization of the SM 
%(agreement probed with a $\chi^2$)
at that point be worse than the one observed.
However, although the CLs have a well defined statistical meaning, 
one must be aware of their strong dependence on the,
to some extent, arbitrary $[\yQCD]$ ranges.
(\cf, Section~\ref{sec_metrology}).
\vs
The results of the global fit in the $\rhoeta$ plane are 
shown in Fig.~\ref{fig_rhoeta} not including (upper plot) 
and including (lower plot) in the fit the world average of $\stb$ 
(see Table~\ref{tab_ckmValues}). The dark, medium and light shaded 
areas correspond to $\ge 90\%$, $\ge 32\%$ and $\ge 5\%$ CL, respectively.
The outer regions with lower probabilities are outside the $\ymodopt$ domain 
where an adjustment of the $\Mu$ parameters can maintain maximal agreement
(\ie, can reproduce the $\ChiMinGlob$ value). Also shown are the 
$5\%$~CL contours of the individual constraints as well as the $\ge 32\%$ 
and $\ge 5\%$~CL regions corresponding to $\stbwa$ (hatched areas).
As described in Section~\ref{sec_metrology}, the CLs 
obtained belong to the metrological phase of the analysis and, 
by construction, do not constitute a test of goodness of the theory. 
A probe of the SM is obtained from the numerical value of $\ChiMinGlob$ 
as discussed in Section~\ref{sec_probingTheSM} and used in 
Section~\ref{sec_compOfTheSM}.
% As a side-remark, in the treatment of $\stb$,
% one may ignore the measurement of $\DmBd$,
% and replace it by its theoretical prediction,
% which is proportional to a $(1-\rhobar)^2 +\etabar^2$.
% The allowed domain in the $\rhoeta$ plane obtained this way
% bears little resemblance with the one obtained with Eq.~(\ref{probstb})
% because the $\stb_\exp$ value should be replaced by 
% a function of $(1-\rhobar)^2 +\etabar^2$. 
% Denoting $\stb_0$ and $\xdo=\DmBdo/\tau_B^0$
% the true values of the two unknown quantities $\stb$ and $\xd$,
% the measurement\footnote
% {
% 	The uncertainty on $\stb$ depends also on the 
% 	value assumed for $\xd$.
% } 
% of $\stb$ can be shown to be (neglecting all detector effects):
% \begin{equation}
% \stb_\exp=\stb_0
% \ {\xdo\over\xd}\
% {1+4\xd^2\over (1+\xd^2+\xdo^2)^2-(2\xd\xdo)^2}~.
% \end{equation}
% It follows that for $(1-\rhobar)^2 +\etabar^2>4$,
% where $\xd$ is significantly larger than $\xdo$,
% the determination of $\stb$ decreases steadily,
% which implies either $\etabar\ll(1-\rhobar)$ or the converse.
% Stated differently, the constraint in the $\rhoeta$ plane 
% set by $\stb$ using Eq.~(\ref{stbtwolines}) is valid,
% but only insofar one wants to dispel utterly the validity of 
% the SM prediction for $\DmBd$.

\subsubsection{Other Two Dimensional Parameter Spaces}
\label{sec_constraintsOthers}

\begin{figure}[t]
  \epsfxsize\largefig
  \centerline{\epsffile{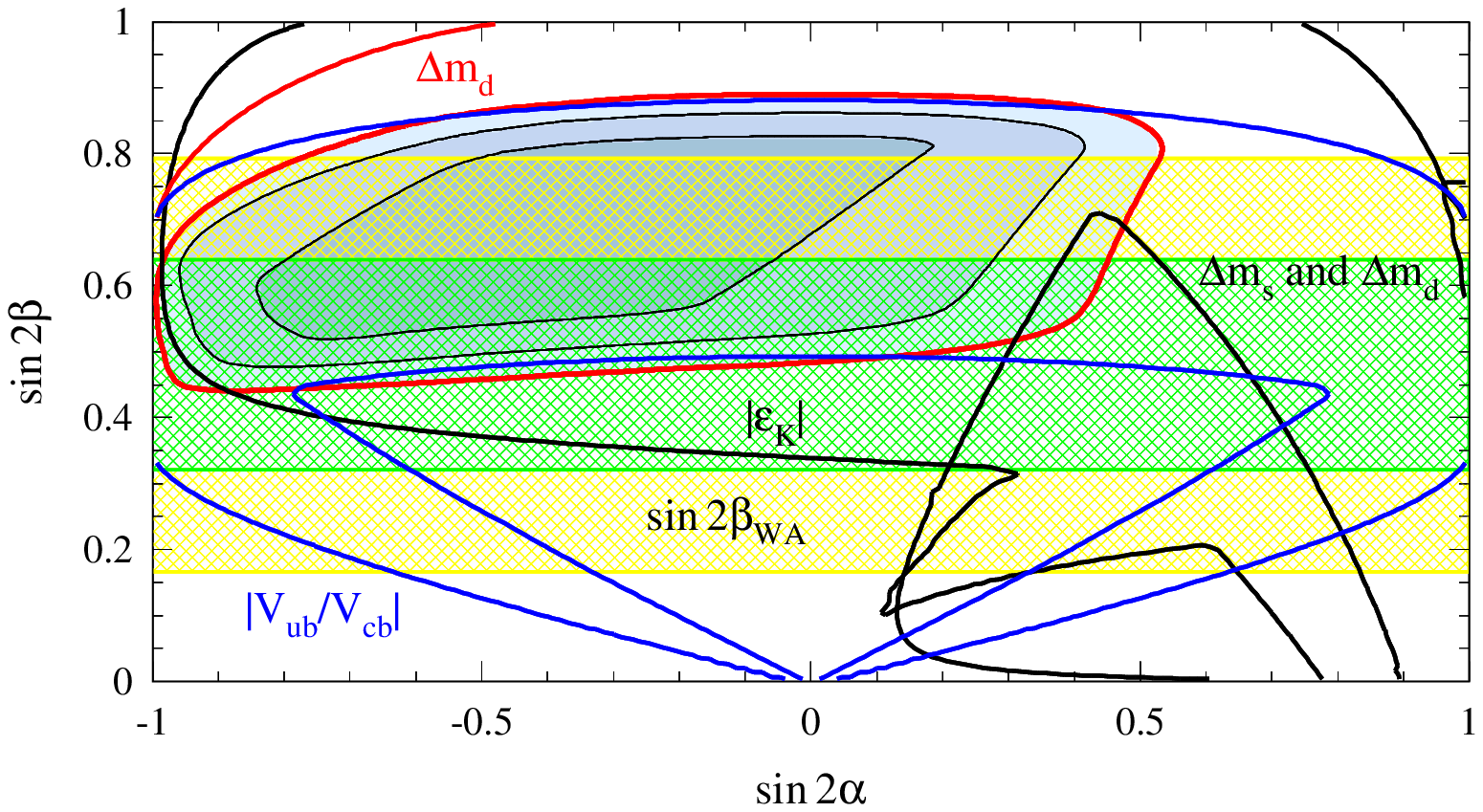}}
  \caption[.]{\label{fig_stastb}\em
	Confidence levels in the $\stastb$ plane for the 
	global CKM fit. The shaded areas indicate the regions of
	$\ge90\%$, $\ge32\%$ and $\ge5\%$ CLs, respectively. 
	Also shown are the $5\%$
	CL contours of the individual constraints. The 
	$\ge32\%$ and $\ge5\%$ CL constraints from the 
	world average of the $\stb$ measurements, 
	not entering the global fit, are given	by the dashed areas.}
\end{figure}

\begin{figure}[p]
  \epsfxsize\largefig
  \centerline{\epsffile{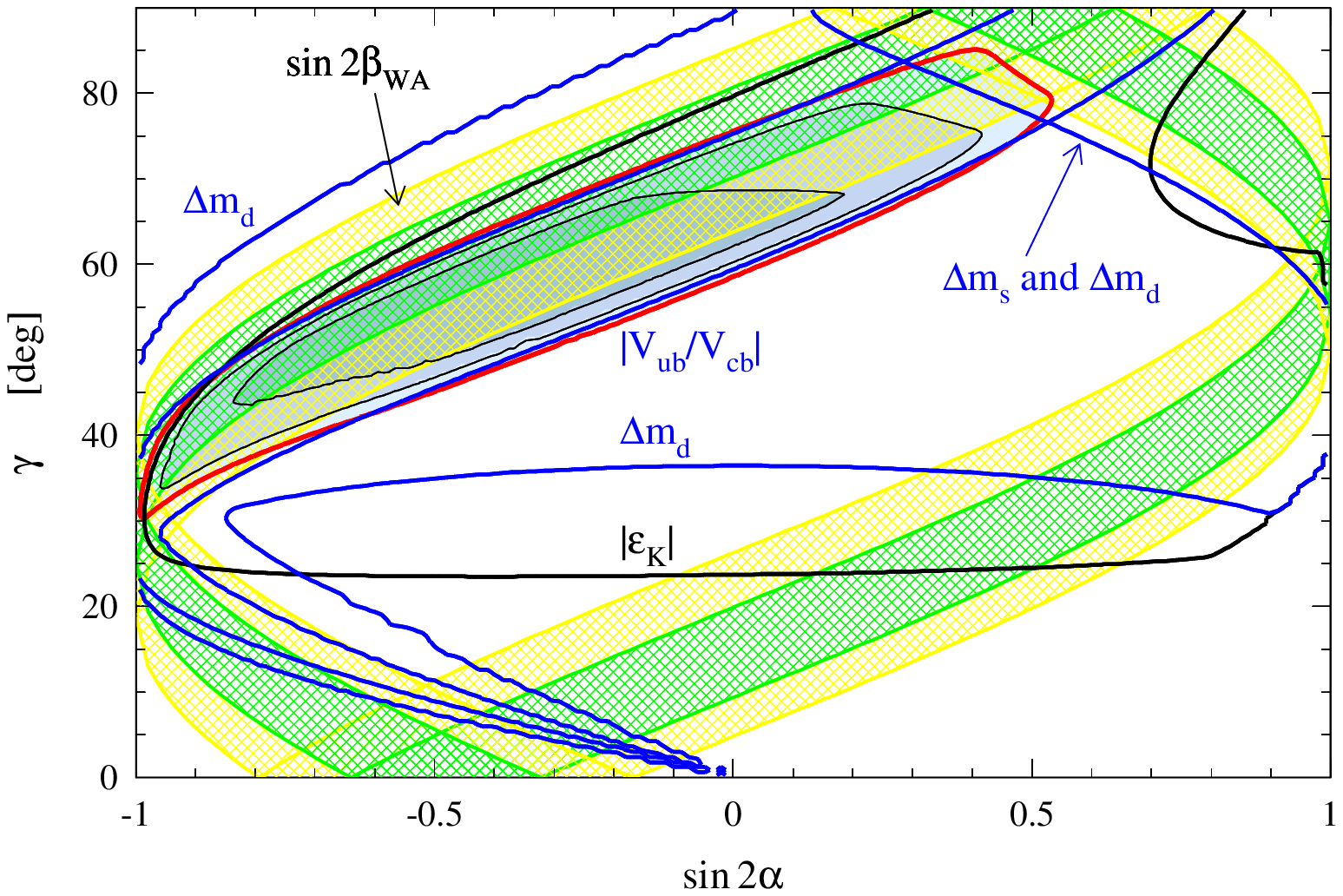}}
  \centerline{\epsffile{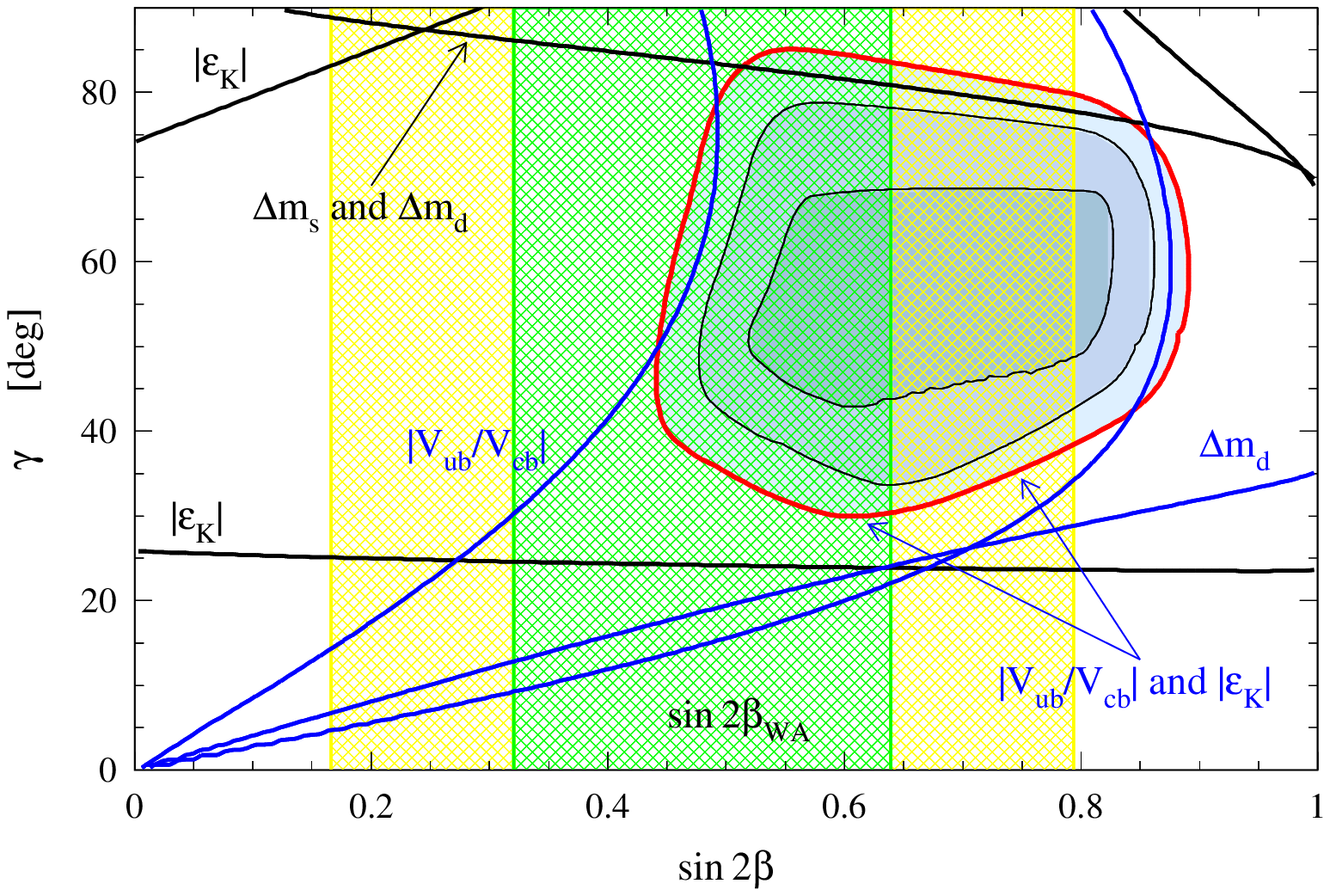}}
  \caption[.]{\label{fig_stgam}\em
	Confidence levels in the $\stagam$ (upper plot)
	and the $\stbgam$ plane (lower plot), obtained from the 
	global CKM fit. The shaded areas indicate the regions of
	$\ge90\%$, $\ge32\%$ and $\ge5\%$ CLs, respectively. Shown in 
	addition are the $5\%$ CL contours of the individual 
	constraints. The 
	$\ge32\%$ and $\ge5\%$ CL constraints from the 
	world average of the $\stb$ measurements are given
	by the dashed areas. It does not enter the global fit.}
\end{figure}

Except for possible multi-valuedness problems\footnote{
For example,
when exploring the $\stastb$ plane,
care should be taken to account for multiple solutions.
A given value of ${\rm sin}2\omega$ 
($\omega=\alpha$ or $\beta$) can be obtained with four values of $\omega$
   ($\omega_1= \frac{1}{2}{\rm arcsin}({\rm sin}2\omega)$,
   $\omega_2 = \frac{\pi}{2} - \omega_1$,
   $\omega_3 = \pi + \omega_1$,
   $\omega_4 = \frac{3\pi}{2} - \omega_1$)
   and corresponds to a pair of curves in the $\rhoeta$ plane
   intersecting on the $\etabar=0$ axis.
Each intersection of one of the $\stb$ curves with one of the $\sta$ curves  
should be considered.   
}, 
it is straightforward to replace the $\rhoeta$ plane by any other 
one, two or higher dimensional parameter constellation.
Figures~\ref{fig_stastb} and \ref{fig_stgam} show the results
from the global fits and for the individual constraints
in the planes $\stastb$, $\stagam$ and $\stbgam$,
respectively. The constraint from $\stb$ does
not enter the fits. As aforementioned, the individual
constraints are given as $5\%$ CL contours, and the
shaded areas depict $\ge90\%$, $\ge32\%$ and $\ge5\%$ CL 
areas.
In general,
for the $\rhoeta$ plane as well as for any parameter spaces, 
the individual inputs are less 
constraining than what they yield in the combined 
fit: \ie,
%This is due to the ambiguities in the inverse of the 
%measured sinus of twice the UT angles,
%\beqn
%   \omega_1 &=& \frac{1}{2}{\rm arcsin}({\rm sin}2\omega)~,  \nonumber\\
%   \omega_2 &=& \frac{\pi}{2} - \omega_1~, \\
%   \omega_3 &=& \pi + \omega_1~, \nonumber
%\eeqn
%for $\omega=0\dots\pi$ and $\omega=\alpha,\;\beta$.
a combination of input
variables can lead to a suppression of solutions thus enhancing the individual constraints. 
An example 
for this is drawn in the lower plot of Fig.~\ref{fig_stgam}:
the individual contribution of $|\epsk|$ is stronger
in combination with $|V_{ub}/V_{cb}|$ (indicated by the arrows).
%suppressing the large 
%ambiguity $\beta_2$. 
The lower plot of Fig.~\ref{fig_stgam}
visualises the complementarity between $|V_{ub}/V_{cb}|$ and $|\epsk|$, constraining
$\stb$, on one hand, and $\dms/\dmd$ and $|\epsk|$, constraining 
$\gamma$, on the other hand.

\subsection{One Dimensional Parameter Spaces}

Following the line of the preceding sections we can derive
one-dimensional constraints for all parameters involved, such
as the various CKM parameters, the moduli of the CKM matrix 
elements, branching ratios of rare $K$ and $B$ meson decays 
as well as theoretical parameters. Consequently, we define the 
parameter we are interested in to be $\a$ and all others to 
be $\Mu$ (\cf, Section~\ref{sec_metrology}), and scan $a$.
Numerical and graphical results are obtained
for CKM fits not including (including) the world average value
of $\stb$ (see Table~\ref{tab_ckmValues} for the input parameters).
\begin{figure}[t]
  \epsfxsize\largefig
  \centerline{\epsffile{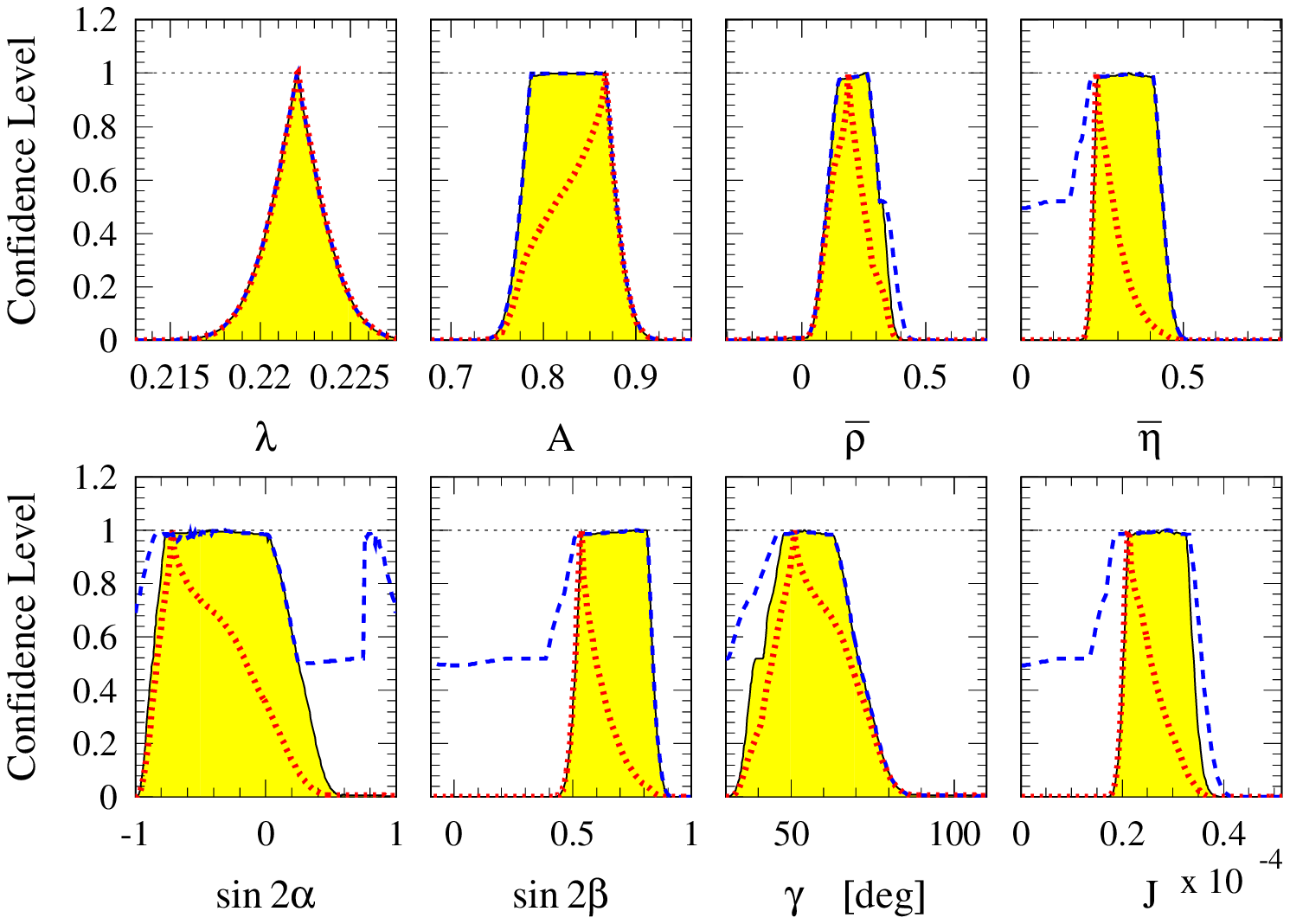}}
  \caption[.]{\label{fig_params1d}\em
	Confidence levels for one-dimensional parameter fits 
	of the CKM parameters, UT angles and 
	the Jarlskog parameter, not including $\stb$ in the fits.
	The solid (dashed) lines give the results with (without)
	the CP-violating $|\epsk|$ as fit input. The fits 
	corresponding to the dotted lines include $\stbwa$ and $|\epsk|$.}
\end{figure}
\vs
As an example, Fig.~\ref{fig_params1d} shows the CLs obtained 
for the Wolfenstein parameters, the UT angles and the Jarlskog
parameter, without (solid line, gray area) and with (dotted line) 
including the world average $\stbwa$ in the global fit. As in the 
two-dimensional case, the CLs shown correspond to the most compatible 
theory for a given point in $\a$. Since the parameter $\lambda$ 
is not significantly constrained by the other inputs,
its CL corresponds to the error function for one degree of freedom.
In contrast, parameters such as, \eg, $\rhobar$, $\etabar$, 
are constrained by observables whose SM predictions are dominated 
by systematic theoretical errors. The positions of the  
flanks of the \CL\  functions are determined by the $[\yQCD]$ ranges, 
whereas their sharp rises are determined by statistical 
errors. Therefore, one should not attribute an absolute meaning 
to the precise locations of the flanks: they are due to the 
{\it assumptions} made to define the $[\yQCD]$ ranges. In particular,
when the world average $\stbwa$ is used, the \CL\ function 
obtained (\cf, Fig.~\ref{fig_params1d}) exhibits a triangular shape:
whereas the fall off on the right hand side of the function is well 
defined, the location of the flank on the left hand side is somewhat 
arbitrary and hence arguable.
\vs
\begin{table}[p]
{%\small
\setlength{\tabcolsep}{0.58pc}
\begin{tabular}{lcccc} \hline
&&&& \\
\rs{Parameter}	& \rs{$\ge32\%$ CL}		& \rs{half width}
			& \rs{$\ge5\%$ CL}	& \rs{half width} \\
\hline \hline
&&&& \\
\rs{$\lambda$}	& \mc{2}{c}{\rs{$0.2221\pm0.0021$}}
		& \mc{2}{c}{\rs{$0.2221\pm0.0041$}} \\
\rs{$A$}	& \rs{0.770 - 0.888}	& \rs{0.059}
		& \rs{0.754 - 0.906}	& \rs{0.076} \\
\rs{$\rhobar$}	& \rs{0.08 - 0.35} 	& \rs{0.14}
		& \rs{0.04 - 0.38}	& \rs{0.17} \\
\rs{$\etabar$}	& \rs{0.22 - 0.46}	& \rs{0.12}
		& \rs{0.21 - 0.49} 	& \rs{0.14} \\ 
\hline
&&&& \\
\rs{$J$ $(10^{-5})$}	& \rs{2.0 - 3.5}
					& \rs{0.8}
		& \rs{1.9 - 3.7}	
					& \rs{0.9} \\
\hline
&&&& \\
\rs{$\sta$}	& \rs{$-0.91$ - 0.34}	& \rs{0.63}
		& \rs{$-0.96$ - 0.49}  	& \rs{0.73} \\ 
\rs{$\stb$}	& \rs{0.50 - 0.86}	& \rs{0.18}
		& \rs{0.47 - 0.89}	& \rs{0.21} \\
\rs{$\alpha$}	& \rs{$80^\circ$ - $123^\circ$}	& \rs{$22^\circ$}
		& \rs{$75^\circ$ - $127^\circ$}	& \rs{$26^\circ$} \\ 
\rs{$\beta$}	& \rs{$15.0^\circ$ - $29.7^\circ$}	
						& \rs{$7.4^\circ$}
		& \rs{$14.0^\circ$ - $31.4^\circ$}	
						& \rs{$8.7^\circ$} \\ 
\rs{$\gamma=\delta$}
		& \rs{$37^\circ$ - $75^\circ$}	& \rs{$19^\circ$}
		& \rs{$34^\circ$ - $82^\circ$}	& \rs{$24^\circ$} \\ 
\hline
&&&& \\
\rs{${\rm sin}\theta_{12}$}
		& \mc{2}{c}{\rs{$0.2221\pm0.0021$}}
		& \mc{2}{c}{\rs{$0.2221\pm0.0041$}} \\
\rs{${\rm sin}\theta_{13}$ $(10^{-3})$}
		& \rs{2.70 - 4.31}
						& \rs{0.81}
		& \rs{2.49 - 4.55}
						& \rs{1.03} \\
\rs{${\rm sin}\theta_{23}$ $(10^{-3})$}
		& \rs{38.4 - 43.2}
						& \rs{2.4}
		& \rs{37.9 - 43.6}
						& \rs{2.8} \\
\hline		
&&&& \\
\rs{$|V_{ud}|$}	& \mc{2}{c}{\rs{$0.97504\pm0.00049$}}
		& \mc{2}{c}{\rs{$0.97504\pm0.00094$}} \\
\rs{$|V_{us}|$}	& \mc{2}{c}{\rs{$0.2221\pm0.0021$}}
		& \mc{2}{c}{\rs{$0.2221\pm0.0042$}} \\
\rs{$|V_{ub}|$ $(10^{-3})$}	
		& \rs{2.70 - 4.31}
						& \rs{0.81}
		& \rs{2.49 - 4.55}
						& \rs{1.03} \\
\rs{$|V_{cd}|$}	& \mc{2}{c}{\rs{$0.2220\pm0.0021$}}
		& \mc{2}{c}{\rs{$0.2220\pm0.0042$}} \\
\rs{$|V_{cs}|$}	& \mc{2}{c}{\rs{$0.97422\pm0.00056$}}
		& \mc{2}{c}{\rs{$0.97422\pm0.00102$}} \\
\rs{$|V_{cb}|$ $(10^{-3})$}	
		& \rs{38.4 - 43.2}
						& \rs{2.4}
		& \rs{37.9 - 43.6}
						& \rs{2.8} \\
\rs{$|V_{td}|$ $(10^{-3})$}	
		& \rs{6.6 - 9.2}	
						& \rs{1.3}
		& \rs{6.3 - 9.6}	
						& \rs{1.6} \\
\rs{$|V_{ts}|$ $(10^{-3})$}	
		& \rs{37.7 - 42.8}
						& \rs{2.6}
		& \rs{37.3 - 43.2}
						& \rs{3.0} \\
\rs{$|V_{tb}|$}	& \rs{0.99907 - 0.99927}	& \rs{$10\times10^{-5}$}
		& \rs{0.99905 - 0.99929}	& \rs{$12\times10^{-5}$} \\
\hline
&&&& \\
\rs{$\dms$ $({\rm ps}^{-1})$}	& \rs{15.5 - 33.7}
						& \rs{9.1} 
 		& \rs{15.0 - 42.0}
						& \rs{13.5} \\
\hline
&&&& \\
\rs{${\rm BR}(K^0_{\rm L}\rightarrow\pi^0\nu\bar\nu)$ $(10^{-11})$}
		& \rs{1.3 - 4.0}	
						& \rs{1.4}
		& \rs{1.2 - 4.4}	
						& \rs{1.6} \\
\rs{${\rm BR}(K^+\rightarrow\pi^+\nu\bar\nu)$ $(10^{-11})$}
		& \rs{5.1 - 9.6}
						& \rs{2.3} 
		& \rs{4.8 - 10.5}
						& \rs{2.9} \\
\rs{${\rm BR}(B^+\rightarrow\tau^+\nu_\tau)$ $(10^{-5})$}
		& \rs{4.6 - 20.0}
						& \rs{7.7}
		& \rs{3.6 - 23.6}
						& \rs{10.0} \\
\rs{${\rm BR}(B^+\rightarrow\mu^+\nu_\mu)$ $(10^{-7})$}
		& \rs{1.8 - 7.9}
						& \rs{3.1}
		& \rs{1.5 - 9.3}
						& \rs{3.9} \\
\hline\hline
&&&& \\
\rs{$\fbdbd$ (MeV)}	
		& \rs{193 - 271}		& \rs{39}
		& \rs{184 - 284}		& \rs{50} \\
\rs{$B_K$}	& \mc{2}{c}{\rs{$>~0.55$}}
		& \mc{2}{c}{\rs{$>~0.50$}}	\\
\rs{$m_t$ (GeV)}	
		& \rs{106 - 406}		& \rs{150}
		& \rs{93 - 565}			& \rs{236} \\
\hline
\end{tabular}}
  \caption[.]{\label{tab_params1d}\em
	Fit results for the various CKM parameters, the CKM
	matrix elements, branching ratios of some rare $K$ and $B$
	meson decays and theoretical quantities. Ranges are quoted
	for the quantities that are limited by systematic 
	theoretical errors. The last three lines give the 
	ranges obtained for chosen theoretical 
	parameters when removing their respective 
	bounds in the fit. }
\end{table}
\begin{table}[p]
{%\small
\setlength{\tabcolsep}{0.40pc}
\begin{tabular}{lcccc|rr} \hline
&&&& \\
\rs{Parameter}	& \rs{$\ge32\%$ CL}		& \rs{half width}
			& \rs{$\ge5\%$ CL}	& \rs{half width} 
	& \rs{\ft $\delta_{32}$}
	& \rs{{\ft $\delta_{5}$}} \\
\hline \hline
&&&& \\
\rs{$\lambda$}	& \mc{2}{c}{\rs{$0.2221\pm0.0021$}}
		& \mc{2}{c|}{\rs{$0.2221\pm0.0041$}} 
	& \rs{\ft 0} & \rs{\ft 0} \\
\rs{$A$}	& \rs{0.782 - 0.888}	& \rs{0.053}
		& \rs{0.758 - 0.906}	& \rs{0.074} 
	& \rs{\ft 10} & \rs{\ft 3} \\
\rs{$\rhobar$}	& \rs{0.09 - 0.29} 	& \rs{0.10}
		& \rs{0.04 - 0.37}	& \rs{0.16} 
	& \rs{\ft 29} & \rs{\ft 6} \\
\rs{$\etabar$}	& \rs{0.22 - 0.32}	& \rs{0.05}
		& \rs{0.21 - 0.42} 	& \rs{0.11} \
	& \rs{\ft 58} & \rs{\ft 21} \\
\hline
&&&& \\
\rs{$J$ $(10^{-5})$}	
		& \rs{2.0 - 2.9}
					& \rs{0.5}
		& \rs{1.9 - 3.5}	
					& \rs{0.8} 
	& \rs{\ft 38} & \rs{\ft 11} \\
\hline
&&&& \\
\rs{$\sta$}	& \rs{$-0.88$ - 0.04}	& \rs{0.46}
		& \rs{$-0.95$ - 0.33}  	& \rs{0.64}  
	& \rs{\ft 27} & \rs{\ft 12} \\
\rs{$\stb$}	& \rs{0.50 - 0.67}	& \rs{0.09}
		& \rs{0.47 - 0.81}	& \rs{0.17} 
	& \rs{\ft 50} & \rs{\ft 19} \\
\rs{$\alpha$}	& \rs{$89^\circ$ - $121^\circ$}	& \rs{$16^\circ$}
		& \rs{$80^\circ$ - $126^\circ$}	& \rs{$23^\circ$}  
	& \rs{\ft 27} & \rs{\ft 12} \\
\rs{$\beta$}	& \rs{$15.0^\circ$ - $21.0^\circ$}	
						& \rs{$3.0^\circ$}
		& \rs{$14.0^\circ$ - $27.0^\circ$}	
						& \rs{$6.5^\circ$}  
	& \rs{\ft 59} & \rs{\ft 25} \\
\rs{$\gamma=\delta$}
		& \rs{$42^\circ$ - $74^\circ$}	& \rs{$16^\circ$}
		& \rs{$34^\circ$ - $82^\circ$}	& \rs{$24^\circ$}  
	& \rs{\ft 16} & \rs{\ft 0} \\
\hline
&&&& \\
\rs{${\rm sin}\theta_{12}$}
		& \mc{2}{c}{\rs{$0.2221\pm0.0021$}}
		& \mc{2}{c|}{\rs{$0.2221\pm0.0041$}} 
	& \rs{\ft 0} & \rs{\ft 0} \\
\rs{${\rm sin}\theta_{13}$ $(10^{-3})$}
		& \rs{2.70 - 4.03}
						& \rs{0.67}
		& \rs{2.49 - 4.38}
						& \rs{0.95} 
	& \rs{\ft 17} & \rs{\ft 8} \\
\rs{${\rm sin}\theta_{23}$ $(10^{-3})$}
		& \rs{38.4 - 43.2}
						& \rs{2.4}
		& \rs{38.0 - 43.6}
						& \rs{2.8} 
	& \rs{\ft 0} & \rs{\ft 0} \\
\hline		
&&&& \\
\rs{$|V_{ud}|$}	& \mc{2}{c}{\rs{$0.97504\pm0.00049$}}
		& \mc{2}{c|}{\rs{$0.97504\pm0.00094$}} 
	& \rs{\ft 0} & \rs{\ft 0} \\
\rs{$|V_{us}|$}	& \mc{2}{c}{\rs{$0.2221\pm0.0021$}}
		& \mc{2}{c|}{\rs{$0.2221\pm0.0042$}} 
	& \rs{\ft 0} & \rs{\ft 0} \\
\rs{$|V_{ub}|$ $(10^{-3})$}	
		& \rs{2.70 - 3.71}
						& \rs{0.51}
		& \rs{2.45 - 4.38}
						& \rs{0.96} 
	& \rs{\ft 37} & \rs{\ft 7} \\
\rs{$|V_{cd}|$}	& \mc{2}{c}{\rs{$0.2220\pm0.0021$}}
		& \mc{2}{c|}{\rs{$0.2220\pm0.0042$}} 
	& \rs{\ft 0} & \rs{\ft 0} \\
\rs{$|V_{cs}|$}	& \mc{2}{c}{\rs{$0.97414\pm0.00049$}}
		& \mc{2}{c|}{\rs{$0.97414\pm0.00098$}} 
	& \rs{\ft 13} & \rs{\ft 4} \\
\rs{$|V_{cb}|$ $(10^{-3})$}	
		& \rs{38.7 - 43.2}
						& \rs{2.3}
		& \rs{38.1 - 43.6}
						& \rs{2.8} 
	& \rs{\ft 4} & \rs{\ft 0} \\
\rs{$|V_{td}|$ $(10^{-3})$}	
		& \rs{7.2 - 9.2}	
						& \rs{1.0}
		& \rs{6.6 - 9.6}	
						& \rs{1.5} 
	& \rs{\ft 23} & \rs{\ft 6} \\
\rs{$|V_{ts}|$ $(10^{-3})$}	
		& \rs{38.0 - 42.7}
						& \rs{2.4}
		& \rs{37.4 - 43.1}
						& \rs{2.9} 
	& \rs{\ft 8} & \rs{\ft 3} \\
\rs{$|V_{tb}|$}	& \rs{0.99907 - 0.99926}	& \rs{$9\times10^{-5}$}
		& \rs{0.99905 - 0.99928}	& \rs{$11\times10^{-5}$} 
	& \rs{\ft 10} & \rs{\ft 8} \\
\hline
&&&& \\
\rs{$\dms$ $({\rm ps}^{-1})$}	
		& \rs{15.5 - 33.7}
						& \rs{9.1} 
 		& \rs{15.0 - 41.3}
						& \rs{13.1} 
	& \rs{\ft 0} & \rs{\ft 3} \\
\hline
&&&& \\
\rs{${\rm BR}(K^0_{\rm L}\rightarrow\pi^0\nu\bar\nu)$ $(10^{-11})$
\hspace{-0.8cm}}
		& \rs{1.2 - 2.6}	
						& \rs{0.7}
		& \rs{1.1 - 3.8}	
						& \rs{1.4} 
	& \rs{\ft 50} & \rs{\ft 13} \\
\rs{${\rm BR}(K^+\rightarrow\pi^+\nu\bar\nu)$ $(10^{-11})$
\hspace{-0.8cm}}
		& \rs{6.6 - 9.5}
						& \rs{1.5} 
		& \rs{5.4 - 10.4}
						& \rs{2.5} 
	& \rs{\ft 35} & \rs{\ft 14} \\
\rs{${\rm BR}(B^+\rightarrow\tau^+\nu_\tau)$ $(10^{-5})$
\hspace{-0.8cm}}
		& \rs{4.6 - 12.4}
						& \rs{3.9}
		& \rs{3.6 - 21.0}
						& \rs{8.7} 
	& \rs{\ft 49} & \rs{\ft 13} \\
\rs{${\rm BR}(B^+\rightarrow\mu^+\nu_\mu)$ $(10^{-7})$
\hspace{-0.8cm}}
		& \rs{1.8 - 4.9}
						& \rs{1.6}
		& \rs{1.4 - 8.3}
						& \rs{3.5} 
	& \rs{\ft 48} & \rs{\ft 10} \\
\hline\hline
&&&& \\
\rs{$\fbdbd$ (MeV)}	
		& \rs{194 - 246}		& \rs{26}
		& \rs{185 - 272}		& \rs{44} 
	& \rs{\ft 33} & \rs{\ft 12} \\
\rs{$B_K$}	& \mc{2}{c}{\rs{$>~0.72$}}
		& \mc{2}{c|}{\rs{$>~0.55$}}	
	& \rs{\ft 31} & \rs{\ft 10} \\
\rs{$m_t$ (GeV)}	
		&  \rs{124 - 406}		& \rs{141}
		&  \rs{102 - 550}		& \rs{224} 
	& \rs{\ft 6} & \rs{\ft 5} \\
\hline
\end{tabular}}
  \caption[.]{\label{tab_params1d_s2b}\em
	Fit results including the world average on $\stbwa$.
	As in Table~\ref{tab_params1d}, ranges are given for 
	the quantities that are limited 
	by systematic theoretical errors. The two right columns
	give the relative improvements (in percent)
	of the $\ge32\%~{\rm CL}$ and $\ge5\%~{\rm CL}$ 
	half widths with respect to
	the fit results without $\stb$ given in  
	Tab.~\ref{tab_params1d}. The last three lines give the 
	ranges obtained for chosen theoretical 
	parameters when removing their respective 
	bounds in the fit. }
\end{table}
The results for all relevant parameters considered in this work
are listed in Table~\ref{tab_params1d}, without $\stb$ in the fit,
and Table~\ref{tab_params1d_s2b} when including $\stbwa$.
Given are the
ranges for $\ge32\%$ and $\ge5\%$ CLs in the case of
theoretically limited quantities and the corresponding
Gaussian errors in the case of experimentally limited
quantities. The $95\%~\CL$ allowed ranges for the CKM 
matrix elements are similar to the ones quoted by the PDG.
Numerical results involving $B\rightarrow \pi\pi/K\pi$ decays
are not presented here\footnote
{
	It proved however straightforward to implement 
	them into the \ckmfit\  scheme of the \CkmFitter\  
	package (see also the discussion in 
	Section~\ref{sec_FutureProspects}).

}.

\subsubsection{Indirect Evidence for CP Violation}
\label{IndirectEvidenceforCPViolation}

It is interesting to investigate the possibility of an indirect 
evidence for CP violation, from the measurements of non CP-violating 
observables.  The dashed curves in Fig.~\ref{fig_params1d} give 
the CLs which are obtained when using neither $\stb$, nor $|\epsk|$ 
in the fits. For $\etabar=0$ (hence no CP violation) \ckmfit\ 
yields $\CL\simeq 50\%$. Therefore, we find that CP conservation 
cannot be excluded without $\stb$ or $|\epsk|$: a better knowledge 
of $|V_{ub}/V_{cb}|$ and $\dmd$ ($\dms$) is needed to draw any 
further conclusions. The large value of ${\rm CL}(\etabar=0)$ 
stems from the fact that the quoted \CLs\ are upper bounds.
There exist realizations of the SM, with $\etabar=0$ and with 
all $\yQCD$ parameters within their allowed $[\yQCD]$ ranges,
which provide a perfectly acceptable description of data 
(without $|\epsk|$ and $\stb$). The realizations of the SM which 
yield the best agreement are chosen to compute ${\rm CL}(\etabar=0)$.
\vs
This result is not in qualitative agreement with the one obtained 
in Ref.~\cite{Achille1}: this illustrates how widely different 
conclusions can be reached depending on the choice made for the 
statistical treatment. The Bayesian approach, while computing 
${\rm CL}(\etabar=0)$, is incorporating in passing,
through the use of PDFs and Eq.~(\ref{BayesianII}),
the "volume" of the domain in the $\yQCD$ space 
(weighted by the theoretical PDFs) where realizations of the SM 
are in agreement with data. The "volume" of this $\yQCD$ domain 
is small, as a result ${\rm CL}(\etabar=0)$ is small.
The frequentist \ckmfit\ scheme does not consider PDFs
for $\yQCD$ parameters as a valid concept, it thus cannot 
define "volume" of a domain in this space deprived of metric:
only the best realizations are retained to define the CL.

\begin{figure}[t]
  \epsfxsize\smallfig
  \centerline{\epsffile{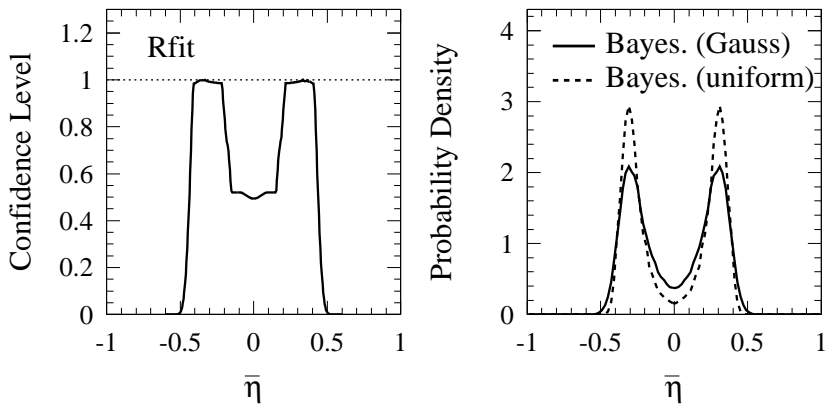}}
  \caption[.]{\label{fig_etatest}\em
    The {\rm \ckmfit} \ $\CL(\etabar)$ function (left hand side)
    and the Bayesian PDF of $\etabar$ (right hand side) 
    obtained using the same inputs.
   The solid (resp. dashed)
line 
is obtained using Gaussian (resp. uniform)
PDFs for the $\yQCD$ parameters. 
        }
\end{figure}

An expanded view of the \ckmfit \ $\CL(\etabar)$ function
is shown on the left hand side of Fig.~\ref{fig_etatest}
in the range $[-1,+1]$. Since no observable sensitive to CP 
violation is incorporated in the fit, $\CL(\etabar)$ is an even 
function. The Bayesian PDF of $\etabar$ (\cf, Eq.~(\ref{BayesianII})) 
is shown on the right hand side of the same figure. The solid (resp. 
dashed) line is obtained using Gaussian (resp. uniform)
PDFs for the $\yQCD$ parameters. One observes that,
independently of the definition used to derive a \CL\ from the PDF,
both Bayesian \CLs\ will be low (at the percent level if one uses 
Eq.~(\ref{eq_CLBayesian})) and most notably the one obtained from 
the uniform PDFs, although the inputs to the fit are identical to 
the ones used by \ckmfit.
 
\subsubsection{Impact of the New \boldmath$\stb$ Measurements}

\begin{figure}[t]
  \epsfxsize\largefig
  \centerline{\epsffile{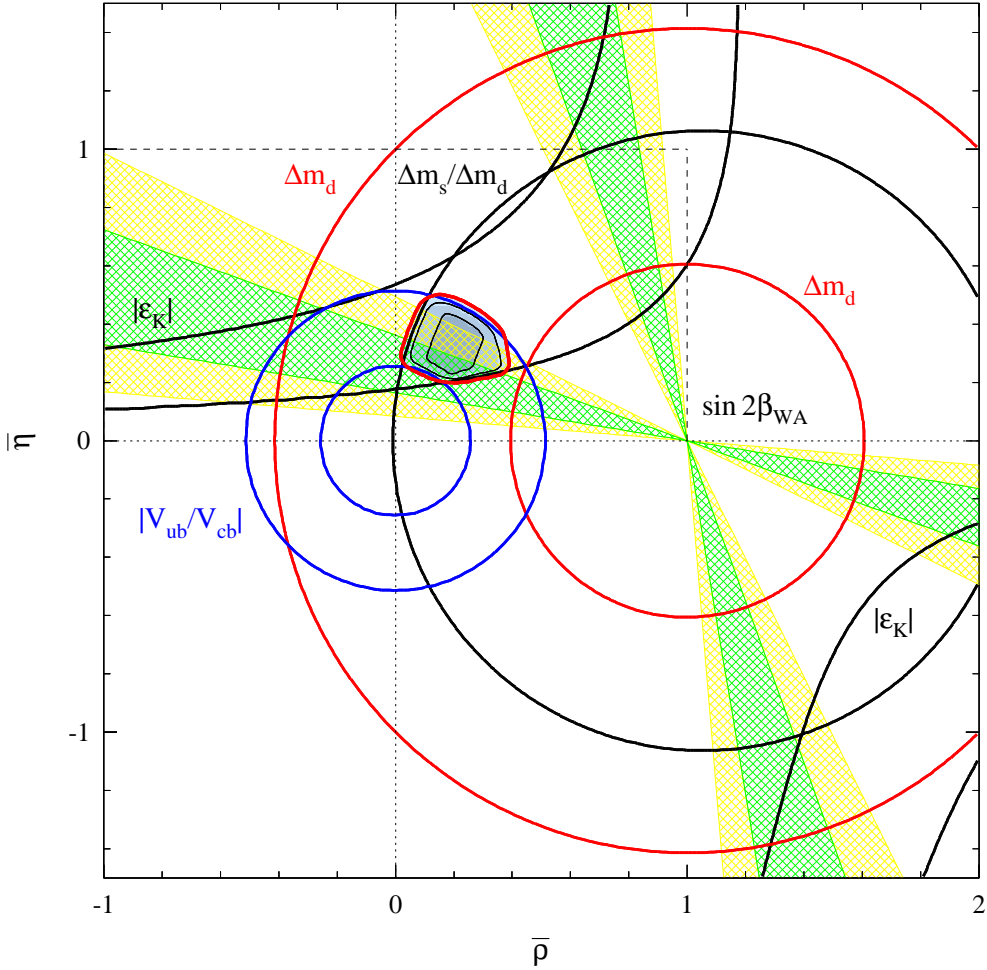}}
  \caption[.]{\label{fig_rhoeta_big}\em
	Confidence levels in the enlarged $\rhoeta$ plane for the 
	global CKM fit. See Fig.~\ref{fig_rhoeta} for a 
	description of the curves shown.
%       The shaded areas indicate the regions of
%	$\ge90\%$, $\ge32\%$ and $\ge5\%$ CLs, respectively. 
%	Also shown are the $5\%$
%	CL contours of the individual constraints. 
	The $\ge32\%$ and $\ge5\%$ CL constraints from the 
	world average of the $\stb$ measurements, not entering the 
	combined fit, are depicted by the dashed areas. 
	All four ambiguities are drawn, three of 
	which are excluded by the Standard Model. }
\end{figure}

\begin{figure}[t]
  \epsfxsize\smallfig
  \centerline{\epsffile{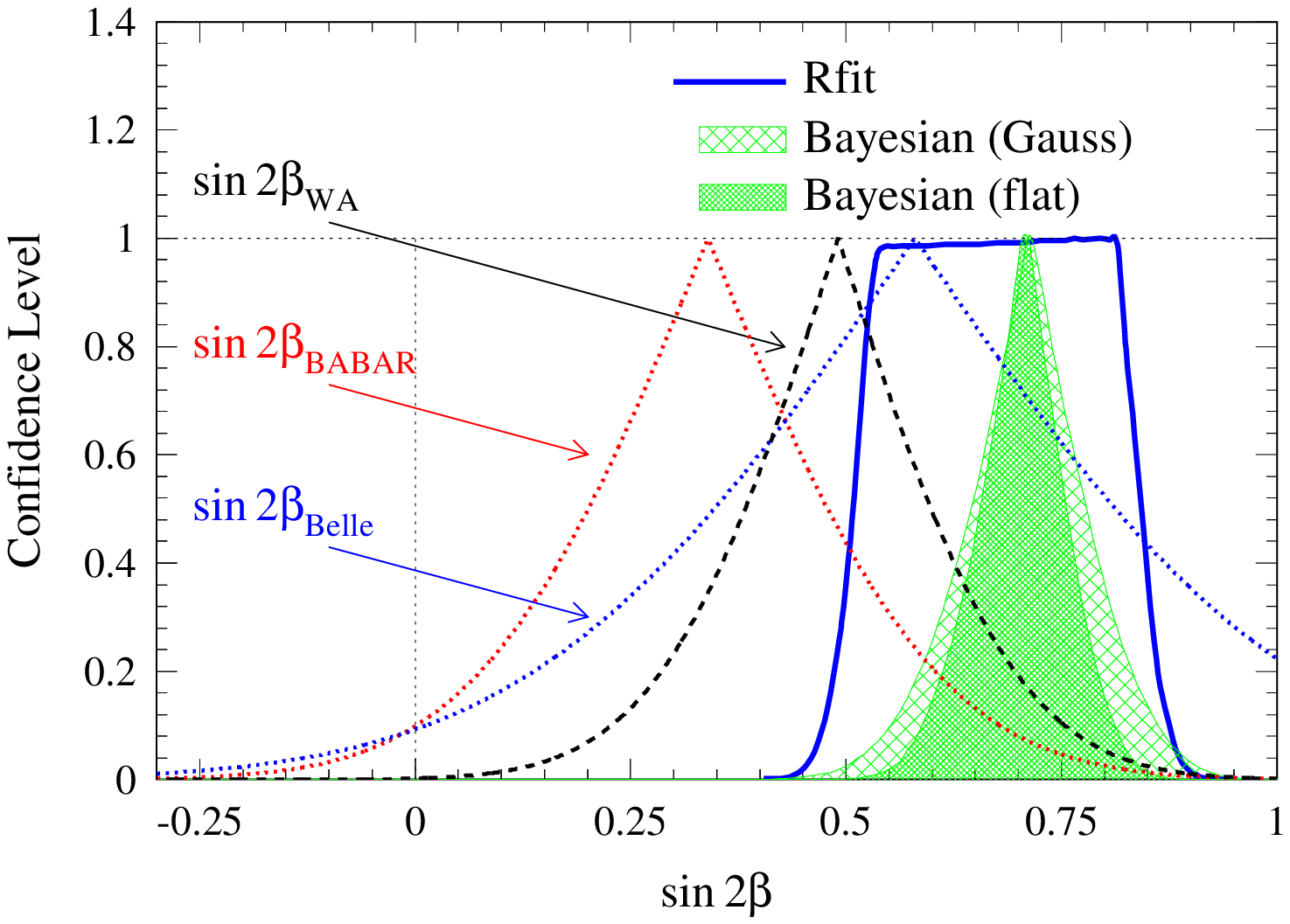}}
  \caption[.]{\label{fig_s2b1d}\em
	Confidence levels of the recent $\stb$ measurements of 
	\babarem~\cite{sin2betaBabar} and Belle~\cite{sin2betaBelle} 
	together with the world average. These \CLs~ are compared to the 
	indirect SM constraints obtained from \ckmfit\ in this 
	work (solid line). Also shown are the results from the 
	integrated PDFs obtained in the Bayesian 
	analysis using Gaussian and uniform PDFs for the systematic 
	theoretical errors.}	 
\end{figure}

The measurement of $\stb$ provides the UT angle $\beta$ up to 
a four-fold ambiguity. To illustrate this, we have enlarged the 
borders of the $\rhoeta$ plane in Fig.~\ref{fig_rhoeta_big}.
Shown are the individual constraints and the result from the  
global fit corresponding to Fig.~\ref{fig_rhoeta}, as well as 
the four solutions from the world average $\stbwa$. It is a 
non-trivial outcome of the SM fit that it leads to an exclusion
of three out of the four ambiguities.
\vs
The confidence levels for the $\stb$ measurements of 
\babar~\cite{sin2betaBabar} and Belle~\cite{sin2betaBelle} 
together with the world average\footnote
{
  As stated before, the measurements are assumed
  to be Gaussian distributed. 
  Therefore, the CLs given are direct confidence levels 
  and not upper bounds,
  as one obtains when theoretical systematics contribute
  significantly to the uncertainty of a quantity.
} 
and the result of \ckmfit\  (without $\stb$) are shown\footnote
{
	The plateau of the {\rm CL} function obtained from \ckmfit\ 
	corresponds to $\stb$ values belonging to the $\ymodopt$
	domain. 
	The {\rm CL} on the plateau is not exactly equal to unity.
	The slight slope which is observed is due to the $\Vcd$ input:
	being a function of $\rhobar$ and $\etabar$,
	and having a statistically dominated uncertainty,
        $\Vcd$ lifts by a very slight amount the degeneracy
	discussed in Section~\ref{sec_metrology}.
} 
in Fig.~\ref{fig_s2b1d}. Given in addition are the results of 
the integrated PDFs obtained in the Bayesian analysis,
when ascribing Gaussian or uniform PDFs to the systematic 
theoretical errors. While they significantly differ in the 
precision they claim for, the \ckmfit\ and Bayesian indirect 
determinations of $\stb$ are both compatible with the world average. 
\vs
The last two columns of Table~\ref{tab_params1d_s2b} give
the relative improvements (in percent) of the parameter
constraints gained by including $\stbwa$ in the CKM fit:
the two quoted numbers $\delta_{32}$ and $\delta_{5}$ refer 
respectively to the ranges allowed at $32\%$ and $5\%$ CLs.
All quantities sensitive to CP violation 
%enjoy 
benefit from significantly smaller $32\%$ allowed ranges,
with a relative reduction of up to 50\%. 
This reduction however 
gets suppressed when going to $5\%$ \CLs. This reduced improvement 
is explained by the fact that a significant fraction of $5\%$~\CL\ 
limits remain determined by the border of the $\ymodopt$ domain,
as can be seen on Fig.~\ref{fig_params1d}.
Although the Gaussian shape of the additional constraint from 
$\stbwa$ leads to significant structures within the allowed ranges 
of CLs, the $[\yQCD]$ ranges still determine part of the 
CL function tails. 
%Clearly, such a behavior is
%artificially provoked by the dictated ranges of the theoretical
%systematics. In the case that one observes SM incompatibilities,
%one therefore needs to employ \ckmfoot-like techniques (see
%Section~\ref{TheExtendedConservativeMethod}) for a further 
%study of the discrepancy.

\subsubsection{Numerical Comparison With Bayesian Results}

\begin{table}[t]
\begin{center}
\setlength{\tabcolsep}{0.93pc}
{\normalsize
\begin{tabular}{lcccc}\hline
  	& 		
	& & \mc{2}{c}{$\Delta{\rm Freq.}/\Delta{\rm Bayes(Gauss)}$} 
	\\ 
 \rs{Parameter}	& \rs{Gaussian}	& \rs{Uniform} & $\ge32\%$ CL	& $\ge5\%$ CL 
	\\	
\hline
&&&& \\
  \rs{$\lambda$}& \rs{$0.2219\pm0.0021$}	
	& \rs{$0.2219\pm0.0021$}& \rs{1.0}	& \rs{1.0}		\\
  \rs{$A$} 		& \rs{$0.832\pm0.040$}		
	& \rs{$0.830\pm0.028$}& \rs{1.5}	& \rs{1.0}	\\
  \rs{$\rhobar$}& \rs{$0.217\pm0.063$}	
	& \rs{$0.203\pm0.048$}& \rs{2.2}	& \rs{1.3}	\\
  \rs{$\etabar$}& \rs{$0.331\pm0.056$}			
	& \rs{$0.330\pm0.039$}& \rs{2.1}	& \rs{1.3}	\\
  \rs{$J$}	& \rs{$(2.70\pm0.36)\times10^{-5}$}
						
	& \rs{$(2.70\pm0.25)\times10^{-5}$}& \rs{2.2}	& \rs{1.3}	\\
  \rs{$\sta$} 	& \rs{$-0.32\pm0.30$}			
	& \rs{$-0.30\pm0.24$}& \rs{2.1}	& \rs{1.2}	\\
  \rs{$\stb$}	& \rs{$0.710\pm0.093$}			
	& \rs{$0.705\pm0.065$}& \rs{1.9}	& \rs{1.1}	\\
  \rs{$\gamma$}	& \rs{$57.0^\circ\pm8.7^\circ$}	
							
	& \rs{$58.5^\circ\pm7.0^\circ$}& \rs{2.2}	& \rs{1.4} \\
\hline
&&&& \\
  \rs{$\fbdbd$}	& \rs{$(230\pm27)~{\rm MeV}$} 
	& \rs{$(227\pm13)~{\rm MeV}$} & \rs{1.4}	& \rs{0.9}\\
  \rs{$B_K$}	& \rs{$0.91\pm0.12$} 
	& \rs{$0.89\pm0.08$} & \rs{-}	& \rs{-}\\
\hline
\end{tabular}}
\caption{\label{tab_params1dBayes} \em
	Results for the CKM parameters, the UT angles and 
	theoretical parameters, using Bayesian statistics with
	Gaussian (second column) or uniform (third column) 
	distributed probability density functions
	for the systematic theoretical part of the 
	input parameters. Note that asymmetric errors have
	been averaged. The constraint from $\stb$ is 
	not used. The fourth and fifth columns give
	the ratios of the half widths of the 
	$\ge32\%$ and $\ge5\%$ CL \ckmfit\
	error intervals of Table~\ref{tab_params1d}, to
	the Bayesian errors given for the Gaussian case in the
	second column.}
\end{center}
\end{table}
\begin{figure}[t]
  \epsfxsize\smallfig
  \centerline{\epsffile{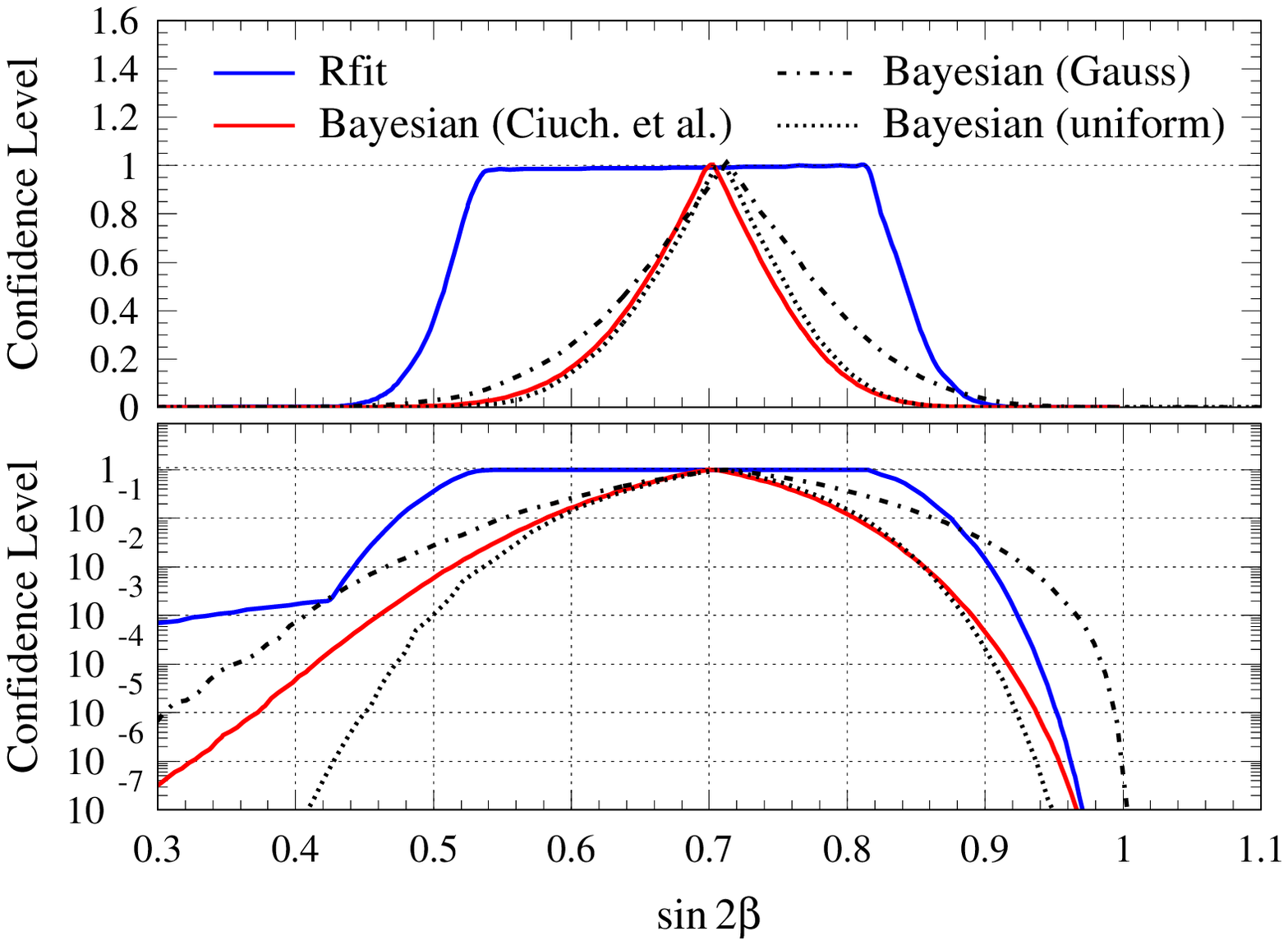}}
  \caption[.]{\label{fig_comprfitbayes}\em
	Comparison between \ckmfit\ (broad solid curve)
	and Bayesian fits for the
	indirect CKM constraint on $\stb$. The lower plot
 	displays the identical curves as in the upper plot
	but in logarithmic scale. For the Bayesian fits:
	Gaussian (uniform) systematic theoretical PDFs are
	depicted as dashed-dotted (dotted) curves. Shown
	in addition is the (integrated) result obtained in the 
	Bayesian analysis of Ref.~\cite{Achille1} (narrow 
	solid curve) for which mostly uniform PDFs were chosen 
	for the dominant theoretical uncertainties.}
\end{figure}
As discussed previously,
the Bayesian treatment identifies experimental and
theoretical likelihoods as probability density functions
which are folded according to Eq.~(\ref{BayesianII})
(\cf, Section \ref{sec_TheBayesianTreatment},
Appendix~\ref{TheBayesianMethodcaughtunder-conservative},
and also Refs.~\cite{Achille1,Achille2}).
In practice, the convolution integrals are solved within 
the \CkmFitter\ package
using Monte Carlo techniques generating some $10^8$ test samples. 
All input quantities fluctuate according to Gaussian distributions, 
using their statistical experimental uncertainties $\sigexp$, 
and to the PDFs ascribed to systematics, characterized by their 
systematical uncertainties $\sigma_\syst$. For the sake of simplicity,
two choices are discussed here for the latter PDFs: all are taken 
to be Gaussians with standard deviations identified to $\sigma_\syst$,
or all are taken to be uniform distributions with half-widths 
identified to $\sigma_\syst$. In the latter case, 
one does not identify the half-widths to $\sqrt{3}\sigma_\syst$ 
in order to use PDFs which (naively) would lead the Bayesian 
approach to yield results the closest to the one of the \ckmfit\ 
scheme. Because of that, in the uniform case the RMS is smaller 
than in the Gaussian case: one thus expects to claim for significantly 
smaller uncertainties for the former choice than for the latter
choice. A comparison of the results for both choices is given in 
Table~\ref{tab_params1dBayes}.
The central values quoted correspond to the mean values of the 
resulting PDFs, while the errors are their RMS (\ie, when present,
asymmetric errors have been averaged).
The corresponding ranges provide a good approximation
of the $68\%$ confidence intervals which can be defined 
from an explicitly asymmetrical integration of the PDFs,
since most of them closely resemble Gaussian distributions. 
As expected, the uncertainties are larger for the Gaussian choice.
We observe a factor of about 2.2 (resp. 2.8) for the ratio between 
the $\ge32\%$ CL intervals of \ckmfit\ (Table~\ref{tab_params1d})
and the Bayesian ranges,
for the Gaussian choice (resp. the uniform choice).
For the Gaussian choice,
this ratio reduces to about 1.3 for the $\ge5\%$ CL.
\vs
Figure~\ref{fig_comprfitbayes} provides a graphical comparison 
between the \ckmfit\ result (the broad solid curve)
and the Bayesian results (the dashed-dotted and dotted curves)
on $\stb$. Shown in addition is the result obtained when 
asymmetrically integrating the output 
PDF obtained from the Bayesian analysis of Ref.~\cite{Achille1} 
(the narrow solid curve) where mostly uniform PDFs were 
chosen for the dominant theoretical uncertainties. The plot 
visualizes the tendency observed in Table~\ref{tab_params1dBayes} 
for the tails of lower CLs from \ckmfit\ and the Bayesian approach 
to evolve towards comparable uncertainty ranges. However, 
the curves do not converge the ones to the others:
the various treatments do not provide identical results,
even for very low CLs.
% but for the trivial null-CL case, 
% where the whole physical parameter range is allowed.

%
% --------------------- Compatibility of the SM -----------------------
%
\subsection{Probing the Standard Model}
\label{sec_compOfTheSM}

We have seen in the introduction that the metrological 
phase is intrinsically unable to detect a failure of 
the SM to describe the data. The interpretation of 
the test statistics $\ChiMinGlob$ is 
performed by means of a toy Monte Carlo simulation as
described in Section~\ref{sec_probingTheSM}.
\begin{figure}[t]
  \epsfxsize\smallfig
  \centerline{\epsffile{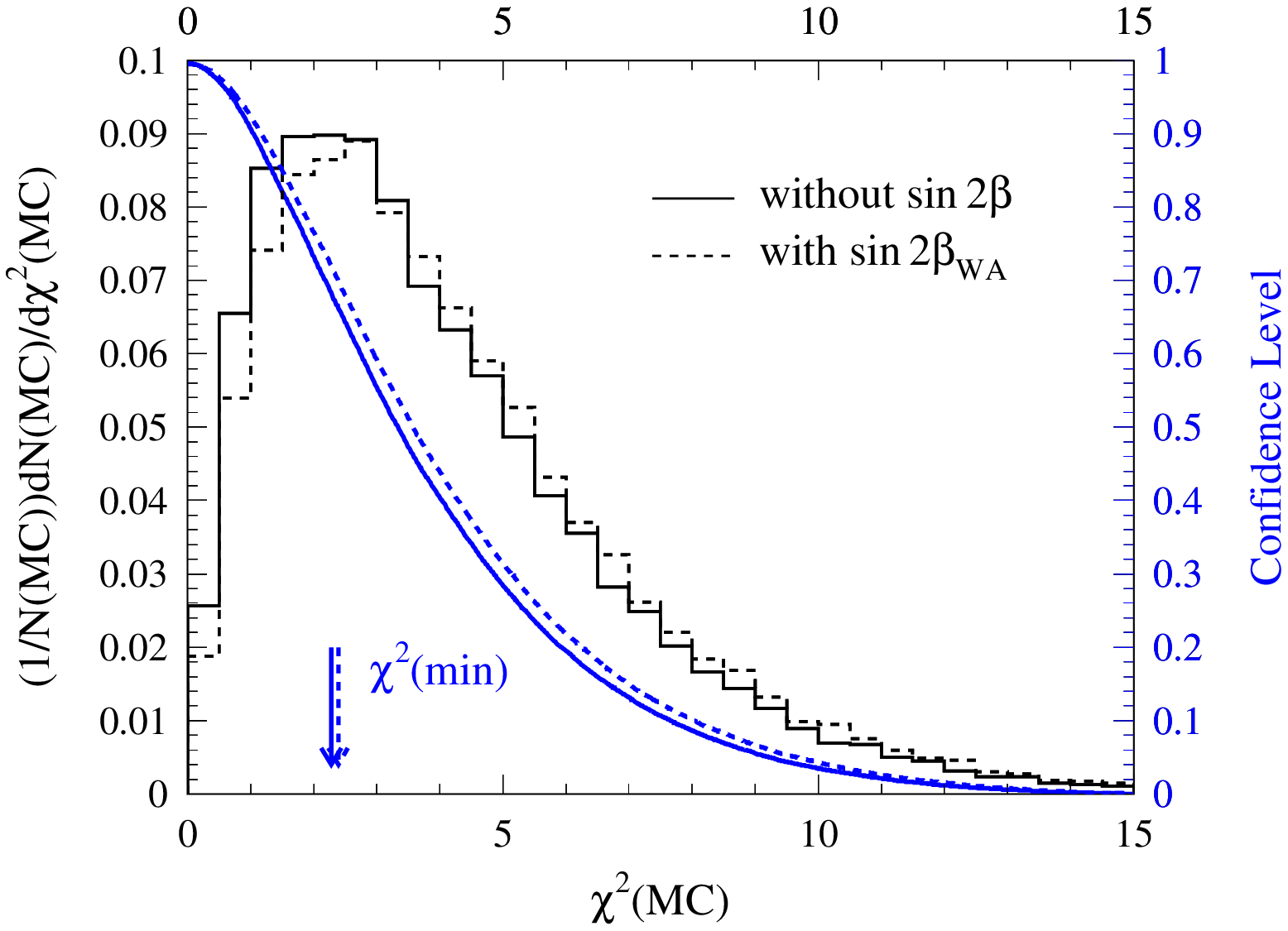}}
  \caption[.]{\label{fig_chi2min}\em
	Simulated $\F(\chi^2)$ distributions and corresponding
	CLs not including (solid lines) and including $\stbwa$ 
	(dashed lines) in the fit. Indicated by the arrows are the 
	corresponding minimal $\ChiMinGlob$ found in the analyses.}
\end{figure}
The fits in the previous section yield for the 
point of best compatibility
\beq
   \ChiMinGlob = 2.3~(2.4)~,
\eeq
for the data set without (with) $\stbwa$. We now generate 
the distribution $\F(\chi^2)$ of $\ChiMinGlob$ by fluctuating the 
measurements and parameters according to their non-theoretical 
errors around the theoretical values obtained using the parameter
set $\ymodopt$ for which is obtained $\ChiMinGlob$.
The resulting toy distributions are shown by the solid 
(with $\stbwa$ in the fit) and dashed (no $\stb$) histograms 
in Fig.~\ref{fig_chi2min}. Integrating the distributions according 
to Eq.~(\ref{eq_monteCarlo}) yields the corresponding
CL (smooth curves in Fig~\ref{fig_chi2min}). We find
\beq
   \Prob({\rm SM}) 
	\le {\rm CL}(\ChiMinGlob) 
	= 69\%~(71\%)~,
\eeq
for the validity of the SM without (with) $\stbwa$. Repeating the
study using the $\stb$ measurement of \babar\ (Belle)
instead of the world average, gives confidence levels of
${\rm CL}(\ChiMinGlob) = 59\%~(77\%)$ for the validity of
the SM.

%
% --------------- The Unitary Triangle and SUSY --------------------
%
\section{Supersymmetric Extensions of the Standard Model}

Having considered both metrology and probing the SM
with the present data set, one is now led to 
attempt an example analysis within an extended theoretical 
framework. This section aims to illustrate the search for 
specific new physics within a simple, predictive 
supersymmetric (SUSY) extension of the SM. 
\vs
There exist a considerable number of SUSY models in which 
new phases appear in the coupling between supersymmetric 
and SM fields. However, these models remain unpredictive as long
as the additional phases are unconstrained. We therefore cannot 
forecast how the shape of the UT is affected by the new 
fields. As a starting point, one can use restrictive assumptions 
which lead to more predictive models. In particular, one
may only retain models which do not involve additional CP 
violating phases, so that flavour-changing processes
are described by the same quark flavor mixing matrix $\VCKM$ 
as in the SM. Supersymmetric contributions to the transitions
between the down-type quarks ($b \rightarrow s$, $b \rightarrow d$, 
$s \rightarrow d$) are then proportional to the SM
CKM matrix elements. This restriction defines the category
of the so-called Minimal Flavour Violation (MFV) models 
which comprise some variants of the Minimal Supersymmetric 
Standard Model (MSSM), as well as the Two Higgs Doublet Models.
\vs
The MSSM has been extensively studied in the literature, 
and next-to-leading order (NLO) corrections to the SM have 
been calculated~\cite{krauss,ciuch}.
In this framework, the SUSY correction to neutral $K$ and $B$ 
meson mixing can be described by a single parameter which scales 
with the Inami-Lim function~(\ref{eq_inamilim})
of the top-quark loops in the box diagrams~\cite{david}:
\beq
        S(x_t) \rightarrow S(x_t)(1+f)~,
\eeq
leading to the following modified expressions
\beqn
\dmd({\rm MSSM}) & = & 
        \dmd({\rm SM})
        \left[S(x_{t}) 
                \rightarrow S(x_{t})(1 + f_d)
        \right]~, \\
\dms({\rm MSSM}) & = &
        \dms({\rm SM})
        \left[S(x_{t}) 
                \rightarrow S(x_{t})(1 + f_s)
        \right]~, \\
\label{eq_epskmssm}
|\epsk|({\rm MSSM}) & = & 
        |\epsk|({\rm SM})
        \left[S(x_{t}) 
                \rightarrow S(x_{t})(1 + f_{\epsilon})
        \right]~.
\eeqn
As pointed out in Refs.~\cite{buras-buras,david}, 
the parameters $f_d$, $f_s$ and $f_{\epsilon}$ belong to the same
\begin{figure}[t]
  \epsfxsize\smallfig
  \centerline{\epsffile{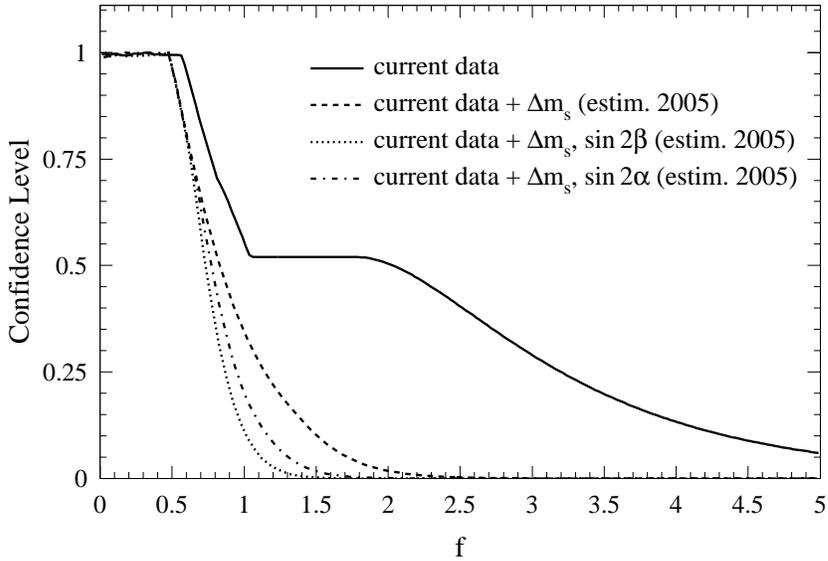}}
  \caption[.]{\label{fig_fsusy}\em
        Confidence level bounds of the global fit versus the
        MSSM offset $f$ (solid line). The ledge at $f\approx1$ 
        starts at the upper frequency limit to which 
        $\dms$ amplitudes have been measured. It continues as 
        a plateau due to the theoretically limited $\dmd$
        information and eventually decreases statistically.
        Also shown are the estimated constraints from 
        additional measurements of $\dms$ (CDF, D0) (dashed line),
        $\stb$ (dotted line) and $\sta$ (dashed-dotted line)
        (both \babarEm, Belle) for the year 2005~\cite{heikobcp4}.} 
\end{figure}
subprocesses so that the equality
\begin{equation}
f \equiv f_d = f_s = f_{\epsilon}~,
\end{equation}
holds in general. The numerical value for $f$ is assumed
to vary within the range
$0.4 \le f \le 0.75$~\cite{david}, 
while other authors find broader ranges
$-0.4 \le f \le 5.5$~\cite{buras-buras,buras,Silvestr}, depending
on whether or not Supergravity constraints (\ie, additional 
relations between masses and other terms of the MSSM 
Lagrangian) are applied to the MSSM.
\vs
The constraint from $|\epsk|$ in the $\bar{\rho}-\bar{\eta}$ plane
follows the form of a hyperbola~\cite{buras-buras}
\beq
	\bar{\eta} \propto \frac{1}{(1-\bar{\rho})S(x_t) + P_c}~,
\eeq
where $P_c = \eta_{ct}S(x_c,x_t) + \eta_{cc}S(x_c)$ 
stems from the charm loop contribution for which the SUSY
contribution is expected to be small. The neutral $B$ mass 
difference $\dmd$ measures the side $R_t$ of the 
UT~(\ref{eq_rt}), where:
\beq
\label{eq_rtil}
        R_t \propto \frac{1}{\sqrt{S(x_t)}}~,
\eeq
so that SUSY will reduce $R_t$ in case of positive $f$.
Using the above formulae, one readily derives the dependence
of ${\rm sin}2\beta$ on the SUSY parameter $f$~\cite{buras-buras}
\beq
        {\rm sin} 2\beta  = \frac{2\bar{\eta}(1-\bar{\rho})}{R_t^2}
        \propto ({\rm const} - \bar{\eta}P_c )~.
\eeq
The first, constant term dominates by a factor of two to three 
the second term, while SUSY modifies the second term \via\
the parameter $\bar{\eta}$ only. Note also that if the charm 
contribution to $\kdo$ mixing were negligeable, SUSY effects 
would be totally absent in  ${\rm sin}2\beta$. It follows from
this that $\gamma$ constitutes the most sensitive UT angle to 
the SUSY contribution $f$.

\subsection{Supersymmetric Fits}

The above SUSY parameterization has been included in 
\CkmFitter\ and constrained fits are performed by setting
$\a=\{f\}$ and $\Mu$ to all other parameters $\ymod$. The resulting 
confidence level bounds for the global CKM fit is shown in 
Fig.~\ref{fig_fsusy} from which one obtains for the present 
data set  the upper limit 
\beq
    f \le 5.2~~~(95\%~\CL)~.
\eeq
Also shown in Fig.~\ref{fig_fsusy} are the improved 
constraints from future measurements assumed to yield
$\dms=(17.0\pm0.9)~{\rm ps}^{-1}$ 
(CDF, D0), $\stb=0.77\pm0.03$ and $\sta=-0.32\pm0.20$ 
(\babar, Belle) (see also the more detailed discussion 
about future precision measurements in Ref.~\cite{heikobcp4}).

% To study the sensitivity of the measured quantities
% to $f$, we calculate numerically the derivatives for 
% $f=0.8$ assuming the above cited future measurement of 
% $\dms$ with an accuracy of $5\%$ and 
% an improved, Gaussian distributed lattice QCD 
% prediction of $\xi=1.16\pm0.03$. This leads to the SUSY sensitivity 
% hierarchy given for the CKM parameters and the UT angles in 
% Table~\ref{tab_SUSYhierarchy}. It turns out that the UT angles
% $\alpha$ and $\gamma$ are more sensitive to SUSY than $\beta$.
% \begin{table}[h]
% \begin{center}
% {\normalsize
% \begin{tabular}{cc}\hline
%   Parameter: $m$      & $\partial m/\partial f$ \\ \hline
%   $\rhobar$           & $0.04$                \\
%   $\etabar$           & $0.07$                \\
%   $J$                 & $-0.56\times10^{-5}$  \\
%   $\sta$              & $-0.51$               \\
%   $\stb$              & $-0.11$               \\
%   $\gamma=\delta$     & $-13^\circ$           \\
%   $|V_{ub}|$          & $-0.36\times10^{-3}$  \\ \hline
% \end{tabular}}
% \caption{\label{tab_SUSYhierarchy} \em
%       SUSY sensitivity hierarchy for $f=0.8$. }
% \end{center}
% \end{table}

%
% ------------------ conclusions --------------------------
%
\section{Conclusions}

We present a new approach to a global fit of the CKM matrix. 
It is denoted \ckmfit\ and is based on frequentist statistics. 
We emphasize the thorough statistical
definition of the method and discuss differences 
from Bayesian statistics~\cite{Achille1,Achille2} 
and from the $95\%$~CL Scan method~\cite{schune}. 
The choice for the fit input parameters,
their values and errors are discussed to some detail;
in cases of doubts we favor the more conservative estimates. 
The CKM analysis is formally subdivided into three distinct phases: 
a metrological phase in which the Standard Model is assumed to be valid 
and confidence levels for the parameters are computed; 
a probing phase addressing the issue of the validity of the 
Standard Model description of data; 
a probing phase for new physics relying on predictive parameterizations. 
For the first phase of the analysis, 
graphical results are displayed in several 
one and two dimensional representations 
and numerical results are given for relevant CKM parameterizations, 
CKM matrix elements, 
Standard Model predictions of rare $K$ and $B$ meson decays, 
and selected theoretical parameters. 
For the parameters related to the CP violating phase of the CKM matrix, 
we find for the different parameterizations 
(the fit includes the present world average of $\stb$ measurements)
\beqns
	J 	&=& 	(1.9\ -\ 3.5)\times10^{-5}~, \\
	\rhobar &=& 	0.04\ -\ 0.37~,		\\
	\etabar	&=& 	0.21\ -\ 0.42~,		\\
	\sta	&=&	-0.95\ -\ 0.33~,		\\
	\stb	&=&	0.47\ -\ 0.81~,		\\
	\alpha	&=&	80^\circ\ -\ 126^\circ~,	\\		
	\beta	&=&	14^\circ\ -\ 27^\circ~,	\\	
	\gamma=\delta	
		&=&	34^\circ\ -\ 82^\circ~,
\eeqns
where the $\ge5\%$~confidence level ranges are quoted.
The second phase of the analysis provides an upper bound for 
the validity of the Standard Model,
\beqns
	{\cal P}({\rm SM})\le71\%~.
\eeqns
A simple predictive supersymmetric extension of the Standard 
Model has been studied in the third analysis phase.

%
% ------------------- Acknowledgements ---------------------------
%

\subsection*{Acknowledgements}

{\small
We gratefully acknowledge the most interesting and helpful
discussions with our colleagues from other active CKM analysis
groups: F.~Parodi, S.~Plaszczynski,
P.~Roudeau, M.H.~Schune and A.~Stocchi, 
who have led pioneering analyses on this subject. We are 
indebted to the advice of D.~Abbaneo,
M.~Artuso, C.~Bernard, I.~Bigi, G.~Boix, A.~Falk,
A.~El Khadra, A.~Kronfeld, Z.~Ligeti,
D.~London, G.~Martinelli, M.~Neubert, J.~Ocariz, H.~Quinn, 
A.I.~Sanda, H.~Wittig 
and many others. We thank our \babar\ collaborators for the many
discussions on this subject and especially 
acknowledge the very fruitful conversations with
G.~Dubois-Felsmann, G.~Hamel de Monchenault, 
H.L.~Lynch and K.~Schubert. Special thanks to
H.~Lynch and K.~Schubert for the careful reading of
this manuscript and their thoughtful and most
constructive comments.
}

%
% ------------ Appendix -----------------------------------------
%

\begin{appendix}

\section{Critical Issues of the Bayesian Approach}
\label{TheBayesianMethodcaughtunder-conservative}

The Bayesian approach injects in the analysis pieces of 
information in the form of probability density distributions 
for the $\yQCD$ parameters. The \ckmfit\ scheme proposed in 
the present paper advocates a non-Bayesian approach because 
most theoretical uncertainties on the $\yQCD$ parameters 
do not stem from statistical fluctuations. The $\yQCD$ 
parameters are not random variables following probability 
density functions: there are poorly known, but fixed, 
parameters. The following examples serve the purpose to 
illustrate in a simplified framework the impact of using the 
Bayesian approach.
\vs
Let $x_i$ denote ${\cal N}$ $\yQCD$ parameters taking their 
values in identical allowed ranges \hbox{$[x_i]=[-\Delta,+\Delta]$.}
These ${\cal N}$ $\yQCD$ parameters are assumed to combine 
to form $T_{\rm P}^{({\cal N})}$, the theoretical prediction 
for an observable, as follows\footnote
{
	This is not an academic exercize: products of $\yQCD$ 
	terms are not rare in the theoretical predictions.
	For example, the $\dmd$ expression 
	(\cf, Eq.~(\ref{eqof-dmd}) of Section~\ref{CPObservablesandMixing})
	involves the product of $\eta_B$, $\fbd^2$ and $B_d$.
	However, the choice of $[x_i]$ ranges containing the 
	origin is not met in practice.
	This choice is made here to highlight the difference between 
	the \ckmfit~scheme and a PDF-based Bayesian approach.
	However, the singularity discussed below is present 
	when Gaussian PDFs are used, as a matter of principle,
	whether or not $[x_i]$ ranges contains the origin.
}
\beq
T_{\rm P}^{({\cal N})}=\prod_{i=1}^{{\cal N}} x_i~.
\eeq
The theoretical prediction $T_{\rm P}^{(\cal N)}$ enters in 
the analysis as an unique $\yQCD$ parameter, which 

\begin{itemize}

\item{} within the \ckmfit\ scheme, is characterized by its 
	allowed range
\beq
[T_{\rm P}^{(\cal N)}]=[-\Delta^{\cal N},+\Delta^{\cal N}]~,
\eeq

\item{} within the Bayesian approach, is characterized by its PDF
\beq
\label{eq_rhoT}
\rho(T) = 
	\intl_{-\infty}^{\infty}
	. . . 
	\intl_{-\infty}^{\infty}
	\prod_{i=1}^{{\cal N}}
	{\rm d}x_i G(x_i)
        \delta(T-T_{\rm P}^{({\cal N})})~, 
\eeq
where we assumed that identical PDFs, $G(x_i)$, were attributed 
to the $x_i$ quantities.
\end{itemize}
Independently of the details of the shape of $G(x)$,
if this PDF is non-zero at the origin,
$\rho(T)$ will exhibit a singularity at $T=0$ with leading term 
$(-\ln T)^{({\cal N}-1)}$. Hence, the larger ${\cal N}$ is,
the more pronounced the peak at the origin of $\rho(T)$ is.
\begin{figure}[t]
  \epsfxsize\mediumfig
  \centerline{\epsffile{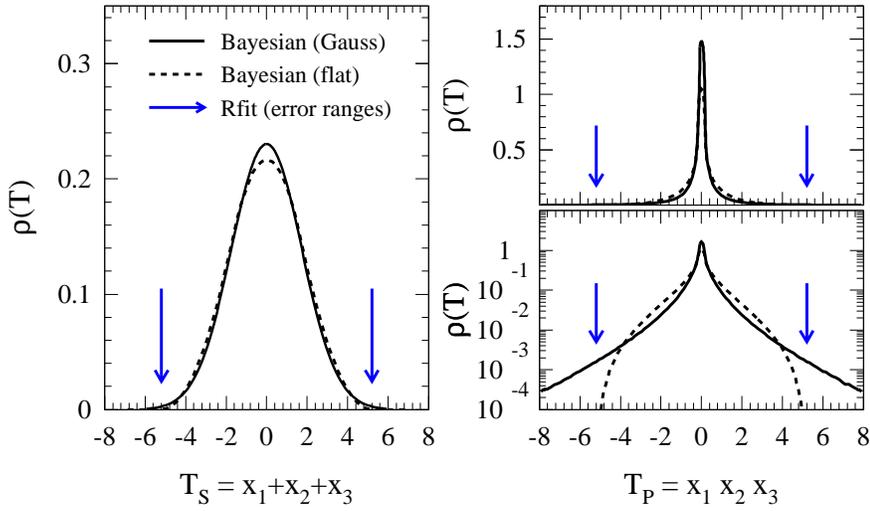}}
  \caption[.]{\label{fig_bayestest}\em
 	Convolution of the sum 
	$T_{\rm S}^{(3)}=x_1+x_2+x_3$ (left hand plot) 
	and the product 
	$T_{\rm P}^{(3)}=x_1x_2x_3$ (right hand plot) of 
	${\cal N}=3$ $\yQCD$ parameters as defined in the text.
	Plotted is
	the PDF $\rho(T)$ obtained from Bayesian statistics 
	using for the $G(x)$ PDF
	a uniform (solid lines, $\Delta =\sqrt{3}$) or  
	a Gaussian (dashed lines, $\sigma=1$)
	distribution.
	Both PDFs $\rho(T)$ of $T_{\rm P}^{(3)}$ present a singularity 
	at the origin which is not shown.
	The \ckmfit\ ranges of $T_{\rm S}^{(3)}$ and $T_{\rm P}^{(3)}$ 
	are indicated by the arrows located in both instances 
	at $\pm 3\sqrt{3}$.}
\end{figure}
If one chooses a uniform distribution $G(x)=1/(2\Delta)$,
that is to say the PDF the ``closest'' to the \ckmfit\ 
assumptions (albeit being a fundamentally different object), 
one obtains a striking result. For ${\cal N}=1$,
the Bayesian approach and the \ckmfit\ scheme are 
equivalent (though not for the associated CLs),
insofar no other variable is involved:
both state that the value of $T_{\rm P}^{(1)}$ is simply within 
the allowed range $[-\Delta,+\Delta]$. However,
though originally no $x_i$ values were favored,
the Bayesian approach departs drastically from \ckmfit\ as soon 
as ${\cal N}\ge 2$: it states that $T_{\rm P}$ is most likely 
close to zero. {\em In effect, when the number of $\yQCD$ parameters 
entering the computation of the theoretical prediction increases,
and hence when our knowledge of the corresponding observable decreases,
the Bayesian approach claims the converse. }
\vs
This is less a consequence of the initial {\em ad hoc} choice 
for the PDF $G(x)$, than a consequence of the inescapable properties 
of Eq.~(\ref{eq_rhoT}), when applied to a product of terms.
For instance, the peaking effect remains present for a sum of terms,
but it is far less pronounced. The above generalities are illustrated, 
for ${\cal N}=3$ and $\Delta=\sqrt{3}$, in Fig.~\ref{fig_bayestest}.
The choice ${\cal N}=3$ is made because the resulting allowed 
ranges for the product $T_{\rm P}^{(3)}=x_1x_2x_3$
and for the sum $T_{\rm S}^{(3)}=x_1+x_2+x_3$ are identical
when using \ckmfit, namely $[-3\Delta,3\Delta]$.
The figure shows the Bayesian PDF
of $T_{\rm S}^{(3)}$ (left hand side) and $T_{\rm P}^{(3)}$ 
(right hand side) assuming for $G(x)$ either a Gaussian distribution 
of standard deviation $\sigma=1$ (the solid lines)
or a uniform distribution in the range $[-\sqrt{3},+\sqrt{3}]$
(the dashed lines) for the three parameters $x=x_i$, $i=1,2,3$. 
The latter case is the closest to the \ckmfit\ 
scheme for which the allowed range is $[T]=[-3\sqrt 3,+3\sqrt 3]$,
indicated by the arrows in Fig.~\ref{fig_bayestest}.
For both, the sum $T_{\rm S}$ and the product $T_{\rm P}$,
the uniform and the Gaussian PDFs yield similar $\rho(T)$ distributions,
because their RMS are chosen identical. The still noticeable 
difference between the two $G(x)$ PDFs is damped away by Eq.~(\ref{eq_rhoT}),
even for such a moderate number like ${\cal N}=3$. As a consequence, 
when following the Bayesian approach, it is not a particularly 
conservative choice to adopt a uniform PDF instead of a Gaussian PDF.
%(even when using ranges scaled to ensure identical RMSs). 
\vs
It is a remarkable difference between the two treatments that, 
whereas within the \ckmfit \ scheme the two theoretical 
predictions $T_{\rm P}$ and $T_{\rm S}$ are genuinely indistinguishable,
they yield sharply different PDFs in the Bayesian approach.
This shows that, even more than the {\it ad hoc} choice made 
to ascribe PDF to the $\yQCD$ parameters, it is the functional 
dependence of the theoretical predictions in these parameters,
and the interplay between the various theoretical predictions 
within the complete CKM analysis, which play the central role.
The attractiveness of the Bayesian approach, offering a 
straightforward procedure for analyses, is fallacious. The 
deep implication of ascribing PDFs to non-random 
variables is hidden inside an apparently innocuous convolution,
the outcome of which reflects, more than anything else,
the mathematical structure of the problem at hand.

%We conclude from this study that the Bayesian approach 
% generates an unavoidable bias,
%totally absent in the \ckmfit\ scheme. 
%The attractiveness of the Bayesian approach, 
%offering a straightforward procedure for the CKM analysis, 
%is fallacious:
%it hides the real-existing difficulties arising from the 
% unthoroughly defined starting 
%conditions of the problem.

\section{Critical Issues of the 95\% CL Scan Method}
\label{DrawbacksoftheSCANmethod}

The 95\%~CL Scan method~\ref{sec_scan}
presents several unwelcome features which are reviewed 
below.

\subparagraph{Drawing features:}

Its final graphical result depends strongly on 
	the choices made for $\CLcut$,
	the selection threshold used to retain {\it models}, 
	and for $\CLcont$,
	the constant defining the contours to be drawn.
	It is not possible to infer how a given drawing is 
	modified by changing these choices.
Regions of weak statistical confidence,
	just barely passing the selection threshold,
	and thus retained,
	bear the same graphical weight than regions of fair \CL.
Similarly, regions of lower \CL, barely not passing the 
	selection threshold, and thus not retained, are ignored.

\subparagraph{Contour features:}

The relevance of drawing the contours is not obvious.
	Whereas a given contour has a clear-cut meaning,
	the message it carries is 
	to show how precise the determination of $\a$ would be,
	in the academic situation 
	where all theoretical uncertainties would supposedly 
	be resolved.
	These contours are not ellipses in general.
	For instance, in the case of the measurement of $\stb$ alone, 
	they are built from straight lines
	(\cf , Section~\ref{sec_Illustrations}).
	They can take complicate shapes,
	depending on the choice made for the $\a$ variables,
	and depending on the presence of sizeable secondary minima.
	In simple situations,
	using the 95\% CL Scan method yields awkward results. 
	For instance,
	in the first example given in Section~\ref{sec_Illustrations},
	where a standard situation is considered, the method 
	would conclude by a set of short intervals in $\a$, 
	each obtained for a fixed $\yQCD$ value
        instead of providing the overall allowed interval,
	\ie, applying the \ckmfit\ scheme.

\subparagraph{Envelope features:}

In the $\a$ space no information can be carried 
	about possibly more or less favored regions:
	a point is either within the envelope of the countours, 
	and thus acceptable, or outside, and thus not acceptable.
The envelope of the contours can be unstable with 
	respect to change of $\CLcut$.
	Indeed, the envelope can be a discontinuous function of 
	this parameter: if lowering $\CLcut$ allows an outsider 
	{\it model} to be selected outside of the envelope, 
	this outsider {\it model} surfaces in the $\a$ space with 
	its full contour and thus lead to an abrupt change of 
	the envelope.
The envelope provided by the 95\% CL Scan method tends to be 
	over-conservative.
	For the example of Section~\ref{sec_Illustrations}, 
	the method yields for 
	the envelope of the $95\%~\CL$ intervals:
	\begin{equation}
		\a_\pm=\a^0\pm 1.96\ {\sigma[\a]\over \sqrt{1-\corcof^2}}
		\left(\corcof+\sqrt{1-\corcof^2}\right)~,
	\end{equation}
	which is always larger than the correct result 
	given by Eq.~(\ref{Correctapm})
	(the two limiting cases $\corcof\rightarrow 0$ and 
	$\corcof\rightarrow 1$ excepted).

\section{Comments on Statistics of Normal Ratios}
\label{sec_RatiosOfBranchingFractions}

\noindent
Ratios of branching fractions of rare $b\rightarrow u$ transitions 
have attracted much attention in recent theoretical and
experimental analyses, by virtue of their potential to constrain 
the unitarity angle $\gamma$ (see Section~\ref{sec_FutureProspects}).
It is shown in this section that the extraction 
of physical observables out of ratios of normally (Gaussian)
distributed quantities, \eg, branching fractions, requires 
some precaution.

\subsection{A Numerical Example}
\label{sec_Example}

The statistical discussion will be accompanied by a numerical
example for rare charmless $B$ decays. It is assumed that 
branching ratios are measured \via\ the relation
${\rm BR}_z = N_z/(\epsilon_z\,\sigma {\cal L})$,
where the reconstruction efficiencies of the two considered 
final states, $z\equiv x,y$, shall be $\epsilon_x=\epsilon_y=13\%$, 
the collected integrated luminosity ${\cal L}=20~{\rm fb}^{-1}$, 
and the production cross section $\sigma=1.1~{\rm nb}$.  
The ``measured'' branching fractions
\beqn
\label{eq_brs}
	{\rm BR}_x &=& (16.0 \pm 2.4)\times 10^{-6}~, \nonumber\\
	{\rm BR}_y &=& ( 8.0 \pm 1.7)\times 10^{-6}~, 
\eeqn
correspond thus to 46 and 23 detected signal events in the 
channels $x$ and $y$, respectively. 
The number of events is sufficient to escape from Poissonian
to normal PDFs for the branching fractions\footnote
{
   As ratios of branching fractions are the subject
   of discussion here, the difference between normal
   and Poissonian statistics is much reduced in the resulting
   PDF, so that the results obtained in this section remain
   approximately valid also for a low number of signal events.
}.
To derive constraints from the above measurements one needs
a predictive theory which, in our example, shall be given by 
the expressions
\beqn
\label{eq_theo}
        {\rm BR}_{x,\,{\rm theo}}(\gamma) &=& |F(0) A_x(\gamma)|^2~, 
	\nonumber\\
        {\rm BR}_{y,\,{\rm theo}}(\gamma) &=& |F(0) A_y(\gamma)|^2~, 
\eeqn
with the ``form factor'' $F(0)$, being identical for both 
final states, and where the ``amplitudes'', which are functions
of the ``angle'' $\gamma$, shall read
\beqn
\label{eq_amp}
     	A_x(\gamma) = 1 + e^{i\gamma} ~, \nonumber\\
     	A_y(\gamma) = 1 - e^{i\gamma} ~.
\eeqn
Our theory depends on the external parameters
$F(0)$ and $\gamma$. The latter provides an example of a quantity
we are interested in ($\a=\{\gamma\}$), while the modulus-squared 
of the first is an example of a $\yQCD$ parameter. In the framework 
of \ckmfit, $| F(0)|^2$ can be eliminated according to 
Section~\ref{sec_MetrologyofRelevantParameters}, using Eq.~(\ref{digest}), 
as illustrated in 
Section~\ref{CombiningStatisticalandSystematicUncertainties}.
It can also be eliminated by taking the ratio 
$R_{\rm theo}(\gamma)\equiv{\rm BR}_{x,\,{\rm theo}}(\gamma)/{\rm BR}_{y,\,{\rm theo}}(\gamma)=|A_x(\gamma)|^2/|A_y(\gamma)|^2$.
Although it appears more straightforward to use the ratio than 
the \ckmfit\ treatment, it is shown below that the converse is true:
using the ratio leads to cumbersome formulae.

\subsection{Probability Density Functions}
\label{sec_PDF}

\begin{figure}[t]
  \epsfxsize\smallfig
  \centerline{\epsffile{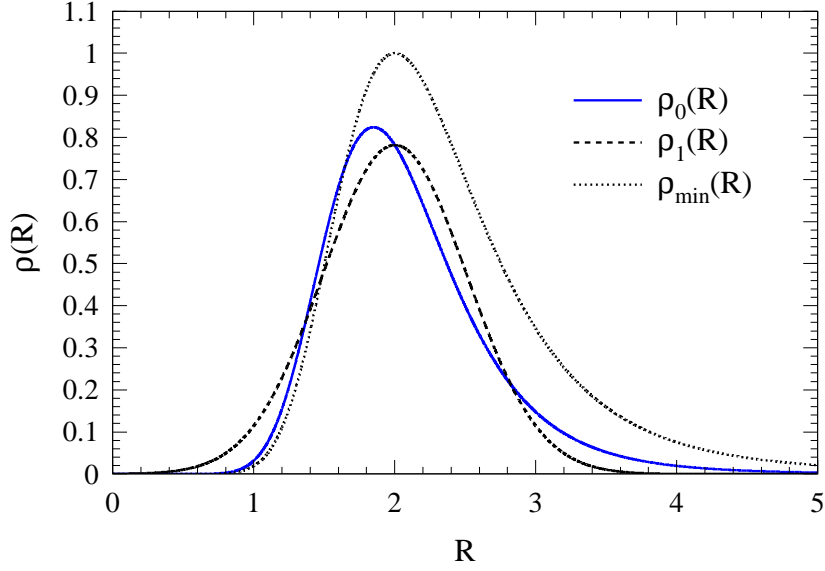}}
  \caption[.]{\label{fig_pdf}\em
	Probability density functions $\rho_0(R)$ (solid line),
	$\rho_1(R)$ (dashed line), and the likelihood
	$\rho_{\rm min}(R)$ (dotted line), where $R=x/y$.
	The numerical values are those chosen for the example (see text).
	The distribution of $\rho_1(R)$ is symmetric by construction.
	The likelihood $\rho_{\rm min}$ is normalized in such a way that its
	maximal value (not its integral) is unity 
	(see Eq.~(\ref{eq_likelmin})).}
\end{figure}

We define a set of two statistically independent measurements
\beqn
	x &=& x_0 \pm \sigma_x~, \\
	y &=& y_0 \pm \sigma_y~, 
\eeqn
obeying normal distributions
\beq
  G(z,z_0,\sigma_z) = \frac{1}{\sqrt{2\pi}\,\sigma_z}\,
			{\rm exp}\left(-\frac{(z-z_0)^2}{2\,{{\sigma_z}}^2}
				 \right)~,
\eeq
with $z\equiv x,y$. For the ratio
\beq
\label{eq_ratio}
	R = \frac{x}{y}~,
\eeq
the marginal PDF obtained from error propagation
\beq
\label{eq_rhoapprox}
	\rho_1(R) \approx 
	G\left(R,\,R_0,\,
		R_0\sqrt{\frac{\sigma_x^2}{x^2}
                               + \frac{\sigma_y^2}{y^2}}
	\right)~,	
\eeq
where $R_0=x_0/y_0$, is only a coarse approximation of 
the exact solution
\beqn
\label{eq_rho}
  \rho_0(R) &=& 
	\intl_{-\infty}^{\infty}\intl_{-\infty}^{\infty}
	\delta\left(R - \frac{x}{y}\right)
	G(x,x_0,\sigma_x) G(y,y_0,\sigma_y)\,dx\,dy~,\nonumber\\
%	&=&
%	\intl_{-\infty}^{\infty}\intl_{-\infty}^{\infty}
%	\delta\left(Ry - x\right)|y|\,
%	G(x,x_0,\sigma_x) G(y,y_0,\sigma_y)\,dx\,dy~,\nonumber\\
%	&=&
%	\intl_{-\infty}^{\infty} 
%	G(Ry,x_0,\sigma_x)G(y,y_0,\sigma_y)\,|y|\,dy~, \nonumber\\
	&=&
	G(0,x_0,\sigma_x) G(0,y_0,\sigma_y)
	\frac{2}{\eta(R)}	
     	\left[ 1 + \sqrt{\pi \xi(R)}
		   \,e^{\xi(R)}
                   \,{\rm Erf}\left(\sqrt{\xi(R)}\right)
	\right]~.
\eeqn
Here, the functions $\xi$ and $\eta$ are defined as
\beqns
  \xi(R) &=& \frac{1}{2\eta(R)}
		\left(R\frac{x_0}{\sigma_x^2}
		      + \frac{y_0}{\sigma_y^2}
		\right)^{\!2}~, \\ 
  \eta(R) &=& \frac{1}{\sigma_x^2}R^2  + \frac{1}{\sigma_y^2}~.
\eeqns
Equations~(\ref{eq_rhoapprox}) and (\ref{eq_rho}) satisfy
\beq
	\intl_{-\infty}^{\infty}\rho_{0,1}(R)\,dR \;=\; 
	2\cdot\!\!\intl_{-\infty}^{R_0}\rho_{0,1}	(R)\,dR \;=\; 1~.
\eeq
The densities $\rho_0(R)$ and $\rho_1(R)$ are plotted
in Fig.~\ref{fig_pdf} for the example of the previous 
section. The maximum of $\rho_0(R)$ is shifted
from the naive expectation, $\Max_R\{\rho_1(R)\}=\rho_1(R_0)$, 
to lower values, while its mean value is larger than the 
naive mean: $\langle\rho_0\rangle=2.10 > \langle\rho_1\rangle=2$.
The root mean square (RMS) of the correct solution
exceeds the one of the naive approximation: 
${\rm RMS}(\rho_0)=0.62> {\rm RMS}(\rho_1)=0.51$, so that 
the use of~(\ref{eq_rhoapprox}) will tend to over-optimistic 
results.

\subsection{Confidence Levels}
\label{sec_CL}

\begin{figure}[t]
  \epsfxsize\smallfig
  \centerline{\epsffile{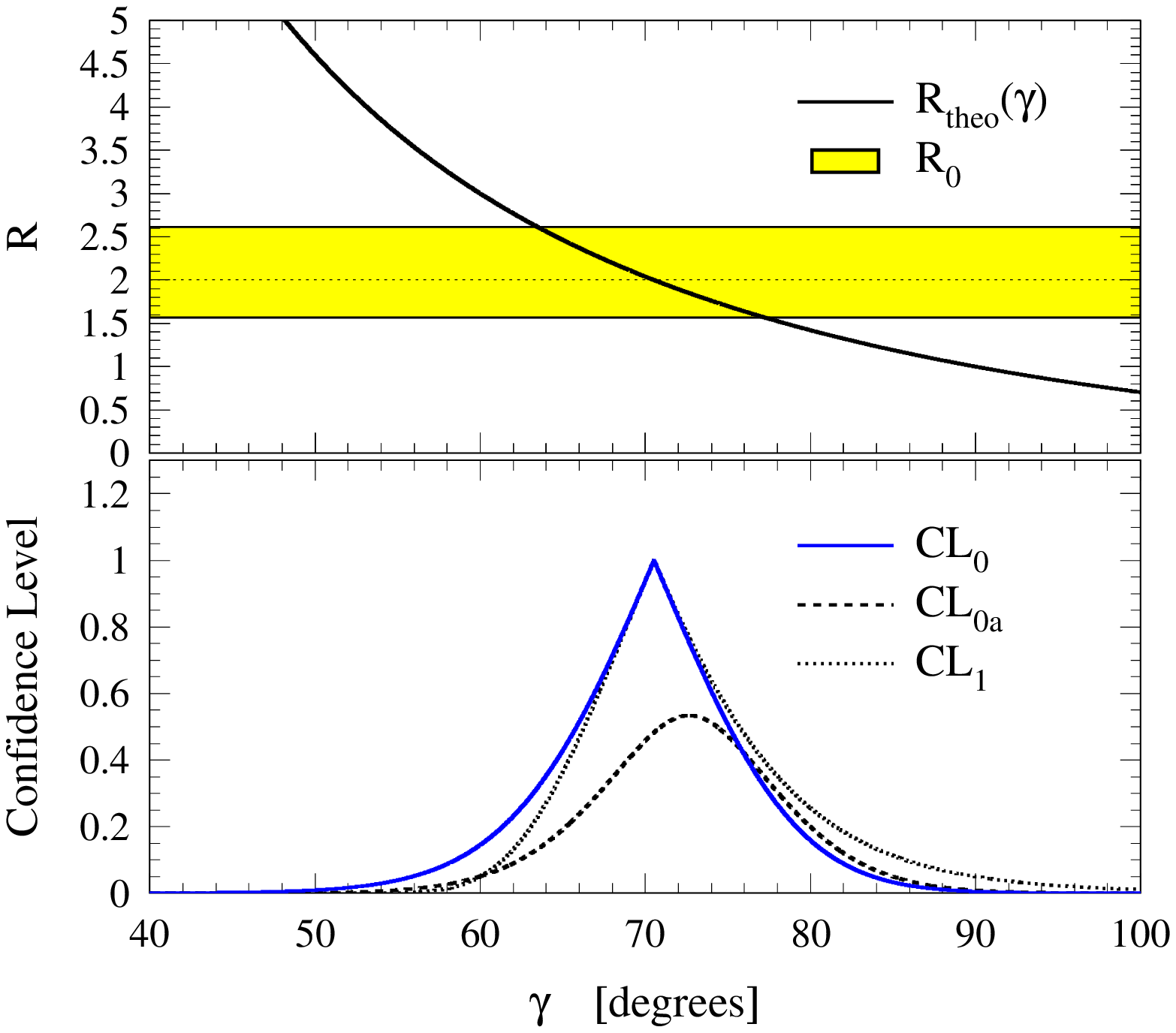}}
  \caption[.]{\label{fig_cl}\em
	\underline{Upper plot:} ``measured'' ratio of branching fractions
	within its asymmetric, one standard deviation error band
	(grey zone), corresponding to the example~(\ref{eq_brs}), 
	and the	``theoretical prediction''~(\ref{eq_theo}) 
	(solid curve) versus $\gamma$. 
	\underline{Lower plot:} confidence levels obtained
	from Eqs.~(\ref{eq_chi21}) (dotted line), (\ref{eq_chi20a})
	(dashed line) (both using Eq.~(\ref{eq_cl1})), and 
	(\ref{eq_cl0}) (solid line), the latter being identical
	to Eq.~(\ref{eq_chi2mincl}).}
\end{figure}

In our example, the analysis of the branching fractions, or 
of their ratio, aims at constraining the physical quantity
$\gamma$. Figure~\ref{fig_cl} shows in its upper plot
the theoretical ratio $R_{\rm theo}(\gamma)$
versus $\gamma$, together with the asymmetric, one standard 
deviation error band (using Eq.~(\ref{eq_cl0})) 
of the ``experimental'' value.
Without loss of generality, we may assume in the following
that, apart from $\gamma$, the theoretical 
predictions~(\ref{eq_theo}) and (\ref{eq_amp}) suffer 
only from the theoretical uncertainty on $F(0)$, 
which shall be maximally unknown:
\ie, we do not assume that $|F(0)|^2$ takes its values
within an {\it a priori} finite range.
An approach to obtain confidence levels for $\gamma$,
that is often met in the literature, is to eliminate the 
factor $F(0)$ by using the ratio~(\ref{eq_ratio}) and 
defining the $\chi^2$ as
\beq
\label{eq_chi21}
	\chi^2_1(\gamma) 
	= \left(\frac{R_0 - R_{\rm theo}(\gamma)}
			    {\sigma_R}
		\right)^{\!2}~,
\eeq
where $\sigma_R$ is the RMS of $\rho_1$ (see 
Eq.~(\ref{eq_rhoapprox})). The associated CL
is then given by the cumulative distribution
of a normal PDF
\beq
\label{eq_cl1}
	{\rm CL_1}(\gamma) 
	= {\rm Erfc}\left(\sqrt{\chi^2_1(\gamma)/2}\right)~,
\eeq
shown versus $\gamma$ by the dotted line in the lower plot 
of Fig.~\ref{fig_cl}.
\vs
Yet, we have seen that the PDF $\rho_1$ is only an
approximation of the correct PDF $\rho_0$ and hence,
one is tempted to build a more accurate $\chi^2$ by means of 
\beq
\label{eq_chi20a}
	\chi^2_{0a}(\gamma)  
	= - 2\,{\rm ln}\rho_0(R_{\rm theo}(\gamma))~,
\eeq
where one may wish to build the corresponding CL, 
${\rm CL}_{0a}(\gamma)$, \via\ Eq.~(\ref{eq_cl1}). However, this 
again constitutes an approximation since the error function
assumes a normal PDF. The dotted curve in the lower 
plot of Fig.~\ref{fig_cl} shows the CL corresponding 
to Eqs.~(\ref{eq_chi20a}) and (\ref{eq_cl1}) as a function
of $\gamma$. Indeed, the correct CL is obtained \via\ an 
asymmetric integration of the PDF $\rho_0$:
\beq
\label{eq_cl0}
	{\rm CL_0}(\gamma)  
	= \left\{\begin{array}{lll}
		2\intl_{-\infty}^{R(\gamma) } 
			\rho_0(R^\prime)\,dR^\prime
		&,& \forall R(\gamma)\le R_0 \\
		2\intl_{R(\gamma) }^{\infty} 
			\rho_0(R^\prime)\,dR^\prime
		&,& \forall R(\gamma)> R_0 
		\end{array}
	\right.
\eeq
plotted versus $\gamma$ 
as solid line in the lower plot of Fig.~\ref{fig_cl}.
\vs
The complication of Eq.~(\ref{eq_cl0}) can be readily 
circumvented when not explicitly using the ratio, but
keeping the original branching fractions 
in the definition of the $\chi^2$:
\beq
\label{eq_chi2right}
	\chi^2(\gamma,\yQCD)
	= \left(\frac{x - \yQCD\cdot | A_x(\gamma)|^2 }{\sigma_x}
	  \right)^{\!2}
	+ \left(\frac{y - \yQCD\cdot | A_y(\gamma)|^2 }{\sigma_y}
	  \right)^{\!2}~.
\eeq
Applying Eq.~(\ref{digest}), hence eliminating $\yQCD$, 
yields the minimum 
%\footnote
%\{
%\  For branching fractions of $N$ distinct
%\  final states, the $\chi^2$:
%\  $
%\  \nonumber
%\    \chi^2(\gamma,C)=\sum_{i=1}^{N}\left[(x(i)-C x_{\rm theo}(\gamma,i))
%\    /\sigma_x(i)\right]^2~,
%\  $
%\  has a minimum for
%\  $
%\	C = \frac{\sum_{i=1}^{N}x(i)x_{\rm theo}
%\			(\gamma,i)/\sigma_x^2(i)}
%\                 {\sum_{i=1}^{N}\left[x_{\rm theo}
%\			(\gamma,i)/\sigma_x(i)\right]^2}~.
%\  $
%} 
\beq
\label{eq_chi2min}
    	\chi^2_{\min ;\yQCD}(\gamma)
	= \frac{(x - y R_{\rm theo}(\gamma))^2}
               {\sigma_x^2 + \sigma_y^2  R^2_{\rm theo}(\gamma)}~,
\eeq
which, by construction, only depends on $R_{\rm theo}(\gamma)$. 
The \CL~ obtained using Eq.~(\ref{eq_cl1})
\beq
\label{eq_chi2mincl}
	{\rm CL_{\min}}(\gamma) 
	= {\rm Erfc}\left(\sqrt{\chi^2_{\min ;\yQCD}(\gamma)/2}\right)~,
\eeq
is identical to the one obtained using Eq.~(\ref{eq_cl0}).
The likelihood corresponding to Eq.~(\ref{eq_chi2min}) 
\beq
\label{eq_likelmin}
	\rho_{\rm min}(R_{\rm theo}(\gamma)) 
	= e^{-\frac{1}{2}\chi^2_{\min ;\yQCD}(\gamma)}~,
\eeq
is shown as the dotted curve in Fig.~\ref{fig_pdf}.
The likelihood is equal to unity for $R_{\rm theo}(\gamma)=x/y$. 
It is worth emphasizing that $\rho_{\rm min}(R)$,
which yields the correct \CL, when using Eq.~(\ref{eq_cl1}), 
is not identical to $\rho_0(R)$, although the latter is the 
correct PDF, which yields the correct \CL, when {\it not} using 
Eq.~(\ref{eq_cl1}), but Eq.~(\ref{eq_cl0}) instead.

\end{appendix}

\end{document}